\documentclass[12pt,preprint]{aastex}

\def\g21{G21.5$-$0.9}
\def\j1833{PSR~J1833$-$1034}
\def\chan{\textit{Chandra}}
\def\xmm{\textit{XMM-Newton}}
\def\xspec{\textit{XSPEC}}

\def\nHunits{$\times$~$10^{22}$~cm$^{-2}$}
\def\fluxunits{erg~cm$^{-2}$~s$^{-1}$}
\def\lumunits{erg~s$^{-1}$}
\def\pownormunits{photons~keV$^{-1}$~cm$^{-2}$~s$^{-1}$}

\shorttitle{\chan Observations of G21.5-0.9}

\begin{document}

\title{The Plerionic Supernova Remnant G21.5$-$0.9 Powered by  PSR~J1833$-$1034: 
 New Spectroscopic and Imaging Results Revealed with the \chan\ X-ray Observatory}
\author{Heather Matheson\altaffilmark{1} and Samar Safi-Harb\altaffilmark{2}} 
\affil{Department of Physics and Astronomy,
University of Manitoba, \\ Winnipeg, Manitoba, R3T 2N2, CANADA}
\altaffiltext{1}{matheson@physics.umanitoba.ca}
\altaffiltext{2}{Canada Research Chair, samar@physics.umanitoba.ca}

\slugcomment{Accepted by ApJ}

\begin{abstract}
In 1999, the \chan\ X-ray Observatory revealed a
150\arcsec--radius halo 
surrounding the 40\arcsec-radius pulsar wind nebula (PWN) \g21. 
  A 2005 imaging study of \g21\ showed
that the halo is
limb-brightened, and suggested this feature is a candidate for the long-sought supernova 
remnant (SNR) shell.  We present a spectral analysis of SNR  \g21,
using the longest effective observation to date (578.6 ks
with the Advanced CCD Imaging Spectrometer (ACIS), 
278.4 ks with the High-Resolution Camera (HRC)) to study unresolved questions about the spectral nature of remnant features, such as the limb-brightening of the X-ray halo
and the bright knot in the northern part of the halo.  
The \chan\ analysis favours the non-thermal interpretation of the limb. Its spectrum is well fit with a power-law model  with a photon index $\Gamma$~=~2.13 (1.94--2.33) and a luminosity
of $L_x$ (0.5--8 keV) = (2.3$\pm$0.6)$\times$10$^{33}$~erg~s$^{-1}$ (at an assumed distance of 5.0 kpc).
An $srcut$ model was also used to fit the spectrum between the radio and X-ray energies. While the absence of
a shell in the radio still prohibits constraining the spectrum at radio wavelengths, we assume a range of spectral indices 
to infer the 1~GHz flux density and the roll-off frequency of the synchrotron spectrum in X-rays,
and find that the maximum energy to which electrons are accelerated at the shock
ranges from $\sim$60--130 TeV ($B$/10$\mu$G)$^{-1/2}$, where $B$ is the magnetic field in units of $\mu$G.
For the northern knot, we constrain previous models and find that a two-component power-law (or $srcut$) + $pshock$ model provides an adequate fit,
with the $pshock$ model requiring a very low ionization timescale and solar abundances for Mg and Si.
Our spectroscopic study  of \j1833,
the highly energetic pulsar powering \g21,
shows that its spectrum is dominated by hard non-thermal X-ray emission with some evidence of a thermal component
that represents $\sim$9\% of the observed non-thermal emission and that
suggests non-standard rapid cooling of the neutron star.
Finally, the ACIS and HRC-I images 
provide the first evidence for 
variability in the PWN, a property observed in other PWNe such as the Crab and Vela.
\end{abstract}


\section{INTRODUCTION}
\label{section:intro}
A pulsar wind nebula (PWN, also `plerion') is a filled-centre
supernova remnant (SNR) which possesses a flat
radio spectrum, centrally peaked radio and X-ray emission,
highly polarized radio emission, and is powered by a rapidly rotating
neutron star.
\g21\ is a particularly interesting PWN since it 
was the first PWN  discovered to be surrounded by an X-ray
halo~\citep{slane2000, warwick2001,
  safi-harb2001} with limb-brightening~\citep{matheson2005, bocchino2005b}.
  
In the 1970s,~\citet{becker1976}\ and~\citet{wilson1976}\ 
mapped \g21\ in the radio and found an 
elliptical brightness distribution with peak brightness near the 
geometric centre of the remnant, similar to the Crab Nebula.
\citet{furst1988}\ performed 22.3~GHz 
observations and found axisymmetric filaments and suggested they resulted 
from a two-sided collimated outflow of particles in precession from a 
central pulsar.  They also found that the X-ray maximum was located in a 
minimum in the small-scale radio emission, suggesting the two types of 
emission originate from the pulsar in a different way.

\begin{figure}
\begin{center}
\plotone{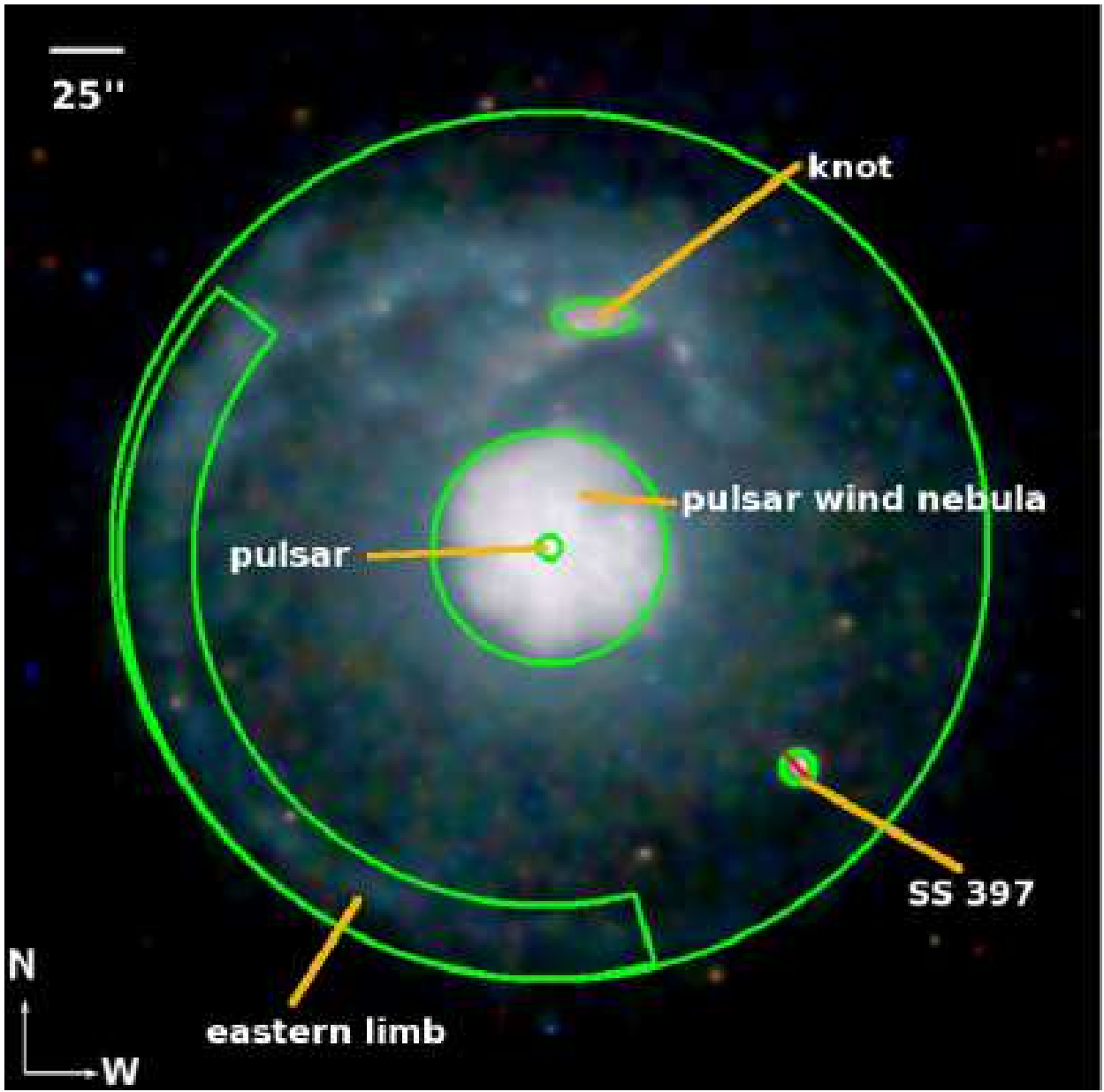}
\caption[Regions used in spectral analysis.]{Regions used in the
  spectral analysis of \g21, with the northern knot and eastern limb highlighted.
  The point source SS~397 was
  excluded from the data prior to extracting spectra.  The image is 377\arcsec~$\times$~377\arcsec\ and is centred on the location of the X-ray peak ($\alpha$(2000) = 18$^{\mathrm h}$33$^{\mathrm m}$33$\fs$54,
$\delta$(2000) = $-$10\arcdeg34\arcmin07\farcs6), which is the pulsar location.  At
  a distance of 5~kpc, the PWN of \g21\ has a diameter of
  1.9~pc and the halo has a diameter of 7.3~pc. 
  Adapted from Matheson \& Safi-Harb~(2005).}
\label{figure:regionpic}
\end{center}
\end{figure}

In 1999, \chan\ revealed a 150\arcsec-radius halo at X-ray wavelengths 
surrounding the 40\arcsec-radius PWN~\citep{slane2000, safi-harb2001}.  
The spectrum of the PWN was well described by a power-law model with
the photon index steepening (increasing) away from the central source.
The X-ray halo was also detected with \xmm\ and found to have a non-thermal
spectrum that is consistent with the \chan\ study (Warwick et al. 2001).
The halo's circular
symmetry, lack of limb brightening, and non-thermal spectrum led 
to the earlier suggestion that the halo was an extension of the PWN rather than a
shell formed by the blast wave of the supernova explosion.  However, the lack of radio emission
from the halo would make this interpretation problematic as the size of the PWN
in X-rays would then exceed the radio size by a factor of $\sim$4, a characteristic
that is at odds with PWN morphology.
Safi-Harb et al. (2001) noted that, since \g21\ is 
bright and heavily absorbed, a significant portion of the emitted X-rays 
could interact with dust and be scattered, forming a halo around the 
bright core.  Bandiera \& Bocchino (2004) modelled the halo 
as an effect of dust scattering in the foreground medium.  While dust 
scattering could not explain the bright knots observed in the north of the remnant
(referred to as the northern knot or `North Spur', see Figure~\ref{figure:regionpic}), 
 a good fit can be obtained to the remainder of the halo.
Matheson \& Safi-Harb (2005) showed \chan\ 
imaging results (modified version in Figure~\ref{figure:regionpic}) 
revealing a candidate for the shell of the supernova remnant \g21\ in the 
form of limb-brightening at the eastern boundary of the X-ray halo.  
They also revealed previously unseen detail in the PWN, and showed that
the power-law photon index increases from $\Gamma$ = 1.61~$\pm$~0.04 (at the centre of the PWN)
to $\Gamma$~=~2.15~$\pm$~0.13 (at a radius 
of 40\arcsec, the edge of the bright PWN), then remains flat at 
$\Gamma$~$\sim$~2.4 inside the halo between a radius of 40\arcsec\ and 150\arcsec\
($N_{\mathrm H}$~=~2.2~\nHunits).
Bocchino et al. (2005) used \xmm\ and \chan\ data to study the X-ray
halo of \g21.  They interpret the diffuse halo as a dust scattering halo, 
 the eastern limb as the location of particle acceleration at the forward shock, and  the
brightest northern knot (or `North Spur') as a possible knot of ejecta in adiabatic expansion.
However, the spectral nature of the northern knot and limb are still under debate.
In particular, two solutions with different spectral properties were found for the northern knot.
As well, the \chan\ data used in Bocchino et al. (2005) had an effective exposure time
that is $\sim$2.5 times smaller than in our spectroscopic study presented here, 
prohibiting a clear detection of the limb-brightened
feature.
This will be addressed with our study presented here (see \S3 and 5).

At higher energies, Bird et al. (2004) identified \g21\ in a soft $\gamma$-ray Galactic
plane survey with \textit{INTEGRAL}.  The flux was measured to be 0.28
$\pm$ 0.02 ct~s$^{-1}$ in 20--40 keV and 0.26 $\pm$ 0.03 ct~s$^{-1}$ in
40--100 keV.   De Rosa et al. (2009) detected \g21\ in the 17--200 keV
range with \textit{INTEGRAL}, showing that the main contribution to
hard X-ray emission is from the PWN, with \j1833\ dominant above
200~keV.  They also present \textit{HESS} observations showing that
\j1833\ is a bright gamma-ray emitter with a 1--10~TeV flux
approximately 2\% that of the Crab.  

In the radio, Gupta et al. (2005) and Camilo et 
al. (2006) independently discovered the long sought pulsar 
(\j1833) associated with \g21.  They found $P$~=~61.86~ms,
$\dot{P}$~=~2.0~$\times$~$10^{-13}$, a surface magnetic field of 
$B$~=~3.6~$\times$~$10^{12}$~G, a characteristic age of 4.8~kyr,
and a spin-down luminosity of 
$\dot{E}$~=~3.3~$\times$~$10^{37}$~erg~s$^{-1}$, making \j1833\ a 
highly energetic Galactic pulsar.  Bietenholz \& Bartel (2008), using VLA observations from
1991 and 2006, derived an expansion speed of 910 $\pm$ 160 km~s$^{-1}$
with respect to the centre of the nebula and estimated the age of
G21.5$-$0.9 as 870$^{+200}_{-150}$~yr, making it one of the youngest known Galactic
PWNe.

Camilo et al. (2006) used a compilation of studies to conclude 
that the best estimate of the distance to \g21\ was 4.7$\pm$0.4~kpc. 
By comparing HI spectra with $^{13}$CO emission spectra, Tian \& Leahy (2008)
also find a kinematic distance for G21.5$-$0.9 of $\sim$4.8 kpc.
In this paper we will adopt 5~kpc as the distance to \g21.

Here we present a detailed spectroscopic analysis (Section~\ref{section:spec}) using the longest effective
exposure time with  \chan\  to date (an extension of the preliminary study in Matheson \&
Safi-Harb (2005) which focused on imaging of \g21) and discuss the 
open questions regarding the spectral nature of the northern knot and the limb-brightening to the east of the SNR.  
We also present the first evidence for thermal emission from the pulsar (Section~\ref{section:psrspec}) and for variability in
the PWN (Section~\ref{section:variable}).

\section{OBSERVATIONS}
\label{section:obs}
\g21\ was chosen as a calibration target for the \chan\ X-ray
Observatory and, as a result, \g21\ has been frequently observed using
the Advanced CCD Imaging Spectrometer (ACIS) and the High Resolution
Camera (HRC).  
Bocchino et al. (2005) made use of 21 \chan\ observations (196.5 ks) to
search for thermal emission
from the northern knots of \g21.  However, multiple \chan\
observations were combined into one dataset prior to extracting the
spectra.  The \chan\ X-ray Center (CXC)
recommends\footnote{``The merged event list should not be used for
  spectral analysis, since it does not contain sufficient information
  to generate correct response files.'' (\url{http://cxc.harvard.edu/ciao/threads/combine/})}
generating spectra for
each observation and analyzing them simultaneously, as we have done in
the following analysis. 
We here made use of a larger set of 65 ACIS observations (1999 August~--~2006 
February, 578.6 ks, Table~\ref{table:exptime}) 
in the variability study (\S4).  Since data observed at 
-100$^{\circ}$C can not
be corrected for charge transfer inefficiency (CTI), we made use of 56 
ACIS observations (480.2 ks) in the spectroscopic analysis presented here,
a factor of $\sim$2.5 deeper than the study by Bocchino et al. (2005) and 
the deepest spectroscopic study to date.  The 
variability study  also made use of 15 HRC-I observations (1999 July~--~2006 February, 278.4 ks,
Table~\ref{table:exptime}).  
For the imaging studies, observations were reprojected to align the
images prior to merging.
The data processing was performed using the CIAO\footnote{Chandra
  Interactive Analysis of Observations (CIAO),
  \url{http://cxc.harvard.edu/ciao/}} software package,
  and the spectral analysis was performed using \xspec
  \footnote{X-ray Spectral Fitting Package (XSPEC),
  \url{http://xspec.gsfc.nasa.gov/}}.

\subsection{Structure of \g21}

The combined \chan\ image in Figure~\ref{figure:regionpic} shows the structure of \g21.  Located at
$\alpha$(2000) = 18$^{\mathrm h}$33$^{\mathrm m}$33$\fs$54,
$\delta$(2000) = $-$10\arcdeg34\arcmin07\farcs6 is a point source
corresponding to the location of the pulsar \j1833.  The PWN is
approximately 40\arcsec\ in radius and is seen to have indentations
in the northwest and southeast.  As well, many filamentary structures
can be seen in the PWN.  The X-ray halo extends to a radius of
153\arcsec\ with limb-brightening observed along the eastern edge. 
The foreground source SS~397 in the southwest portion of the halo was
removed from the data prior to any spectral analysis.  The northern
portion of the halo is dominated by bright knots which appear to merge
with the limb in the northeast.  The brightest knot is located north
and slightly to the west of the PWN.  As mentioned earlier, this will be referred to as the `northern knot'
throughout the paper, and corresponds to the `North Spur' studied by Bocchino et al. (2005).
The open questions on these regions are studied
further in Section~\ref{section:spec}.

Figure~\ref{figure:xrayradio} zooms on the PWN and compares the \chan\
data to radio data.   Despite targeted searches for the SNR shell
in the radio, the X-ray limb and halo have not
yet been detected at radio wavelengths, except for the northern knot
(Bietenholz et al. 2010).  In Figure~\ref{figure:xrayradio}a, the contours
from the X-ray image are
overlaid on radio data taken with the Nobeyama Millimeter-Wave Array 
at 22.3 GHz (F\"urst et al. 1988). 
Figure~\ref{figure:xrayradio}b combines the 0.2--10.0 ACIS X-ray data (blue)
with the 22.3 GHz radio data (red) to show the similarity of the structure at
both wavelengths.  In Figure~\ref{figure:xrayradio}c, 4.75~GHz radio data from the VLA (see Bietenholz
\& Bartel (2008) for details) is coloured red and the 0.2--10.0 keV X-ray
data is again coloured blue.  The X-ray and radio structure are remarkably
similar, with the radio following the shape of the X-ray PWN along the
edges of the PWN, including indentations in the northwest
and southeast.   The PWN appears slightly larger at 4.75~GHz than at 22.3~GHz,
but the images at 22.3~GHz and 0.2--10 keV are comparable in size,
which indicates a small magnetic field in the PWN. Indeed, a low magnetic field
estimate ($B$$\sim$25~$\mu$Gauss, see also \S5.2) has been inferred from modeling 
G21.5--0.9 (de Jager et al. 2008).
The most prominent difference between the radio and X-ray
emission is in the centre of the PWN.  The pulsar is seen in a location
that peaks in X-rays but has a minimum in radio emission.
The X-ray emission traces the particles freshly injected by the pulsar,
whereas the radio emission traces the older population characterized
by a much larger synchrotron lifetime.
The location of minimum radio emission could be indicative of a magnetic
field direction along the line of sight, since the synchrotron emissivity
scales as $\varpropto$ $B_{\perp}^{\alpha+1}$ (where $B_{\perp}$ is the magnetic field's component
that is perpendicular to the line of sight, and  $\alpha$ is the spectral index).
This interpretation is consistent with the radial magnetic field distribution inferred
from polarization studies (F\"urst et al. 1988).

\section{SPECTROSCOPY}
\label{section:spec}

\subsection{Spectra Creation}
\label{section:createspec}
For each region shown in Figure~\ref{figure:regionpic}, weighted spectra 
were extracted for every observation using the CIAO tools \textit{dmcopy} and \textit{dmextract}.
The background chosen for each region is described in the section
specific to that region. 

To compensate for the effects of cosmic radiation damage, a charge 
transfer
inefficiency (CTI) correction was applied to the ACIS data (CIAO CTI
correction\footnote{\url{http://cxc.harvard.edu/ciao/why/cti.html}}
for ACIS-I, -120$^{\circ}$C data, Penn State CTI
correction\footnote{\url{http://www.astro.psu.edu/users/townsley/cti/}}
\citep{townsley2000} for ACIS-I,
-110$^{\circ}$C and ACIS-S, -110$^{\circ}$C and -120$^{\circ}$C data). 
The -100$^{\circ}$C data cannot be corrected for CTI 
and was therefore omitted from the spectral analysis.  

For the observations that were CTI corrected using the
Penn State CTI corrector, 
RMF files were provided with the corrector.
For the observations that were CTI corrected with CIAO 
(\textit{acis\_process\_events}), the tool \textit{mkacisrmf} was 
used to create weighted RMF files.

The regions used in the following spectral analysis are listed in
Table~\ref{table:regions} and shown in Figure~\ref{figure:regionpic}. 
The spectra were individually binned using the
FTOOL\footnote{\url{http://heasarc.gsfc.nasa.gov/ftools/}
  \citep{blackburn1995}} \textit{grppha} to improve the 
signal-to-noise ratio.  The minimum number of counts per bin for the
various regions are shown in 
Tables~\ref{table:fitresults1} and~\ref{table:fitresults2}. 
All spectral models used contain a
component which accounts for absorption along the line of sight ($wabs$ in $XSPEC$).  A 
power-law was used to model synchrotron emission
from high energy electrons in a magnetic field.  A blackbody model was used
to study any thermal emission from the neutron star directly.  A
\textit{pshock} model (a plane-parallel non-equilibrium ionization model with different ionization ages and a constant electron temperature,
Borkowski et al. 2001) was used to search for thermal 
emission from shock-heated ejecta or interstellar matter.  The
\textit{pshock} model is characterized by the ionization timescale,
$\tau$~=~$n_{e}t$, where $n_e$ is the post-shock electron density and $t$ is the
time since the passage of the shock.  
The \textit{vpshock} model is a \textit{pshock} model which also accounts for variable abundances of metals.  The \textit{srcut} model
was used to model the non-thermal component of the SNR associated with
electrons accelerated by the SNR shock (see e.g. Reynolds \& Keohane 1999). 
 In order to use complete
  response information included with each observation, 
all spectra for a particular region were fit 
simultaneously.  All errors are reported to the 90\% confidence level.

\subsection{\j1833}
\label{section:psrspec}
Only observations with an
off-axis angle less than 3\arcmin\ were used to study the compact
source (Table~\ref{table:exptime}).  For an off-axis angle of
3\arcmin, at 1.5 keV, 90\% of the energy of a point source
is contained within 1.5\arcsec\ (within 2.5\arcsec\ at 6.4 keV).  Therefore a circular region with 2\arcsec\ radius, centered
at $\alpha$(2000) = 18$^{\mathrm h}$33$^{\mathrm m}$33$\fs$54,
$\delta$(2000) = $-$10\arcdeg34\arcmin07\farcs6, was defined as the extraction region for each dataset.
The spectra were each grouped to have
a minimum of 50 counts per bin and cover the energy range 0.5--8.0~keV.   To remove contamination
from the PWN, the background chosen was an
annulus centred on \j1833\ with radius 2\arcsec--4\arcsec.  

The best fit using an absorbed power-law model yields
a column density of $N_{\mathrm H}$ = 2.24$^{+0.09}_{-0.10}$~\nHunits, a
photon index of $\Gamma$~=~1.47$^{+0.05}_{-0.06}$, and an observed
flux in the 0.5--8.0 keV band of (3.2$\pm$0.3) $\times$ 10$^{-12}$~erg~cm$^{-2}$~s$^{-1}$
($\chi^2$~=~1287.9 and $\nu$~=~1047 degrees of freedom,
see Table~\ref{table:fitresults1} for a list of parameters). This column
density is consistent with previous work on \g21\ and with our global fit
to the PWN, and will be used in
the remainder of the spectral analysis. 
Fitting with an absorbed blackbody alone 
gives a low column density of $N_{\mathrm
  H}$~=~0.78$^{+0.05}_{-0.06}$~\nHunits\ and a temperature 
of $kT_{bb}$~=~1.34$^{+0.03}_{-0.03}$~keV = 1.55$^{+0.03}_{-0.03}$~$\times$~10$^7$~K
  (reduced $\chi^2_{\nu}$~=~1.51 and $\nu$~=~1047).  
Freezing the column density to the acceptable value of 2.24~\nHunits\ gives an unacceptable fit with 
a temperature of $kT_{bb}$~$\sim$~1.0~keV and $\chi^2_{\nu}$~$\sim$~2.6 for a single component blackbody.
As expected, the blackbody fit alone is poor at low and high energies, indicating the
need for a non-thermal component. 

To test for a combination of thermal + magnetospheric emission, we
freeze the hydrogen column density to $N_{\mathrm H}$~=~2.24~\nHunits\
and fit
the pulsar's emission with an absorbed \textit{power-law+blackbody} model 
(Figure~\ref{figure:psrpower}).  The
best fit ($\chi^2$=1271.6, $\nu$=1046) was obtained for a
hard photon index of $\Gamma$~=~1.14$^{+0.05}_{-0.07}$  and a temperature of
$kT_{bb}$~=~0.52$^{+0.03}_{-0.04}$~keV~=~6.0$^{+0.3}_{-0.5}$~$\times$~10$^{6}$~K.
This is an improved fit over the power-law fit alone,
with an F-test probability of $\sim$2.6$\times$10$^{-4}$.
Additional observations of the pulsar will help confirm or constrain this emission.

The additional blackbody component suggests the first evidence for thermal emission from the pulsar,
a result that is observed in other young neutron stars.
  The observed thermal flux (2.6~$\pm$~1.0~$\times$~10$^{-13}$~\fluxunits) was found
to be $\sim$9\%
of the non-thermal flux (3.0~$\pm$~0.6~$\times$~10$^{-12}$~\fluxunits)
in the 0.5--8.0~keV range.  
Converting the unabsorbed thermal flux
(7.4~$\pm$~3.0~$\times$~10$^{-13}$~\fluxunits, $\sim$16\% of
unabsorbed non-thermal flux)
to luminosity, and assuming a blackbody, we can derive the size of the emitting area
using the blackbody formula: $L$~=~4$\pi$$R^2$$\sigma$$T_{bb}^4$~=~4$\pi$$D^2$$F$
(where $\sigma$ is the Stephan-Boltzmann constant and $F$ is the observed flux).
   The radius of the emitting region is found to be $R = 0.49 \pm 0.15$~km for a
temperature of 0.52~keV.
This is an unreasonably small radius for the size of a neutron star, suggesting that we may be 
instead be observing emission from a small ``hot spot'' on the surface of the neutron star. 
Detecting the X-ray pulsations is needed to confirm this interpretation\footnote{Past searches for X-ray pulsations
have only led to an upper limit on the pulsed fraction (Camilo et al. 2006, La Palombara \& Mereghetti 2004, Safi-Harb et al. 2001)}.
 Alternatively, constraining the radius of the emitting region to
12~km and assuming a distance of 5~kpc to \g21, we use the
\textit{bbodyrad} model to then calculate the surface temperature of the
neutron star, assuming the thermal component is from the entire
surface.  In this case, the power-law model parameters are the same as those
for the power-law model alone (with a photon index $\Gamma$=1.47) but we derive a lower temperature 
$kT$ = 0.11 ($<$0.14) keV, corresponding to an
effective temperature of 1.28~($<$1.57)~$\times$~10$^6$~K which is closer to what
has been observed in other young neutron stars. This result is further discussed in \S5.2.

\subsection{X-ray Halo (r = 45\arcsec--153\arcsec)}
\label{section:halospec}
Since studying the emission from the limb requires subtracting the emission from the halo (to remove the contamination by the dust scattering halo),
we here briefly summarize our spectral fit to the X-ray halo.

The X-ray halo of \g21\ is visible in Figure~\ref{figure:regionpic} at a radius of 45\arcsec~--~153\arcsec\ from the location of \j1833.  Since the northern half of the halo is dominated by emission from the
knots, the southern half of the X-ray halo was selected for study.  Emission from the
eastern limb (defined between 125\arcsec\ and 153\arcsec, see \S3.4)
and SS~397 was removed from the data prior
to extracting the spectra.  The background used was the southern half
of an annulus with radius 153\arcsec~--~175\arcsec.   The halo was fit with a power-law with the column density fixed at the best fit
value from \j1833 ($N_{\mathrm H}$ = 2.24$^{+0.09}_{-0.10}$~\nHunits).  We find a
photon index of
$\Gamma$~=~2.50$^{+0.05}_{-0.05}$, $\chi^2_{\nu}$~=~1.08, and $\nu$~=~865 
 (Figure~\ref{figure:halopower}).  
The fit is improved by the 
addition of a thermal 
$pshock$ component with a temperature of $kT$~=~0.33$^{+0.12}_{-0.08}$ keV
and an ionization timescale of $n_et$~=~1.0$^{+12.5}_{-1.0}$
$\times$10$^8$~cm$^{-3}$~s ($\Gamma$ =
2.25$^{+0.11}_{-0.18}$, $\chi^2_{\nu}$=1.04, $\nu$=862).  This fit to the halo is used in Section~\ref{section:limbspec} to subtract the halo component from the limb region.

\subsection{Eastern Limb (r = 125\arcsec--153\arcsec)}
\label{section:limbspec}

The eastern limb (Figure~\ref{figure:regionpic}, radius = 125\arcsec--153\arcsec \footnote{Bocchino et al. (2005) chose a region between 115\arcsec\ and
138\arcsec, with the outer radius located near the midpoint of the limb shown in our $Chandra$ data.}) was studied to search for emission characteristic of a SNR shell.
Due to the lower count rate in the limb, the background spectrum was
extracted from a region outside the halo and therefore the data contains a component due to
the halo.  To correct for the halo emission we add a model component
equal to the best fit to the halo (Section~\ref{section:halospec}),
with the normalization scaled to the area of the limb.  The parameters
for this component are all frozen and the fits presented below are
emission from the limb only.  

The \textit{pshock} model was first used to search for thermal emission from
interstellar matter shock-heated by the forward shock.  As shown in
Table~\ref{table:fitresults2}, the \textit{pshock} model provides an
adequate fit ($\chi^2_{\nu}$~=~0.538, $\nu$~=~909); however with an
unrealistically high temperature (7.5$^{+6.9}_{-2.5}$~keV), suggesting
that the X-ray emission is likely non-thermal.   

The limb of \g21\ can not be explained by a dust scattering halo nor by shock-heated ejecta
(e.g. Bocchino et al. 2005) and must have another origin. 
Shocks in young supernova remnants have been known to accelerate electrons 
to TeV energies where they produce synchrotron X-rays (Reynolds 1998).  To 
determine if the non-thermal component of the limb is due to this 
acceleration we used the \textit{srcut} model in \xspec.  Since the limb 
has not yet been observed in the radio (Bietenholz et al. 2010), we do 
not know the radio spectral index ($\alpha$, where $S \propto \nu^{-\alpha}$) for the limb and consider a range for
$\alpha$ between 0.3 and 0.8,
which covers the range observed for other supernova remnants (Green 2009).  
Similarly, without a radio observation of the shell, we 
do not know the 1~GHz radio flux density and so we leave it as a free parameter to 
find an estimate of the radio flux density.  
For $\alpha=0.5$, which is typical for SNR shells, and $N_{\mathrm
  H}$~=~2.24~\nHunits, we find a rolloff frequency of
$\nu_{\mathrm{rolloff}}$~=~4.3$^{+9.2}_{-2.2}$~$\times$~$10^{17}$~Hz,
and a 1~GHz flux density of 4.9$^{+0.7}_{-0.8}$~$\times$~$10^{-3}$~Jy,
corresponding to a surface brightness of
2.3$^{+0.3}_{-0.4}$~$\times$~10$^{-22}$~W~m$^{-2}$~Hz$^{-1}$~sr$^{-1}$
for a solid angle $\Omega$~=~2.1~$\times$~10$^{-7}$~sr
($\chi^2_{\nu}$~=~0.541, $\nu$~=~910, Figure~\ref{figure:limbsrcut}).
This estimate of the surface brightness is a factor of $\sim$17 times smaller than the upper limit
obtained by Slane et al. (2000) and a factor of $\sim$3 times smaller than the most recent upper limit of 
7~$\times$~10$^{-22}$~W~m$^{-2}$~Hz$^{-1}$~sr$^{-1}$ obtained by
Bietenholz et al. (2010) using a new sensitive 1.43-GHz observation with  the Very Large Array (VLA).

For $\alpha$~=~0.3 and $N_{\mathrm H}$~=~2.24~\nHunits\ with the model
\textit{srcut}, we find a rolloff frequency of
$\nu_{\mathrm{rolloff}}$~=~2.2$^{+2.5}_{-1.0}$~$\times$~$10^{17}$~Hz,
and a 1~GHz
flux density of 1.4$^{+0.2}_{-0.3}$~$\times$~$10^{-4}$~Jy,
corresponding to a surface brightness of 6.7$^{+0.9}_{-1.4}$~$\times$~10$^{-24}$~W~m$^{-2}$~Hz$^{-1}$~sr$^{-1}$
($\chi^2_{\nu}$~=~0.542, $\nu$~=~910), approximately two orders of magnitude smaller than
the observed recent upper limit (Bietenholz et al. 2010).  
For $\alpha$~=~0.8 and $N_{\mathrm H}$~=~2.24~\nHunits\ with the model
\textit{srcut}, we find a rolloff frequency of
$\nu_{\mathrm{rolloff}}$~=~8.7$^{+2400}_{-3.3}$~$\times$~$10^{17}$~Hz,
and a 1~GHz
flux density of 1.4$^{+0.3}_{-0.2}$~Jy, corresponding to a surface
brightness of 6.7$^{+1.4}_{-1.0}$~$\times$~10$^{-20}$~W~m$^{-2}$~Hz$^{-1}$~sr$^{-1}$
($\chi^2_{\nu}$~=~0.542, $\nu$~=~910).  This estimated surface brightness is two orders of magnitude higher than the 
recent upper limit
by Bietenholz   et al. (2010), suggesting the spectral index is at the lower end of
  the 0.3--0.8 range.

The quality of fit for each of the \textit{power-law}, \textit{pshock}, and
\textit{srcut} models are similar however the temperature for the thermal
model is too high, confirming that the limb is dominated by non-thermal emission
and strengthening the previous suggestion that it's a possible site
for cosmic ray acceleration (Bocchino et al. 2005, see \S5).  A
two-component (thermal + non-thermal) model does not give well constrained parameters and so
we have not included the results here.

\subsection{Northern Knots}
\label{section:northspec}

Bocchino et al. (2005) examined the brightest knot of emission to the
north of the PWN (`North Spur'). Using a two-component power-law plus thermal component
(with the latter modelled by the $vnei$ model, a non-equilibrium ionization model
with a single ionization age and temperature, and with variable metal abundances), 
they found two solutions for the 
metal abundances with a similar quality of fit.  The first solution is characterized by enhanced 
abundances for Mg and Si,
with a Mg abundance of 0.6--3
times solar, a Si abundance of 2--20 times solar, a temperature of
$kT$$\sim$0.17~keV, and an ionization timescale of
$n_et$$\sim$7~$\times$~10$^{11}$~cm$^{-3}$~s.  The second solution is characterized by
solar abundances, a temperature of $kT$~$\sim$~0.30~keV, and a lower ionization
timescale of $n_et$$\sim$1~$\times$~10$^{10}$~cm$^{-3}$~s.  Here we
explore the same region with more data in an attempt to better
constrain the parameters.

Before attempting two-component models however, we first fit with single component models:
a power-law or $srcut$ model for a non-thermal interpretation and a $pshock$\footnote{we favour the $pshock$ model over the $nei$ model used by
Bocchino et al. (2005) since the former includes a range of ionization timescales, which is
more reasonable for an extended region like the northern knot.} model for a thermal
interpretation, and found the following.
An absorbed power-law fit to the brightest knot in the north
(Figure~\ref{figure:regionpic}, background inside the halo) produces an artificially
low hydrogen column density ($N_{\mathrm 
H}$~=~0.98$^{+0.15}_{-0.15}$~\nHunits) with $\chi^2_{\nu}$~=~0.98 
($\nu$~=~605). Freezing $N_{\mathrm H}$ to the best-fit value for \j1833\ (2.24~\nHunits)
gives a photon index of $\Gamma$~=~2.72$^{+0.09}_{-0.09}$, an unabsorbed
flux of 6.3~$\times$~10$^{-13}$~\fluxunits, and an estimated
X-ray luminosity (0.5--8.0 keV) of 1.9~$\times$~10$^{33}$~\lumunits\
($\chi^2_{\nu}$~=~1.18, $\nu$~=~606, assumed $D$~=~5~kpc) for the knot.  
While the fit is statistically acceptable, the residuals show evidence of line emission and
the fit is improved by adding a thermal $pshock$ component, as discussed below.
Similarly, an $srcut$ model alone is not favoured based on the evidence of thermal
emission lines in the spectrum. 

As shown in Table~\ref{table:fitresults2}, the \textit{pshock}
 model alone with a
column density $N_{\mathrm H}$~=~2.24~\nHunits\
gives a temperature
$kT$~=~4.9$^{+0.9}_{-0.8}$~keV, and an
ionization timescale $n_et$~=~2.0$^{+0.6}_{-0.4}$~$\times$~10$^{10}$~cm$^{-3}$~s
with $\chi^2_{\nu}$~=~0.959 ($\nu$~=~605).
Allowing the abundances of Mg, Si and S to vary we find that
Mg~=~1.08~(0.95--1.21), Si~=~0.96 (0.79--1.13), and  S~=~0.52 (0.12--0.93);
all consistent with solar abundances.
While the fit is statistically acceptable, the shock temperature is unrealistically high, suggesting
that the spectrum is dominated by non-thermal emission.
  
 We therefore confirm the need for a two-component, thermal + non-thermal, model
  to fit the northern knot. As well we rule out an equilibrium ionization models (such as $MEKAL$
  used by Bocchino et al. 2005)
  for the thermal component,
  since the fit requires a low ionization timescale ($<$10$^{12}$~cm$^{-3}$~s).
   Next, we explore the two solutions discussed in Bocchino et al. (2005).
  
Fitting a two-component model 
(\textit{power-law+pshock}) to the northern knot ($N_{\mathrm H}$~=~2.24~\nHunits), we find
a photon index of $\Gamma$~=~2.21$^{+0.15}_{-0.15}$, a 
temperature of $kT$~=~0.20$^{+0.04}_{-0.06}$~keV, and an
ionization timescale of $n_et$~=~7.1$^{+7.1}_{-3.4}$~$\times$~10$^{9}$~cm$^{-3}$~s ($\chi^2_{\nu}$~=~0.96, 
$\nu$~=~603, Figure~\ref{figure:northpowerpshock}). The observed thermal flux is (9.3$\pm$3.2)$\times$10$^{-15}$~\fluxunits,
which is only  $\sim$6\% that of the non-thermal flux ((1.6$\pm$0.3)$\times$10$^{-13}$~\fluxunits) in the 0.5--8.0~keV range,
again confirming that the spectrum of the northern knot is dominated by non-thermal emission.
Figures~\ref{figure:knottemptau} and~\ref{figure:knotnormtau} show the range of values allowed by the above fit for
the parameters of the \textit{pshock} component.
Figure~\ref{figure:knottemptau} shows the relationship between $kT$ and
$n_et$ and Figure~\ref{figure:knotnormtau} shows the relationship
between the emission measure and $n_et$.   
Allowing the abundances of Mg, Si and S to vary we find that
Mg~=~0.72 (0.40--1.06), Si~=~0.84 (0.32--1.35), and S~=~107.1
(3.9--210.3).  Figure~\ref{figure:knotabundances} shows that Mg and
Si are consistent with solar, in agreement with solution 2 of Bocchino
et al. (2005).  We do not observe a solution with overabundances of Mg and Si, that is we rule out
solution 1 of Bocchino et al. (2005).
However we note that S (although poorly constrained) appears overabundant in our fits.
While additional observations will help constrain the S abundance, we note that the apparent high S abundance might be an
 artifact of the model used, because the S lies in the energy range where X-ray spectra from the $pshock$ and power-law components 
 overlap.

Finally, a two-component \textit{srcut+pshock} model was used to further study the non-thermal emission from the northern knot
that could be due to particle acceleration at a shock.  We fix the column density, $N_{\mathrm H}$, to
2.24~\nHunits\ and the radio spectral index, $\alpha$,  to
0.5 (as for the limb, we also attempted other values for $\alpha$ in the range that brackets all possible spectral indices--see the results summarized in Table~\ref{table:fitresults2}).  
We find 
a temperature of $kT$~=~0.21$^{+0.04}_{-0.04}$~keV,  an
ionization timescale of $n_et$~=~5.8$^{+7.1}_{-1.8}$~$\times$~10$^{9}$~cm$^{-3}$~s,
a rolloff frequency of $\nu_{\mathrm{rolloff}}$~=~4.3$^{+4.1}_{-2.1}$~$\times$~$10^{18}$~Hz,
and a 1~GHz flux density of 2.6$^{+1.5}_{-0.9}$~$\times$~$10^{-3}$~Jy
 ($\chi^2_{\nu}$~=~0.963,
$\nu$~=~603).  The thermal flux
(1.0$^{+1.5}_{-0.5}$~$\times$~10$^{-14}$~\fluxunits) is again $\sim$7\%
of the non-thermal flux (1.5$^{+0.9}_{-0.5}$~$\times$~10$^{-13}$~\fluxunits)
in the 0.5--8.0~keV range. Varying the abundances again yields Mg and Si abundances that are
consistent with solar, but indicates enhanced (and poorly constrained) abundance for S;
a result that is consistent with the power-law+$pshock$ model above.
We note that the derived 1~GHz flux density (for $\alpha$=0.5) is
about an order of magnitude smaller than the flux density of 19~$\pm$~7 mJy inferred
from the radio detection of the northern knot with the VLA  (Bietenholz et al. 2010).
Freezing the $srcut$ normalization to the measured value of $\sim$20 mJy while fitting for the corresponding spectral index yields
$\alpha$ = 0.61$^{+0.03}_{-0.02}$, $kT$ = 0.21$^{+0.03}_{-0.04}$ keV, $n_et$ = 6.0$^{+12.2}_{-2.6}$$\times$10$^9$~cm$^{-3}$~s,
 and a roll-off frequency $\nu_{roll}$ = 6.9$^{+11.3}_{-3.7}$$\times$10$^{17}$~Hz ($\chi^2_{\nu}$~=~0.962,
$\nu$~=~603).
Due to the large uncertainty in the VLA spectral index measurement, our fitted value for $\alpha$ 
could not be excluded by the radio study and remains to be confirmed.

To conclude, the X-ray emission from the bright northern knot is dominated by non-thermal emission, suggesting cosmic ray acceleration at 
a shock. The thermal component does not require enhanced abundances of Mg and Si, suggesting
that the second solution of Bocchino et al. (2005) is more reliable.  The inferred low temperature, ionization timescale
and solar abundances have been interpreted as evidence for possible interaction
between ejecta and the H envelope of a type IIP SN.

\section{VARIABILITY in the PWN}
\label{section:variable}
Variability has been previously observed in PWNe such as the Crab 
and Vela nebulae.   
The Crab nebula shows wisps
(moving outward from the Crab pulsar with a velocity
$\sim$0.5\textit{c}) and knots (that do not have the outward
motion) which brighten quickly and fade over approximately one
month~\citep{hester2002}.  
The Vela nebula has PWN features whose surface brightness changes up
to $\sim$30\% and move with speeds up to $\sim$5000~km~s$^{-1}$ \citep{pavlov2001}.
The Vela PWN also shows bright compact
blobs moving at 0.3\textit{c}--0.6\textit{c} which brighten and disappear over a couple
of weeks~\citep{pavlov2003}.  These remnants are located at
approximately 40\% and 10\%, respectively, of the distance to \g21,
making changes in their morphology easier to detect than for 
\g21.				

Using the distance estimate  
of $D$~=~5~kpc to \g21, 
we estimate that features moving at a constant 0.5\textit{c} would be expected 
to cover
$\theta \sim 
6.32$~\arcsec~yr$^{-1}$.
Since the spatial resolution of \chan\ is $\sim$0.5\arcsec\ 
and we have observations obtained over a period 
of 6 years, we can search for variability in \g21.

Figures~\ref{figure:variable} (ACIS) and~\ref{figure:variablehrc}
(HRC-I) show the evolution of \g21\ over more than 6 years of \chan\
observations (from
1999 Aug to 2006 Feb).  These images are presented as movies in the
online version of the journal.    
Observations with the same observing date were combined after reprojecting
their coordinates to align the images, smoothed with a
Gaussian of $\sigma=1$\arcsec, and normalized to
an effective exposure time of 20~ks.  
We can see knots that appear to change position between images.
However, we have months between observations and previous observations
of pulsar wind nebulae~\citep{hester2002, pavlov2003} have indicated
that knots brighten and fade on a timescale of weeks.  Therefore, we can
not be sure that we are seeing the motion of knots rather than new knots
that have appeared after the originals have faded.  

Figure~\ref{figure:hrc_se} shows HRC-I images of the PWN and
demonstrates the high degree of variability in the nebula.  We see
that either there are knots in motion or old knots disappear and new
knots appear between observations. 
Several clumps of emission are marked and we see that they either fade
or move to a new location in the subsequent image.  Assuming the knots
persist, the knots labelled 6, 12, 13, and 14 can be located in the
next image in the sequence and appear to have velocities
$\sim$0.1$c$--0.3$c$.  These knots are visible in the unsmoothed
images, ensuring that the highlighted features are not an artifact of
the smoothing.

Sample difference images created by
subtracting one of the images in Figure~\ref{figure:variable} from the
following image in the same sequence are presented in
Figure~\ref{figure:differences}.  The emission near the pulsar was
omitted to highlight the variability in the PWN.  The background level
in individual images was approximately 0.5 counts per pixel
immediately surrounding the PWN.  Ignoring regions in the range -0.5 
to +0.5 counts per pixel, we see that white and red
indicate regions that are brighter in the later image while black, purple, and blue
indicate regions that are brighter in the earlier
image. To the north of the pulsar location in
Figure~\ref{figure:differences}b we see a region of increased emission
with a region of decreased emission to the west, spaced by
6.0\arcsec.  For a distance to \g21\ of 5~kpc, and assuming that we
are observing the motion of a single knot, this corresponds to a
knot velocity of $\sim$0.5c.

The expansion of the nebula is not detectable in the observations
presented here.  This may be due to the fact that the outer limit of the
nebula ($\sim$40\arcsec) is not well defined and there are wisps
that appear and disappear.  
However, the expansion velocity of the PWN was estimated as 910~$\pm$~160~km~s$^{-1}$ 
by Bietenholz \& Bartel (2008).  This velocity corresponds to an expansion 
of 0.038\arcsec~$\pm$~0.007\arcsec\ (less than 0.1  ACIS pixels) per year, 
suggesting the expansion of the PWN may not be detected with \chan\ for several more years.
In addition, the expansion of the shell is not detectable in this study,
as a large number of observations are required to   
improve the signal-to-noise ratio
so that the limb is more visible.

\section{DISCUSSION}
\label{section:discussion}

\subsection{The SNR}

The absence of shells around Crab-like SNRs has been a mystery
for decades. Recent deep or targeted searches have been however in some cases
successful in revealing the long sought SNR shells around PWNe. Aside from G21.5$-$0.9,
shells have been found in 3C~58~\citep{gotthelf2007}\footnote{We note
that the shell seen in 3C~58 has an extent that is smaller than the PWN,
suggesting a shocked-ejecta origin, which is also strengthened by
the evidence of enhanced abundances in the X-ray spectrum.} and most
recently G54.1+0.3~\citep{bocchino2010}\footnote{For G54.1+0.3, the presence
of diffuse emission out to $\sim$10~pc has been interpreted as evidence for the SNR shell.
However, the available data did not yet reveal limb-brightening as expected from SNR shells.}.
A shell has been unsuccessfully searched for around the Crab Nebula~\citep{frail1995,seward2006}.  
A deep search with $Chandra$ revealed a dust-scattered halo with an intensity of 5\% that of the
Crab, out to a radial distance of 18\arcmin.  However, there was no
evidence of emission from shock-heated material in the form of a shell.
The luminosities for the shells seen in 3C~58 and \g21 are below the
current upper limit of $L_{\mathrm{X}}$~$\sim$~10$^{34}$~erg~s$^{-1}$
for the Crab Nebula~\citep{seward2006}\footnote{For G54.1+0.3, using the unabsorbed
flux of 4.7$\times$10$^{-12}$~\fluxunits\ tabulated in Table~2 of Bocchino et al. (2010), we obtain a luminosity of 2.2$\times$10$^{34}$~erg~s$^{-1}$
at an assumed distance of 6.2~kpc.}. 
The low luminosity of the shell in \g21\  ($\sim$2$\times$10$^{33}$~erg~s$^{-1}$ in the non-thermal interpretation) 
could be attributed to the fact that the SNR is young and expanding 
in a low-density medium, and so the shell may
only now be coming into view. Furthermore, the heavy absorption towards \g21\ ($N_{\mathrm H}$~=~2.2~\nHunits) explains the difficulty in
detecting the SNR shell which became clearly visible (although only the eastern side) after accumulating
$\sim$500~ksec of $Chandra$ time.
Next, we will comment on the age of the SNR and the ambient density of the ISM in which it is expanding.

To estimate the density of the ISM in which the SNR is expanding,
we use the \textit{pshock} fit to the limb\footnote{Since we believe the X-ray spectrum of the limb
is dominated by non-thermal emission, the derived estimate of the ambient density
should be only viewed as an upper limit.}
which yields an emission measure of 
$EM$~=~3.7~$\times$~10$^{-4}$~cm$^{-5}$.  This implies that $\int n_e
n_{\mathrm H} dV 
\sim f n_e n_{\mathrm H} V$~=~10$^{14}$$\left( 4 \pi D^2\right) \left(
EM \right)$~=~1.1~$\times$~10$^{56}$~cm$^{-3}$ (assuming $D$~=~5~kpc),
where $f$ is the volume filling factor, $n_e$ is the post-shock
electron density, and $n_{\mathrm H} \sim n_e / 1.2$.  Using a volume
of 3.3~$\times$~10$^{56}$~cm$^3$ (assuming the limb is a shell and the
region we studied is $\sim$10\% of the shell volume), $n_e \sim 0.63
f^{-1/2}$~cm$^{-3}$, which implies an upstream
ambient density $n_0 = n_e / 4.8 = 0.13 f^{-1/2}$~cm$^{-3}$.  For
filling factors $f \sim 0.1-1.0$ this corresponds to $n_0 \sim 0.13-0.42$~cm$^{-3}$ (2.2--7.1~$\times$~10$^{-25}$~g~cm$^{-3}$ when
$n_0$ includes only hydrogen).  
This estimate of the density is consistent with the upper limit of
0.65~cm$^{-3}$ derived by Bocchino et al. (2005), and confirms that
\g21\ is expanding into a low-density medium.

We subsequently comment on the non-thermal interpretation of the limb
as evidence of cosmic ray acceleration to TeV energies.
Young SNRs have been shown to accelerate cosmic rays to TeV energies
at shock fronts 
(e.g. SN1006, Koyama et al. (1995)).    
The energy at which the electron energy distribution steepens from its slope at radio-emitting energies is given by
\begin{equation} \label{eqn:emax} E_{\mathrm{max}} = 10 \left( \frac{10
  \mu G}{B}\right)^{1/2}\left(\frac{\nu_{\mathrm{rolloff}}}{0.5 \times
    10^{16}}\right)^{1/2} \mathrm{TeV} 
\end{equation}
\citep{reynolds1999}.  
 Fitting the limb with the \textit{srcut} model with a range $\alpha$~=~0.3--0.8, we found that 
$\nu_{\mathrm{rolloff}}$~$\sim$~2--9~$\times$~10$^{17}$~Hz,  
which implies
$E_{\mathrm{max}}\left(B/\left(10\mu G\right)\right)^{1/2}$~$\sim$~60--130~TeV.
This range (60--130 TeV, for a magnetic field of 10$\mu$G) implies that shells in
plerionic SNRs could be important sites for cosmic ray acceleration to TeV energies.
We note that the lower values of $\alpha$ give an upper limit on the maximum
electron energy which is within the range
of values found by Reynolds \& Keohane (1999) for young shell-type supernova
remnants.  
The predicted $E_{\mathrm{max}}$ is dependent on
$\nu_{\mathrm{rolloff}}$, which is very dependent on $\alpha$ that
remains to be measured in the radio. The magnetic field estimate
in the region surrounding the limb is also needed to refine the estimated maximum energies for
the accelerated particles.

Finally, we note that the non-thermal component of the northern knot 
also implies cosmic ray acceleration at a shock--possibly resulting from
the interaction between ejecta and the H envelope of a type IIP SN (see \S3.5). Using the $srcut$ model
component with $\alpha$=0.6 (the likely value based on its radio detection, see \S3.5), 
the derived roll-off frequency is $\nu_{roll}$ =  6.9 (3.2--18.2)$\times$10$^{17}$~Hz, 
which yields $E_{max}$ = 117 (80--191) TeV (again assuming $B$ = 10$\mu$G).

\subsection{The PWN and pulsar}
Camilo et al. (2006) show that inside the elliptical emission region
containing the pulsar, the compact source is offset to the northeast
from the centre of
the ellipse.  The HRC-I image shown in Figure~\ref{figure:psronly}
confirms this offset.  The image is centred on \j1833, smoothed by 2
pixels (0.2635\arcsec), and is 9\arcsec~$\times$~9\arcsec.  
Using the ACIS imaging data used to produce
Figure~\ref{figure:regionpic}, we measured the distance between the
location of the pulsar and the centre of the arc traced by the eastern
limb.  \j1833 is found to
be offset from the centre of the shell of \g21 by 7.5\arcsec, which is equivalent
to a 0.18~pc offset for a distance to \g21\ of 5~kpc.  For a remnant
age of 870$^{+200}_{-150}$, this corresponds to a 2-D pulsar velocity of
200~$\pm$~40~km~s$^{-1}$.  This velocity is comparable with the average 2-D
speed of 246~$\pm$~22~km~s$^{-1}$ derived for `normal' pulsars by Hobbs et al. (2005). 

The PWN of \g21\ demonstrates the type of variability previously
observed in other PWNe, including bright knots
which appear, move at speeds of $\sim$0.1$c$--0.3$c$, and fade with
time.  There is no evidence of expansion of the nebula, nor do we
expect to observe any with this sequence of observations (\S\ref{section:variable}). 

Using $B_{eq} \sim 0.3$~mG and $\dot{E} = 3.37 \times
10^{37}$~erg~s$^{-1}$ for \g21, Camilo et al. (2006) 
calculate the characteristic scale of the wind termination shock around 
\j1833\ to be $\sim$1\farcs6, which corresponds to 
$r_t$~=~1.2~$\times$~10$^{17}$~cm~=~0.04 pc for a distance of 5~kpc to
\g21 (Section~\ref{section:intro}).  For an off-axis angle
of 3\arcmin, at 1.5 keV, 90\% of the encircled
energy of a point source is contained within 2.5\arcsec\ (within
1.5\arcsec\ at 6.4 keV) with ACIS.  
For the HRC data, the 90\% encircled energy radius is 0.9\arcsec\ at 1.5 keV and 2.0\arcsec\ at 6.4 keV (for an off-axis angle of 0.3\arcsec).
The estimated termination shock radius  
is then comparable to the point-spread function,
making it difficult to distinguish the termination shock from the
pulsar emission.  Recent estimates of the magnetic field strength in
the nebula predict a field strength which is an order of magnitude
below equipartition.  de Jager et al. (2008) estimate the current
field strength as 25~$\mu$G and Tanaka \& Takahara (2010) estimate
64~$\mu$G.

In Section~\ref{section:psrspec} we presented the spectroscopic analysis
of \j1833, the pulsar powering G21.5$-$0.9. The X-ray spectrum is dominated
by hard non-thermal emission, with evidence of a thermal blackbody component that
represents $\sim$9\% of the non-thermal flux observed from the pulsar in the 0.5--8.0 keV range.
While this additional component needs to be confirmed/constrained with additional observations of the pulsar,
we comment next on the implication of this detection.
Tsuruta et al. (2002) used X-ray observations of neutron stars to
compare NS cooling theories.  They found that less massive stars
appear to cool by standard cooling (conventional slower neutrino
cooling mechanisms: modified Urca, plasmon neutrino, bremsstrahlung)
while more massive stars cool by non-standard cooling (exotic,
extremely fast cooling processes: direct Urca processes involving
nucleons, hyperons, pions, kaons, \& quarks).  For an age of
870$^{+200}_{-150}$ years~\citep{bietenholz2008} and a luminosity of 
$L_\infty$~$\sim$~1.3$^{+0.5}_{-0.5}$~$\times$~10$^{33}$~erg~s$^{-1}$ (\textit{blackbody} component
of the \textit{power-law+blackbody} fit, see Section~\ref{section:psrspec}),
\j1833\ falls between the non-standard and standard cooling curves.
It is however closer to the non-standard curve than the standard
curve, which imply luminosities of $L_\infty$
$\sim$5~$\times$~10$^{32}$~erg~s$^{-1}$ and
$\sim$9~$\times$~10$^{33}$~erg~s$^{-1}$, respectively, for a 870
year-old neutron star.  Fixing the radius to 12~km, the luminosity of
the \textit{bbodyrad} component is
$\sim$5.9~$\times$~10$^{32}$~erg~s$^{-1}$, consistent with the
non-standard cooling curve of Tsuruta et al. (2002).
Other young neutron stars observed to have thermal components in their spectra also point to rapid
cooling by non-standard processes \citep{slane2004}. 
The blackbody fit presented here for \j1833
yields a temperature that is higher or comparable to that inferred
for other young pulsars
($T_{\infty}$$\sim$5~$\times$~10$^6$~K, reduced to
$\sim$1~$\times$~10$^6$~K when fixing the size of the emitting region to that of the neutron star surface). For
example, PSR~J0205+6449 ($\sim$830 year-old based on its probable association with SN~1181) in
3C~58 has an upper limit on its temperature of
1.02~$\times$~10$^6$~K~\citep{slane2004}, while the highly magnetized and
young pulsar PSR~J1119-6127 (with a characteristic age of $\sim$1600 years)  in G292.2$-$0.5 has a temperature of
(1.2--2.5)~$\times$~10$^6$~K~\citep{safi-harb2008}.

\section{CONCLUSIONS}
We have made use of 6 years of archival \chan\ data to
demonstrate the importance of deep searches for shells
surrounding plerionic supernova remnants.  The technology incorporated into
\chan\ made it possible to study the SNR \g21\ in detail at a high
spatial and spectral resolution by combining
many short observations into one effective exposure of 578.6~ks with
ACIS and one of 278.4~ks with HRC.

The spectroscopic analysis presented here extends the study by Matheson \& 
Safi-Harb (2005).
The \chan\ data clearly reveal a partial shell on the eastern side
of the SNR, suggesting the detection of the long sought SNR shell.
This (partial) shell surrounding the PWN has a radius of $\sim$3.6~pc and is centred
7.5\arcsec\ from \j1833.   
Its spectrum is equally well fit by thermal (\textit{pshock})
and non-thermal models (power-law or $srcut$--a curved model with a cutoff
in the relativistic electron distribution).
However, the thermal model gives an unreasonably high temperature, indicating that
the non-thermal interpretation is more physical.  While a two-component model
can not be ruled out, the quality of the data is not sufficient to reliably fit the spectrum of the limb.
In the non-thermal interpretation, the shell has a luminosity of $\sim$2--3$\times$10$^{33}$~erg~s$^{-1}$
and may be the site of cosmic ray acceleration
up to $\sim$60--130 TeV energies (for a magnetic field of 10$\mu$G).
The derived upper limit for the upstream density is $\sim$0.1--0.4~cm$^{-3}$,
confirming that \g21\ is expanding in a rarefied medium.
Additional X-ray observations of \g21\
 will help  better
constrain any thermal emission from the limb and may potentially
detect the western side of the shell.  

In addition, we addressed the spectral nature of the northern knot (also known
as North Spur studied by Bocchino et al. 2005) detected in the X-ray halo between the PWN
and the outer limb.
We confirm that its spectrum is best fit with a combination of
non-thermal and thermal components; however with the thermal
component characterized by a low ionization timescale and solar abundances
for Mg and Si. This favours the interpretation of the northern knot
resulting from interaction between ejecta and the H envelope
of a type IIP SN. Furthermore, the non-thermal component of the emission from the knot could be
well fit with the $srcut$ model, characterized by a roll-off
frequency of 6.9$^{+11.3}_{-3.7}$~$\times$~10$^{17}$ Hz 
for a radio spectral index $\alpha$=0.6.
Constraining the spectral index of the knot, detected with the VLA,
is needed to confirm this result.

Our spectroscopic study of the powering engine of \g21, the highly energetic 61.86~ms pulsar \j1833,
showed that its X-ray spectrum is dominated by non-thermal emission, with
evidence for thermal X-ray emission from
the neutron star. The observed thermal component, yet to be confirmed with additional
spectroscopic and timing observations of the pulsar,  represents $\sim$9\% the non-thermal emission
and suggests non-standard rapid cooling of the neutron star.

Finally, the \chan\ data also allowed us to demonstrate for the first time the
variable nature of the PWN, identifying knots that appear and disappear
with time.  However, due to the time between observations we can not
be sure we are observing one knot over time rather than a new one that
has appeared after the original faded.

\begin{figure}
\begin{center}
\includegraphics[width=2.5in]{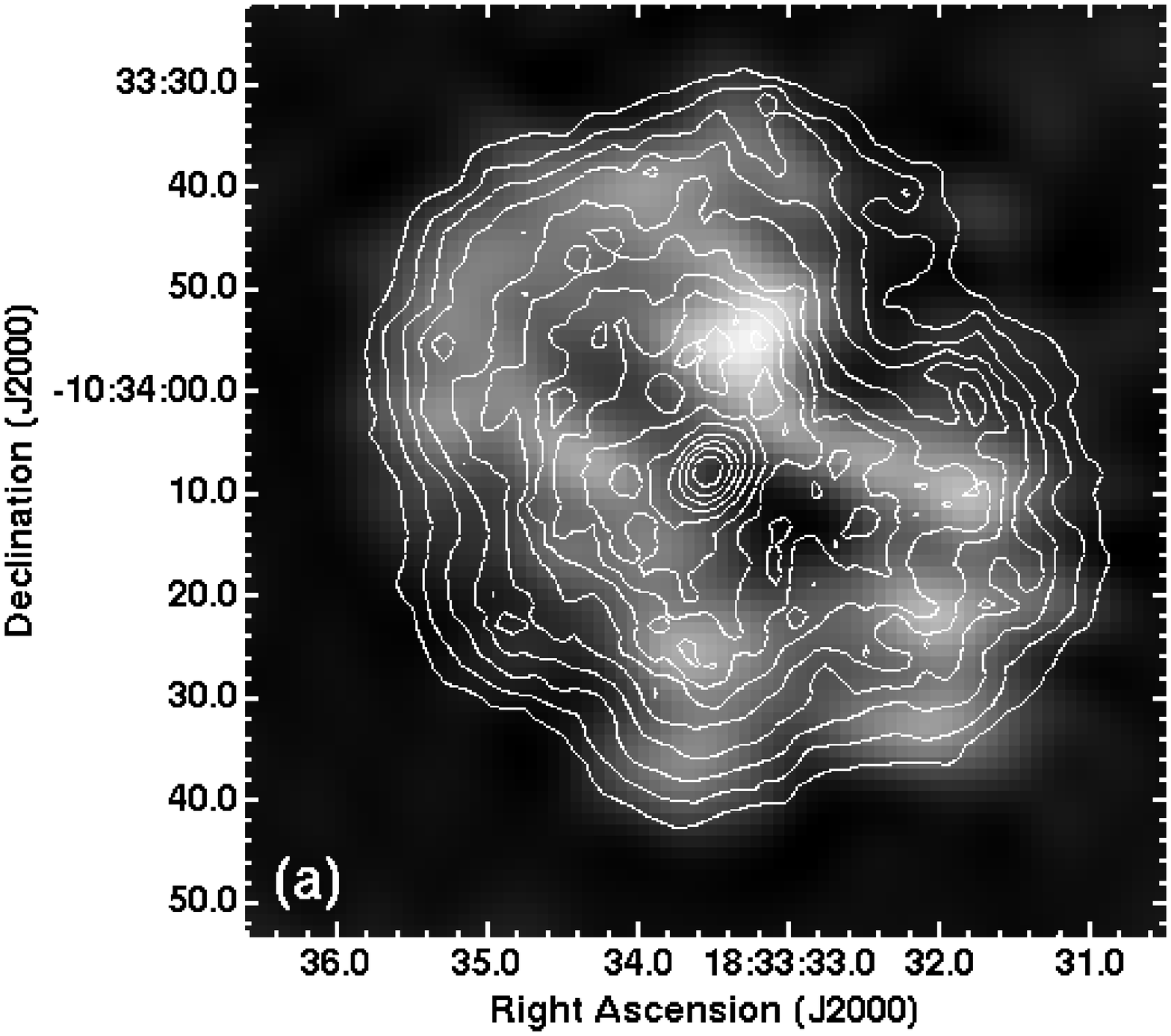}
\includegraphics[width=2.5in]{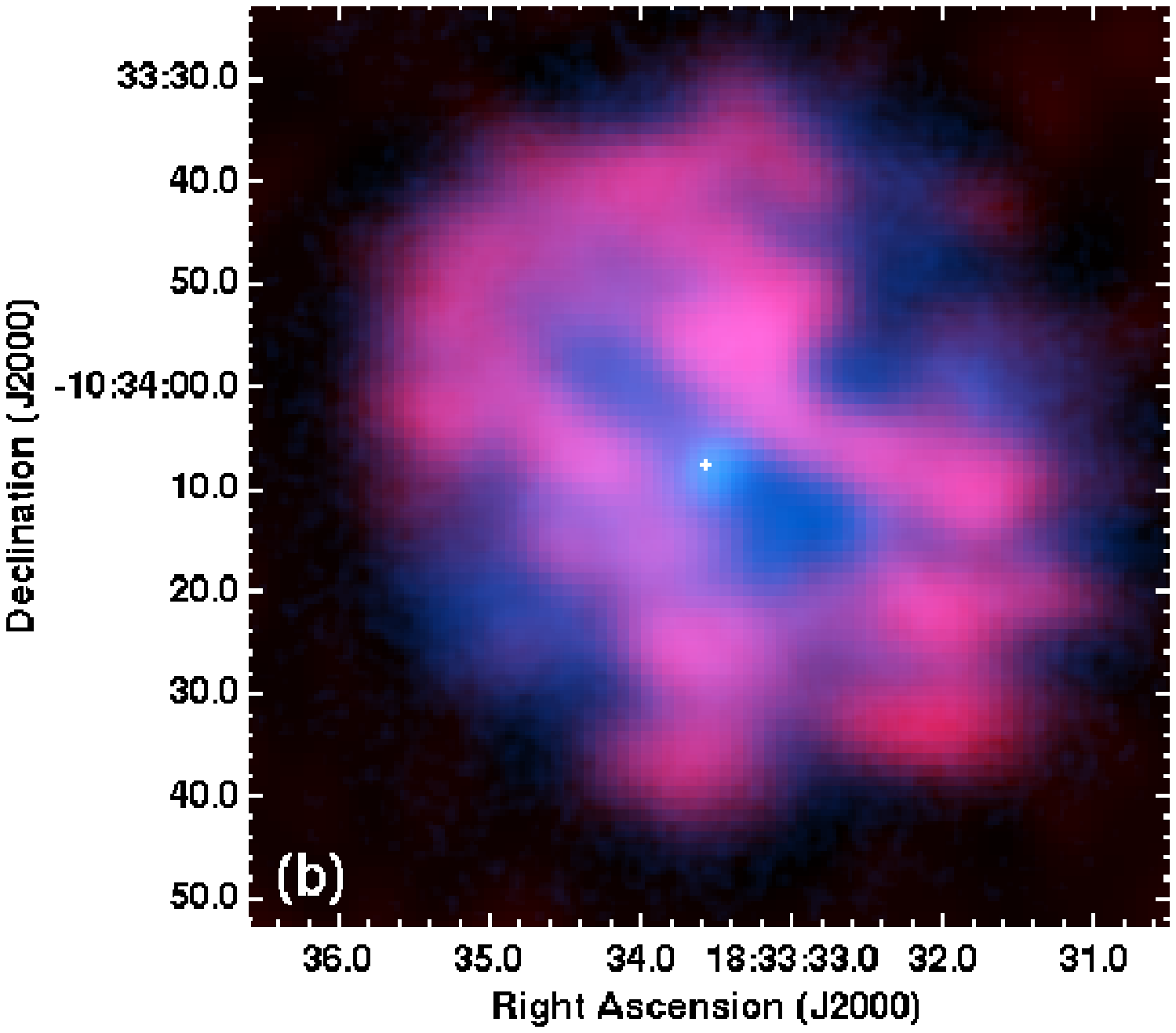}
\includegraphics[width=2.5in]{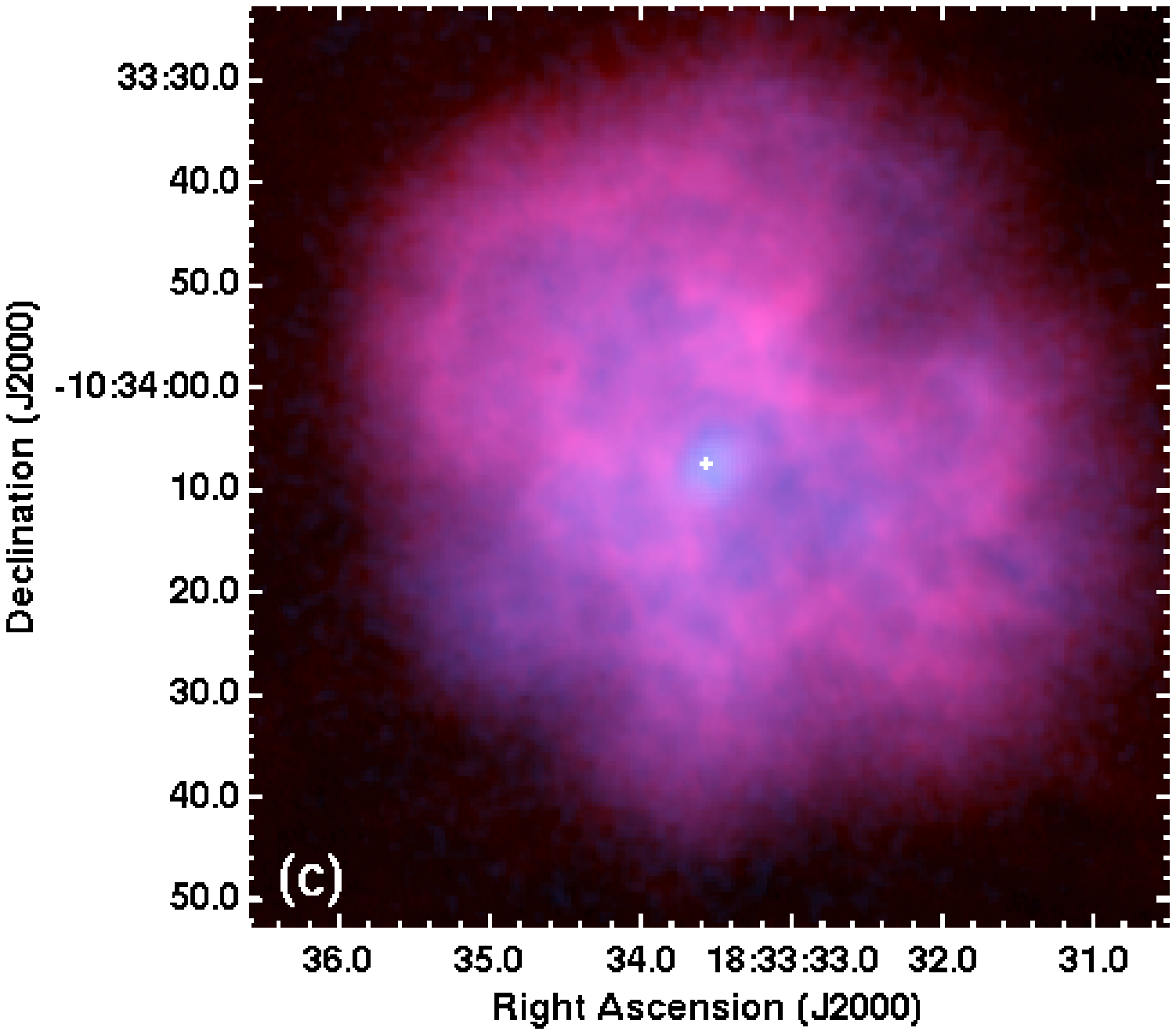}
\caption[Multi-wavelength images.]{(a) 22.3~GHz radio data from the Nobeyama Millimeter-Wave Array
(resolution  8\arcsec, F\"urst et al. 1988), overlaid with \chan\ ACIS 
  X-ray (0.2--10.0 keV) contours.  Both the X-ray and radio data are in a log
  scale. (b) The 22.3~GHz radio data are here shown in red and overlaid with the 0.2--10.0 keV X-ray data
  shown in blue.
 The cross indicates the position of \j1833. (c) 4.75~GHz radio
  data from the VLA (see Bietenholz \& Bartel (2008) for details)  coloured in red
  overlaid with the  0.2--10.0 keV X-ray data (again coloured in blue).
  The FWHM of the radio beam was 0.82\arcsec~$\times$~0.53\arcsec\ at p.a. 10\arcdeg.  
 See \S2.1 for   details.  }
\label{figure:xrayradio}
\end{center}
\end{figure}

\begin{figure}
\begin{center}
\plotone{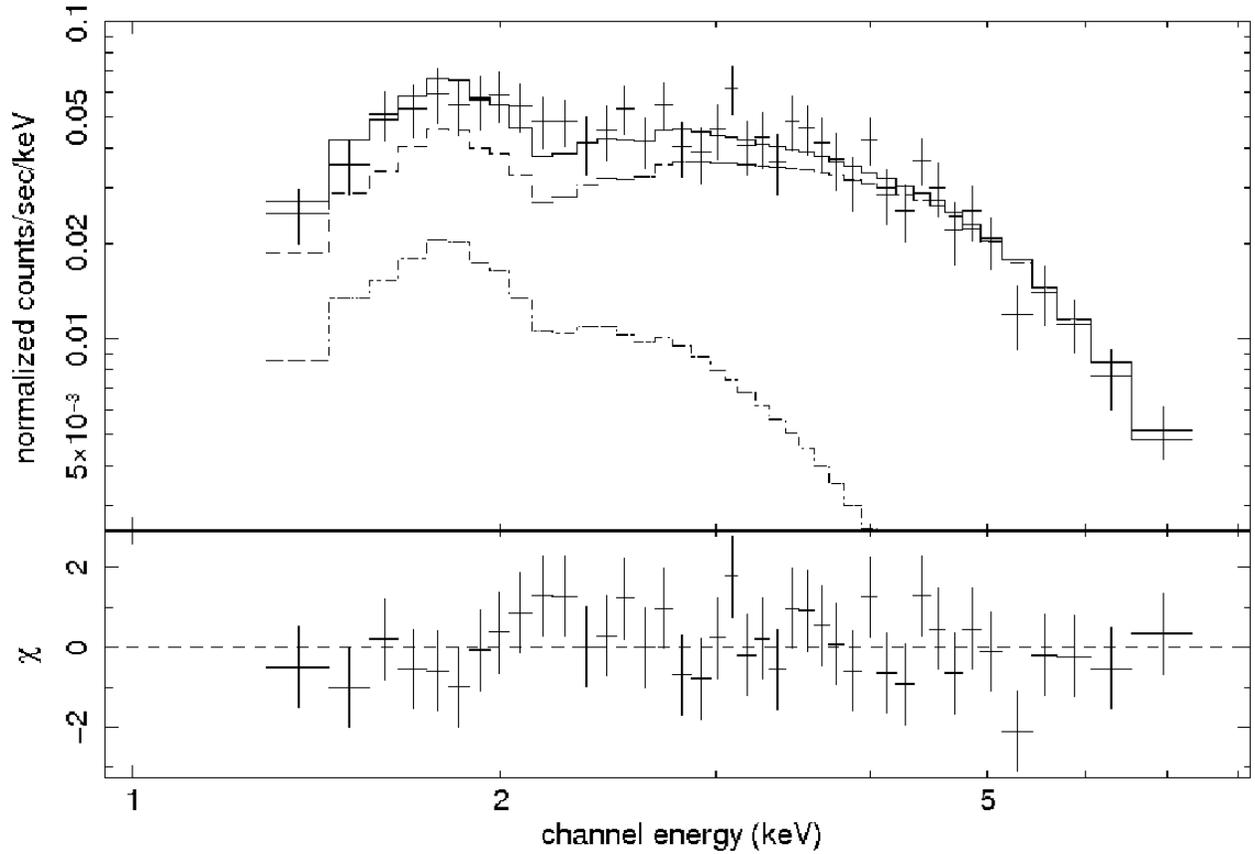}
\caption[Sample spectrum of the pulsar \j1833.]{Sample spectrum 
(obsID 6071)
  of the pulsar \j1833, showing the data (crosses), model (solid
  line), \textit{power} and \textit{bbody} components (dashed lines),  
and
  residuals (bottom panel) of the fit to
  an absorbed power-law + blackbody (\textit{wabs*(power+bbody)} in \xspec).}
\label{figure:psrpower}
\end{center}
\end{figure}

\begin{figure}
\begin{center}
\plotone{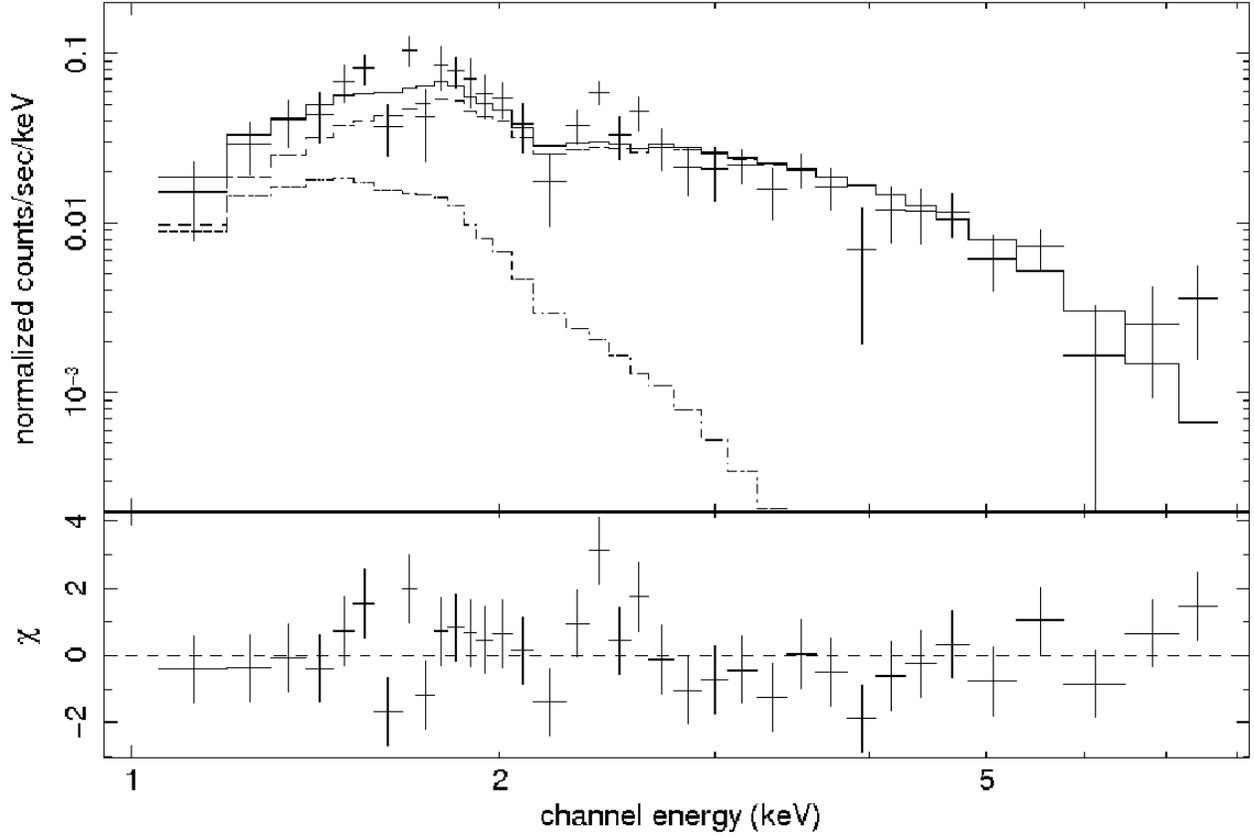}
\caption[Sample spectrum of the X-ray halo (r = 40\arcsec\ - 
  153\arcsec).]{Sample spectrum (obsID 6071)
  of the southern half of the X-ray halo (r = 40\arcsec\ - 153\arcsec), showing the data
  (crosses), model (solid line), model components (dotted lines), and residuals (bottom panel) of the
  fit to
  an absorbed power-law + pshock model with $N_{\mathrm H}$~=~2.24~\nHunits\ (\textit{wabs*(power+pshock)} in \xspec).}
\label{figure:halopower}
\end{center}
\end{figure}

\begin{figure}
\begin{center}
\plotone{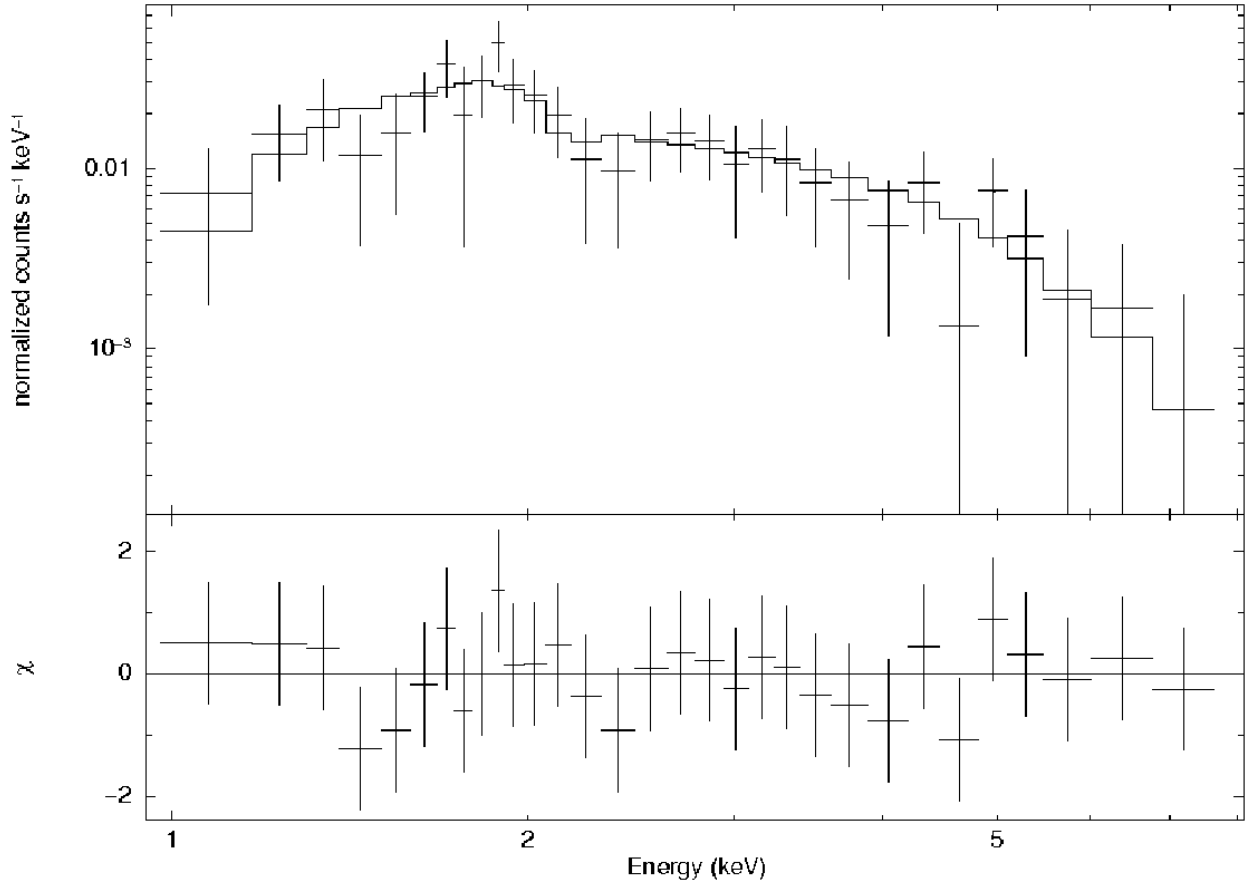}
\caption[Sample spectrum of the eastern limb.]{Sample spectrum
(obsID 6071)
  of the eastern limb, showing the data (crosses), model
  (solid line), and residuals (bottom panel) of the fit to an absorbed
  srcut model with $N_{\mathrm H}$~=~2.24~\nHunits\ and $\alpha$~=~0.5 (\textit{wabs*srcut} in \xspec).}
\label{figure:limbsrcut}
\end{center}
\end{figure}

\begin{figure}
\begin{center}
\plotone{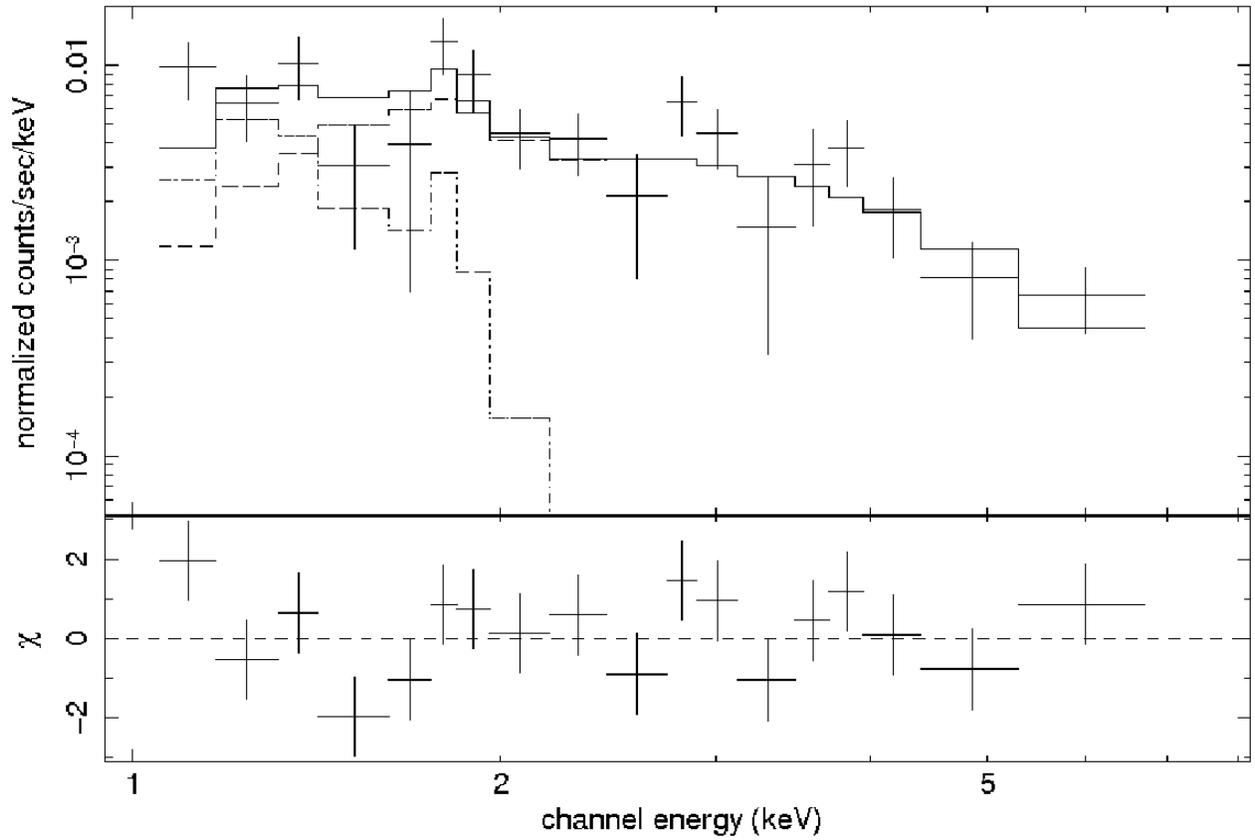}
\caption[Sample spectrum of the northern knots.]{Sample spectrum 
(obsID 6071)
  of the brightest northern knot, showing the data (crosses), model (solid
  line), and residuals (bottom panel) of the fit to
  an absorbed power-law + pshock model with $N_{\mathrm H}$~=~2.24~\nHunits\ (\textit{wabs*(power+pshock)} in \xspec).}
\label{figure:northpowerpshock}
\end{center}
\end{figure}

\begin{figure}
\begin{center}
\plotone{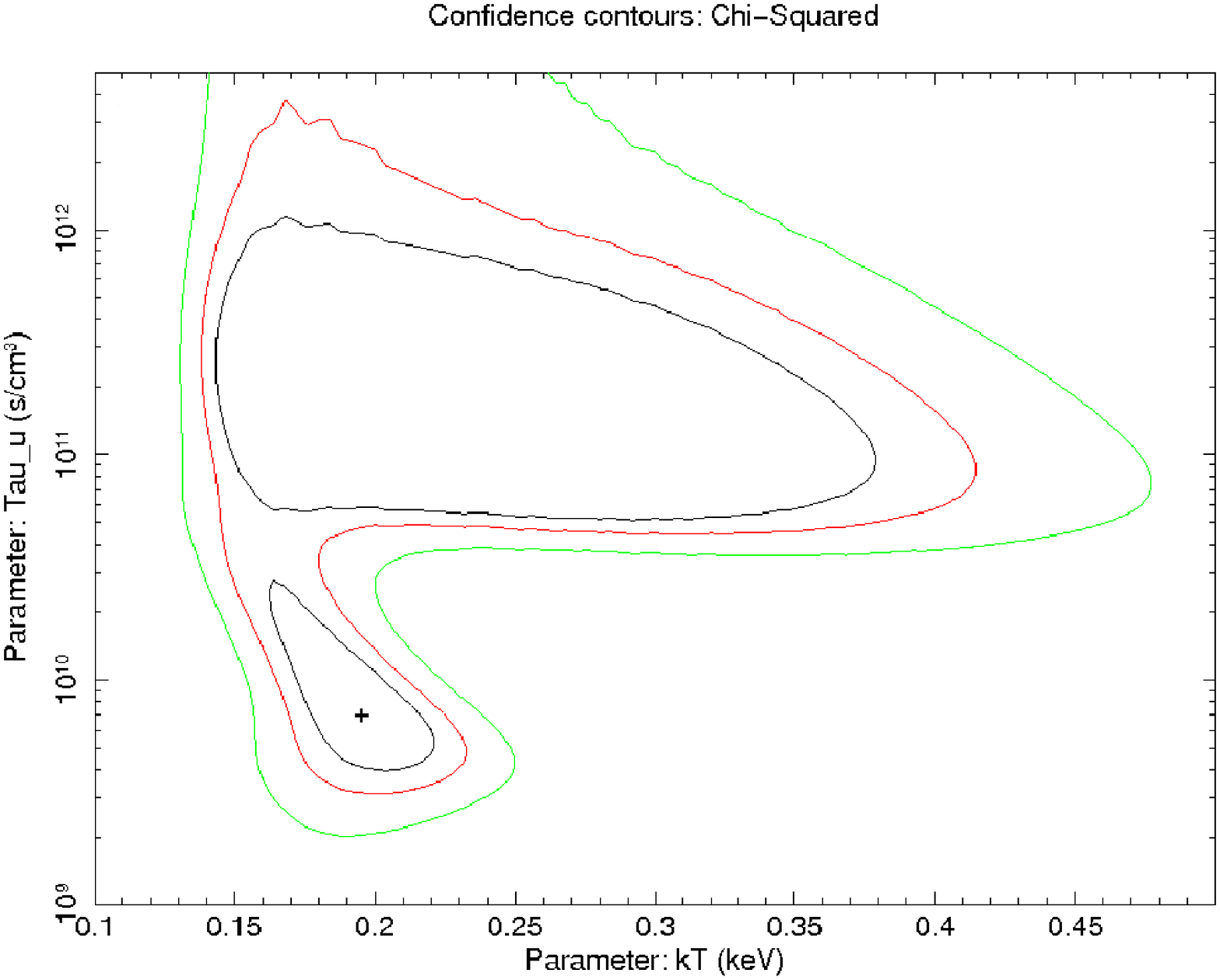}
\caption[]{$n_et$ and $kT$ confidence contours for the
  \textit{power+pshock} fit to the brightest knot north of the PWN.}
\label{figure:knottemptau}
\end{center}
\end{figure}

\begin{figure}
\begin{center}
\plotone{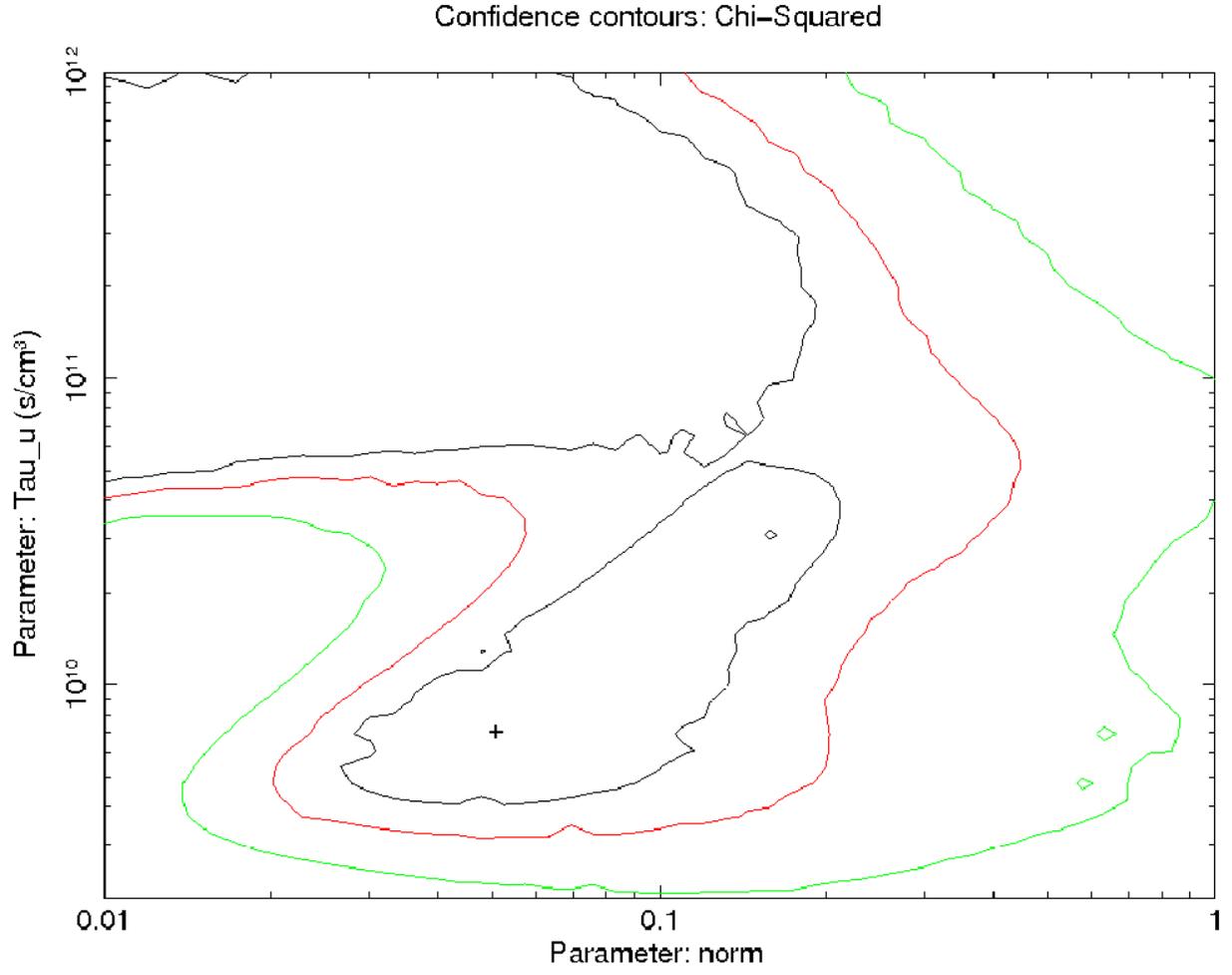}
\caption[]{Emission measure and $kT$ confidence contours for the
  \textit{power+pshock} fit to the brightest knot north of the PWN.}
\label{figure:knotnormtau}
\end{center}
\end{figure}

\begin{figure}
\begin{center}
\plotone{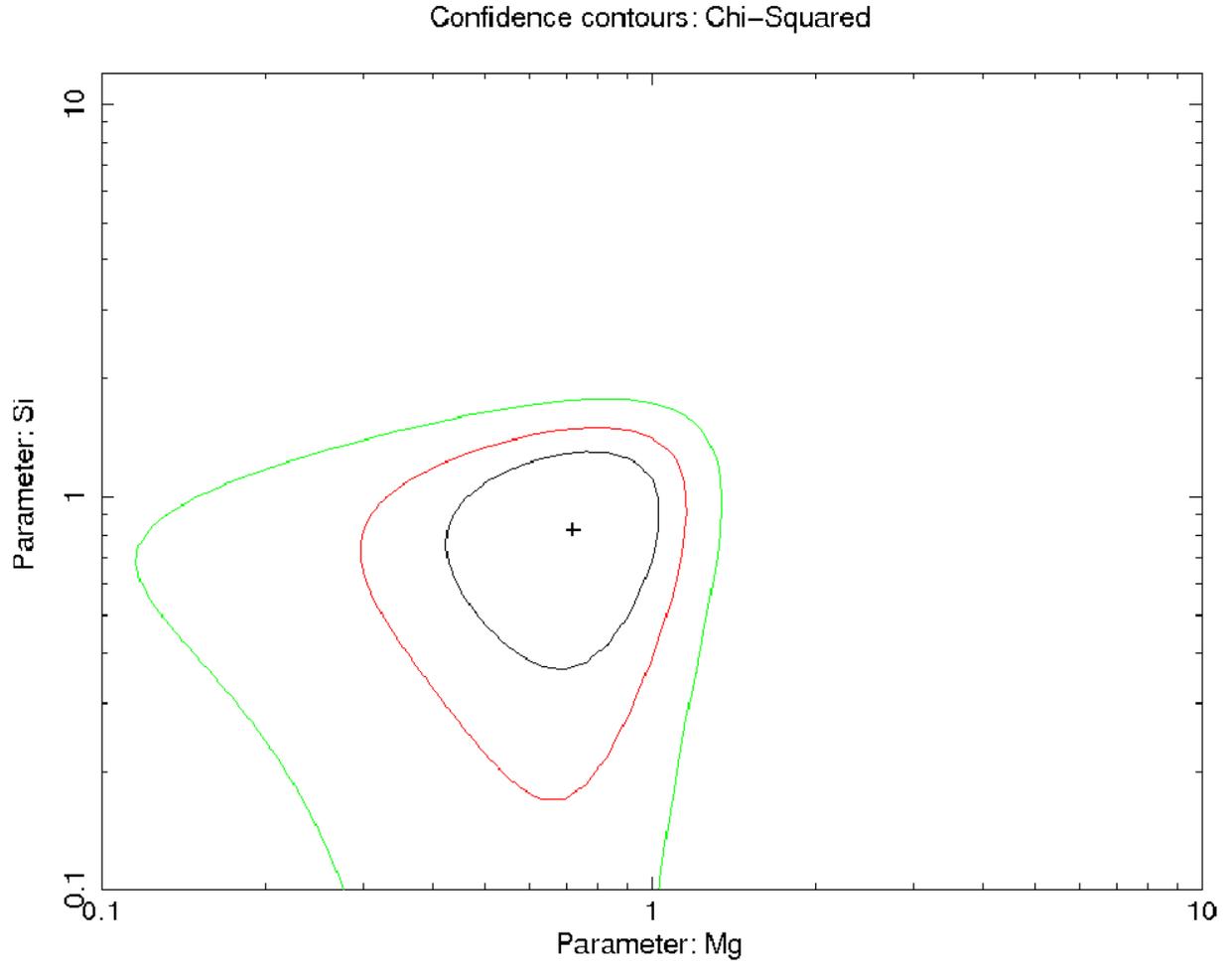}
\caption[Elemental abundances in the brightest knot.]{Elemental abundances in the brightest knot to the north of the PWN. The axes range shown is the same as that of Bocchino et al. (2005).}
\label{figure:knotabundances}
\end{center}
\end{figure}

\begin{figure}
\begin{center}
\includegraphics[width=1.5in]{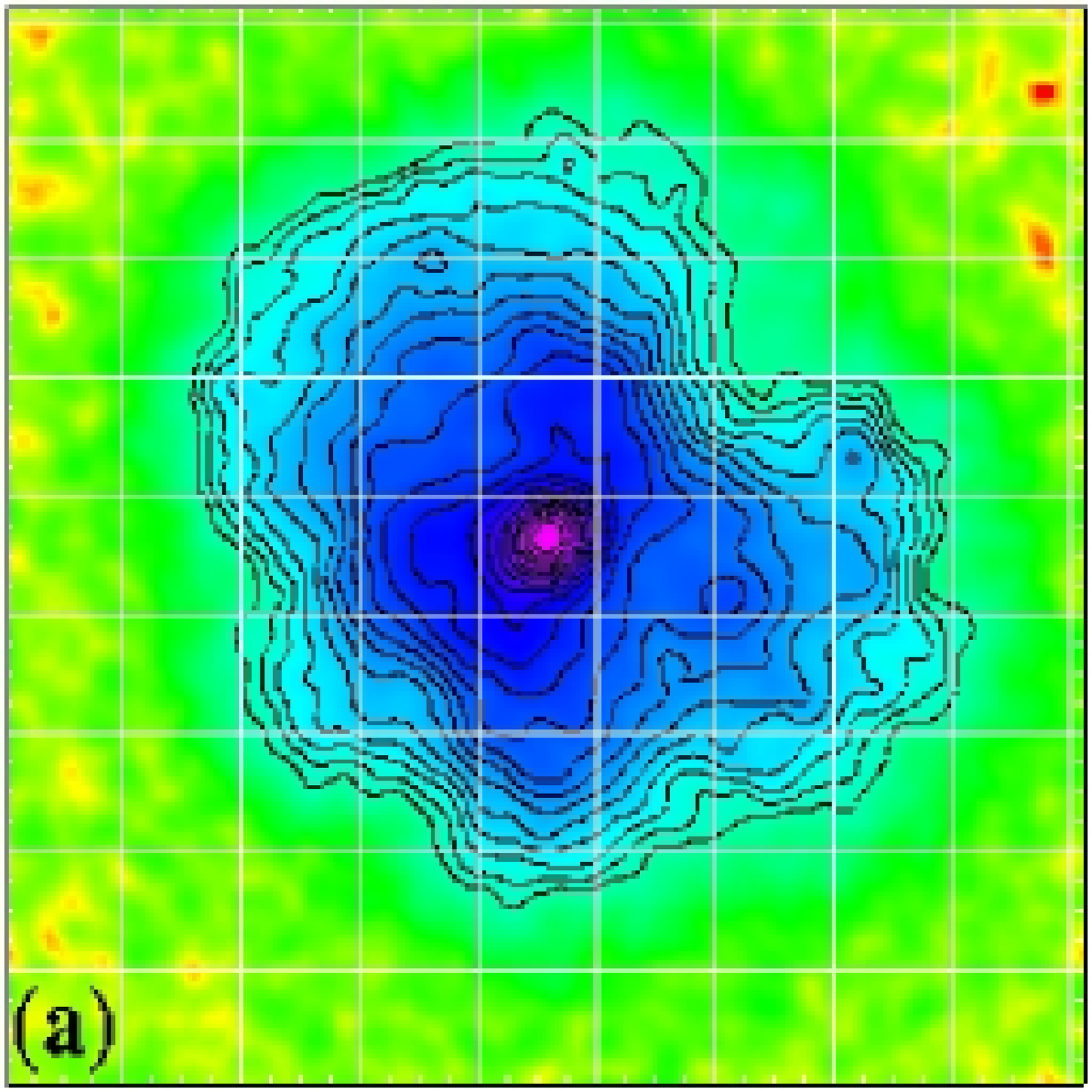}
\includegraphics[width=1.5in]{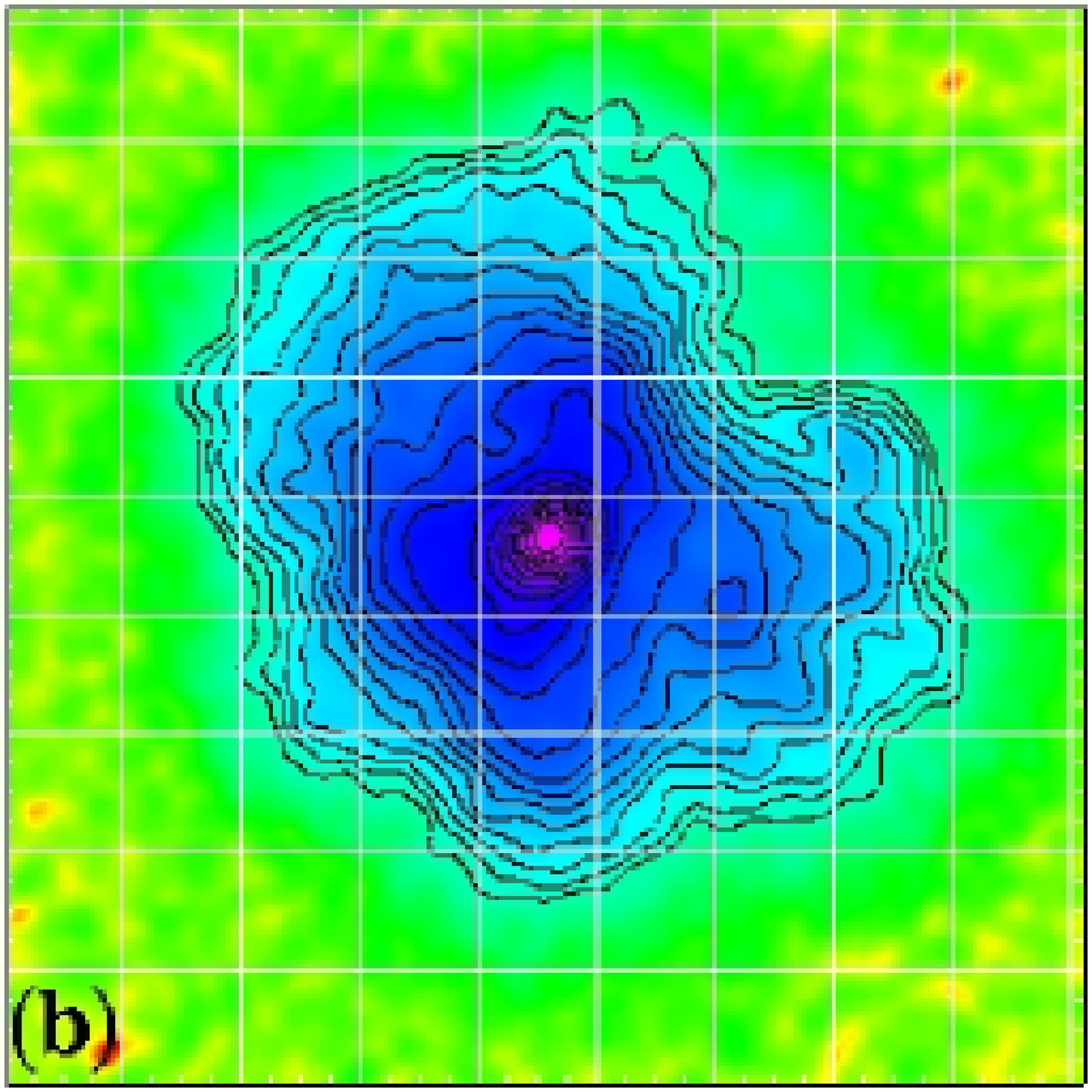}
\includegraphics[width=1.5in]{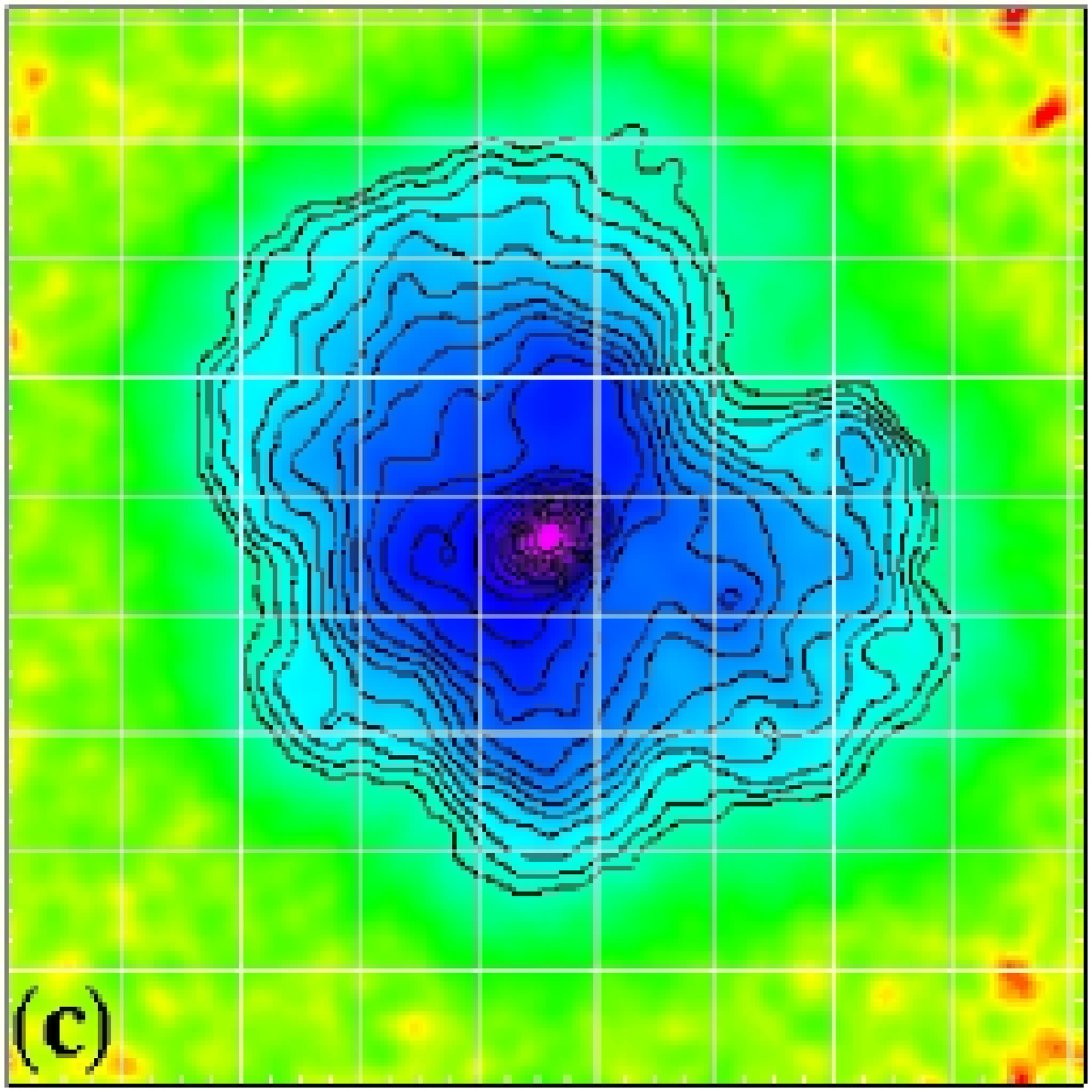}
\includegraphics[width=1.5in]{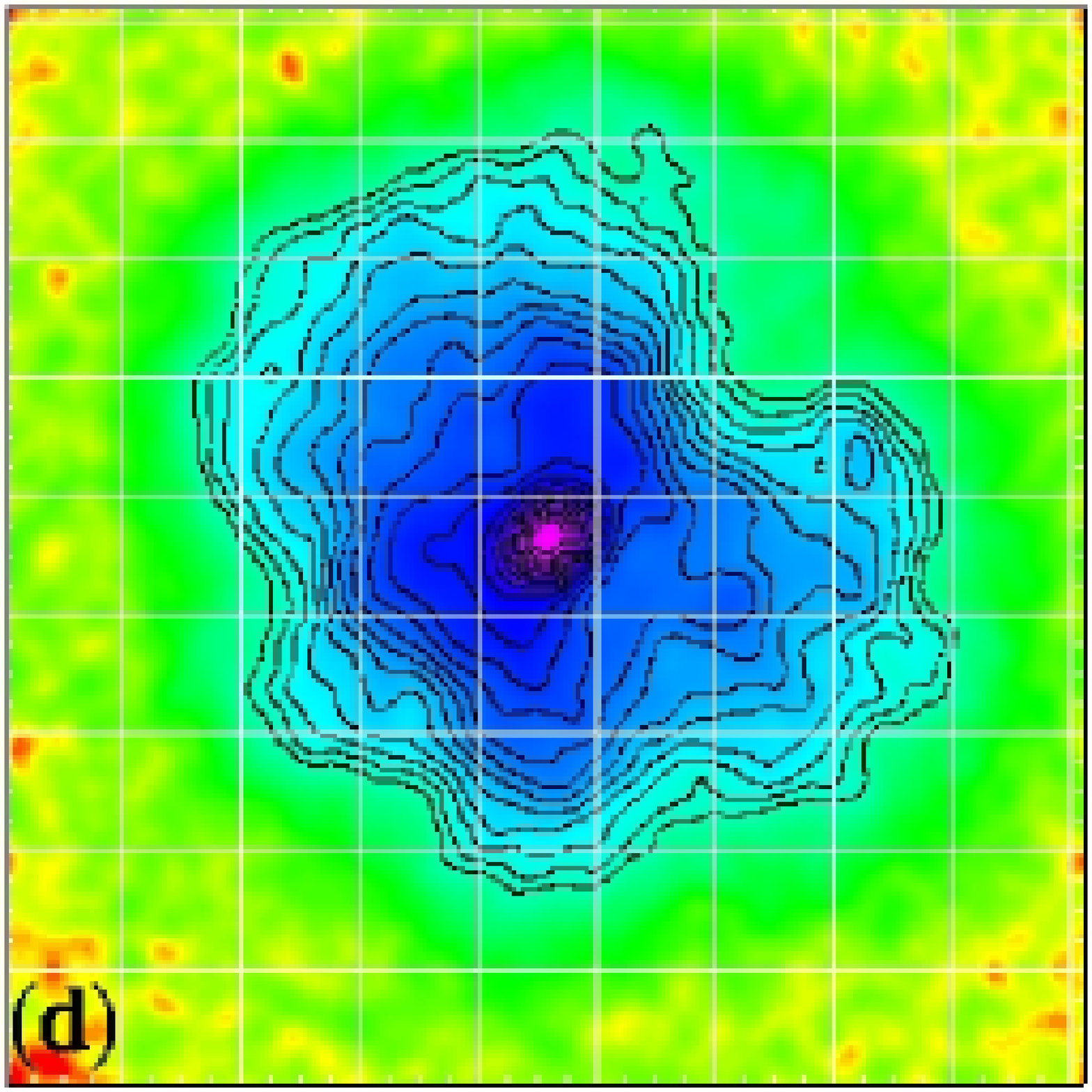}
\includegraphics[width=1.5in]{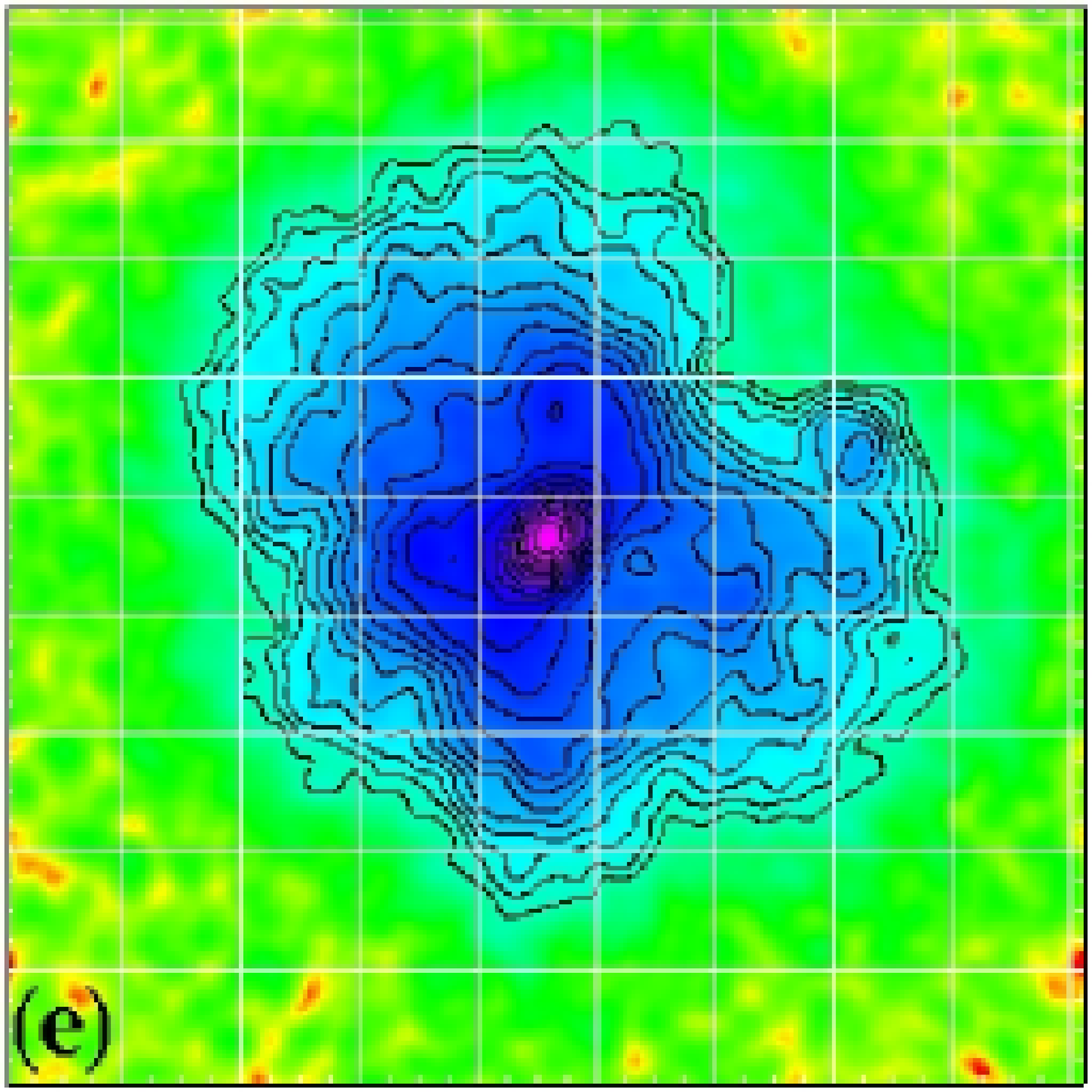}
\includegraphics[width=1.5in]{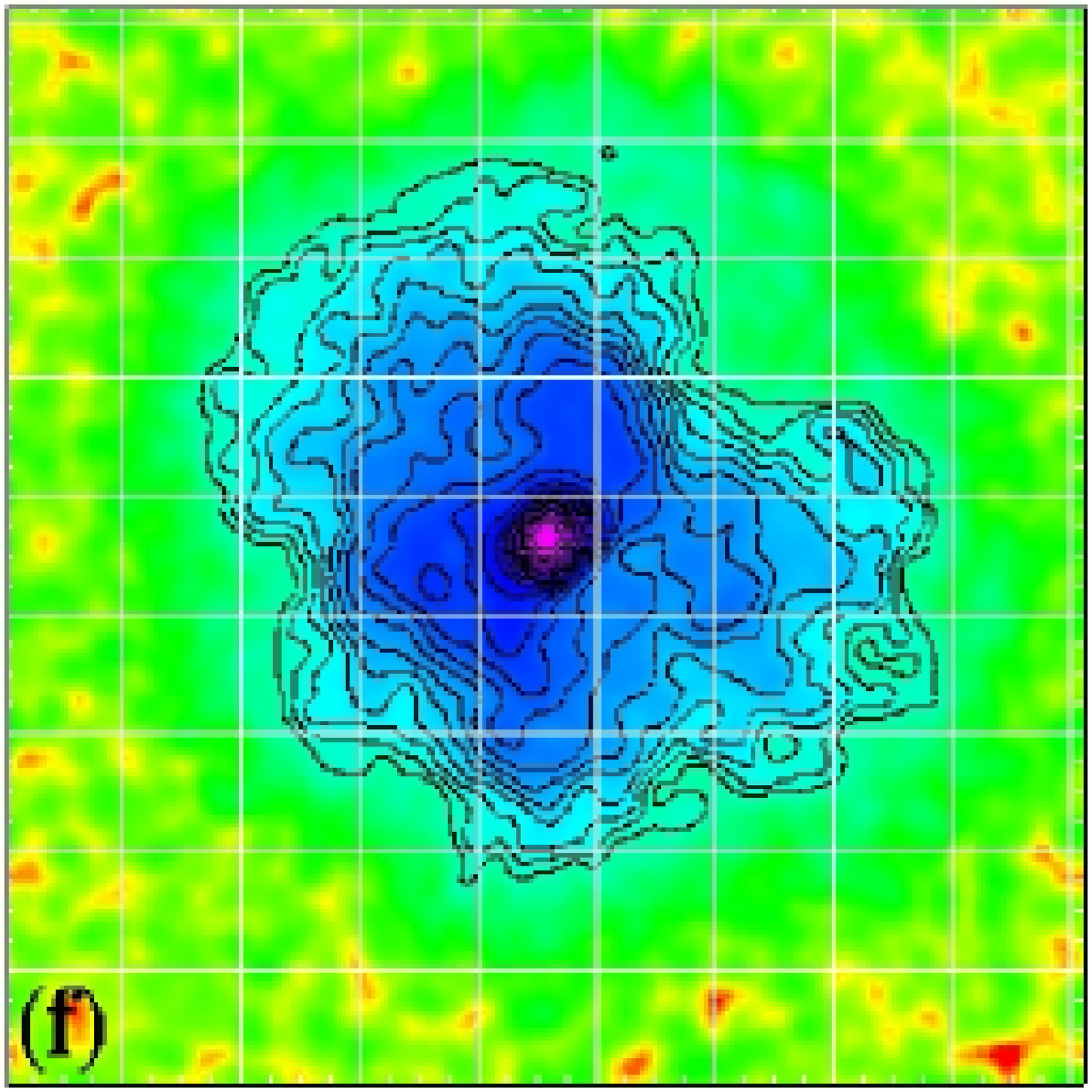}
\includegraphics[width=1.5in]{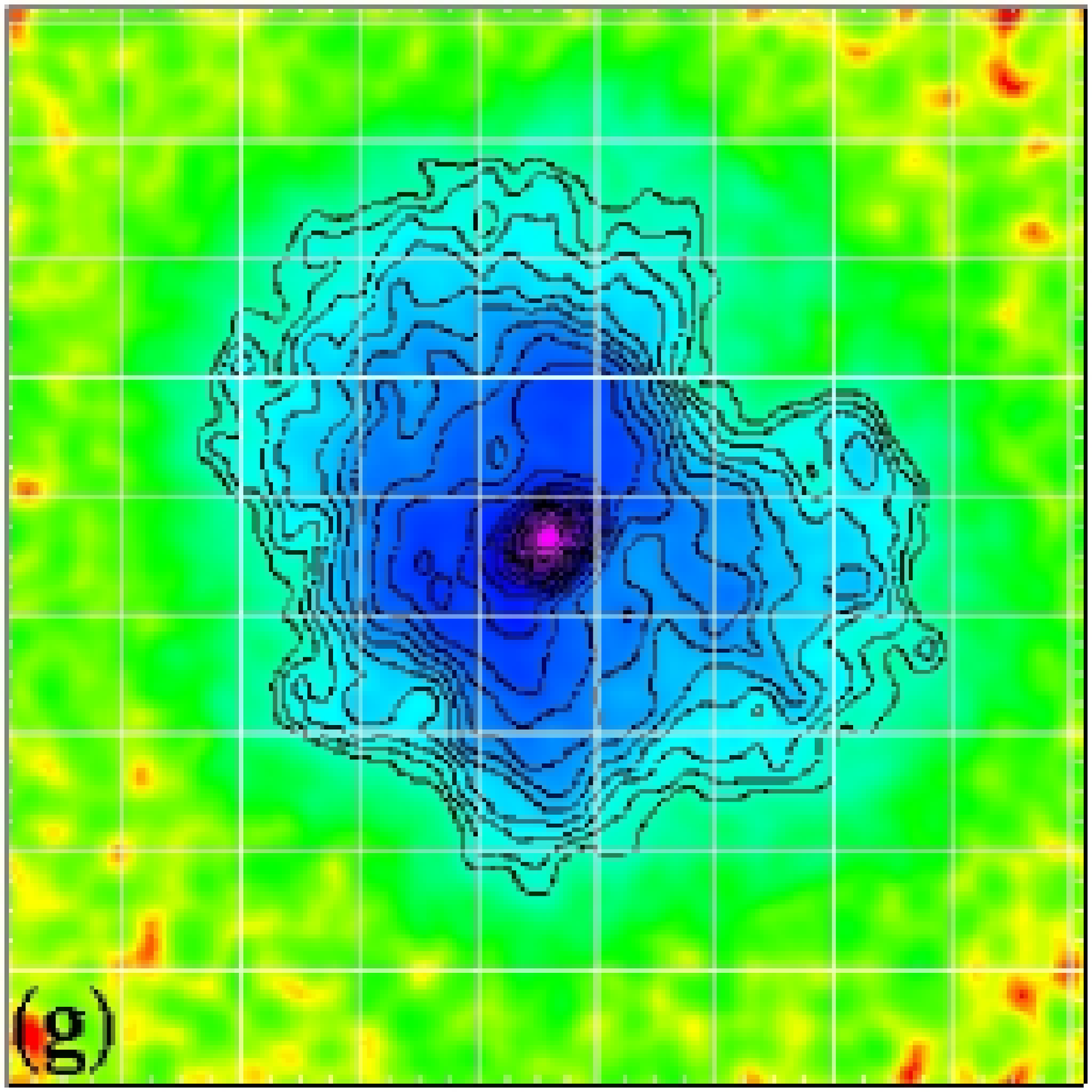}
\includegraphics[width=1.5in]{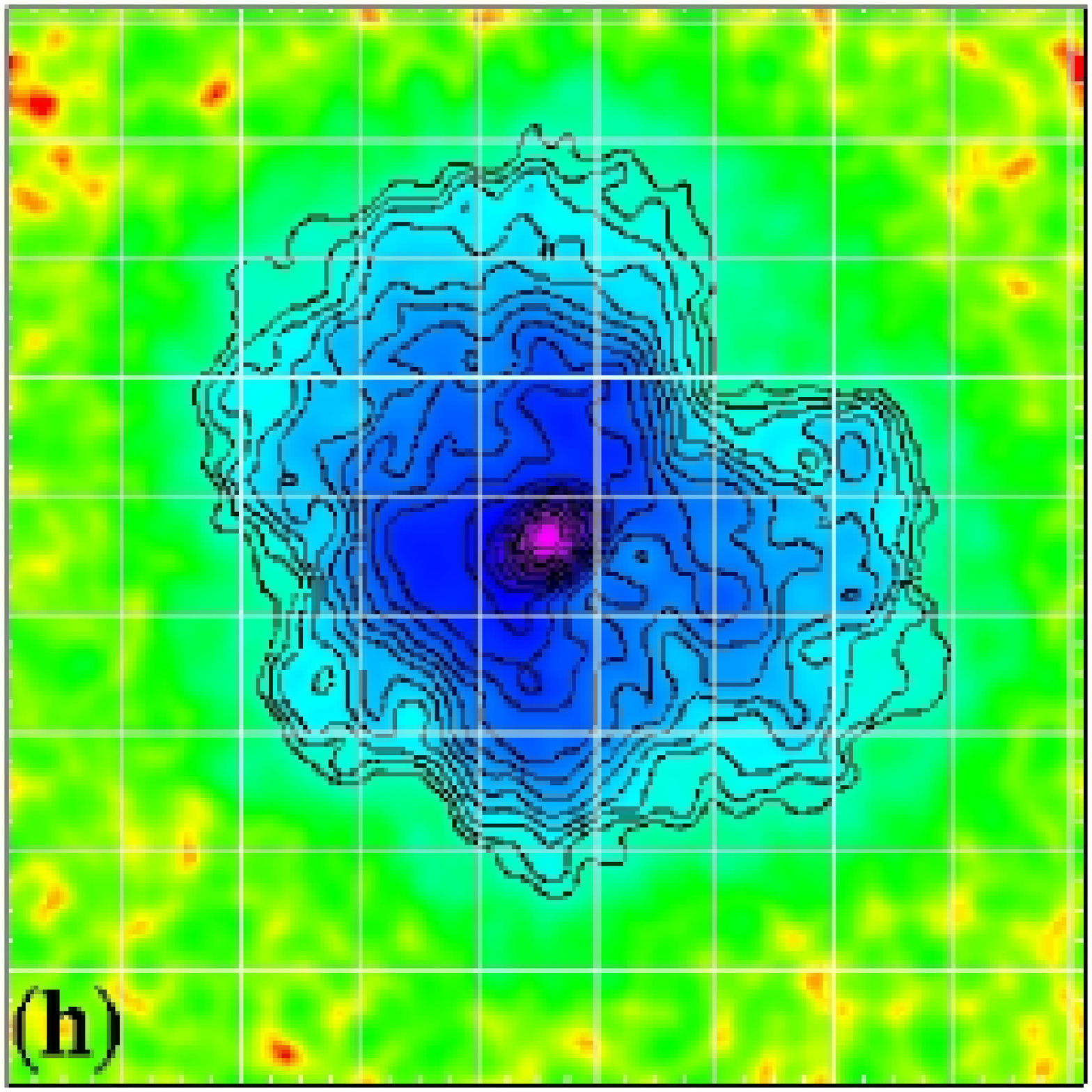}
\includegraphics[width=1.5in]{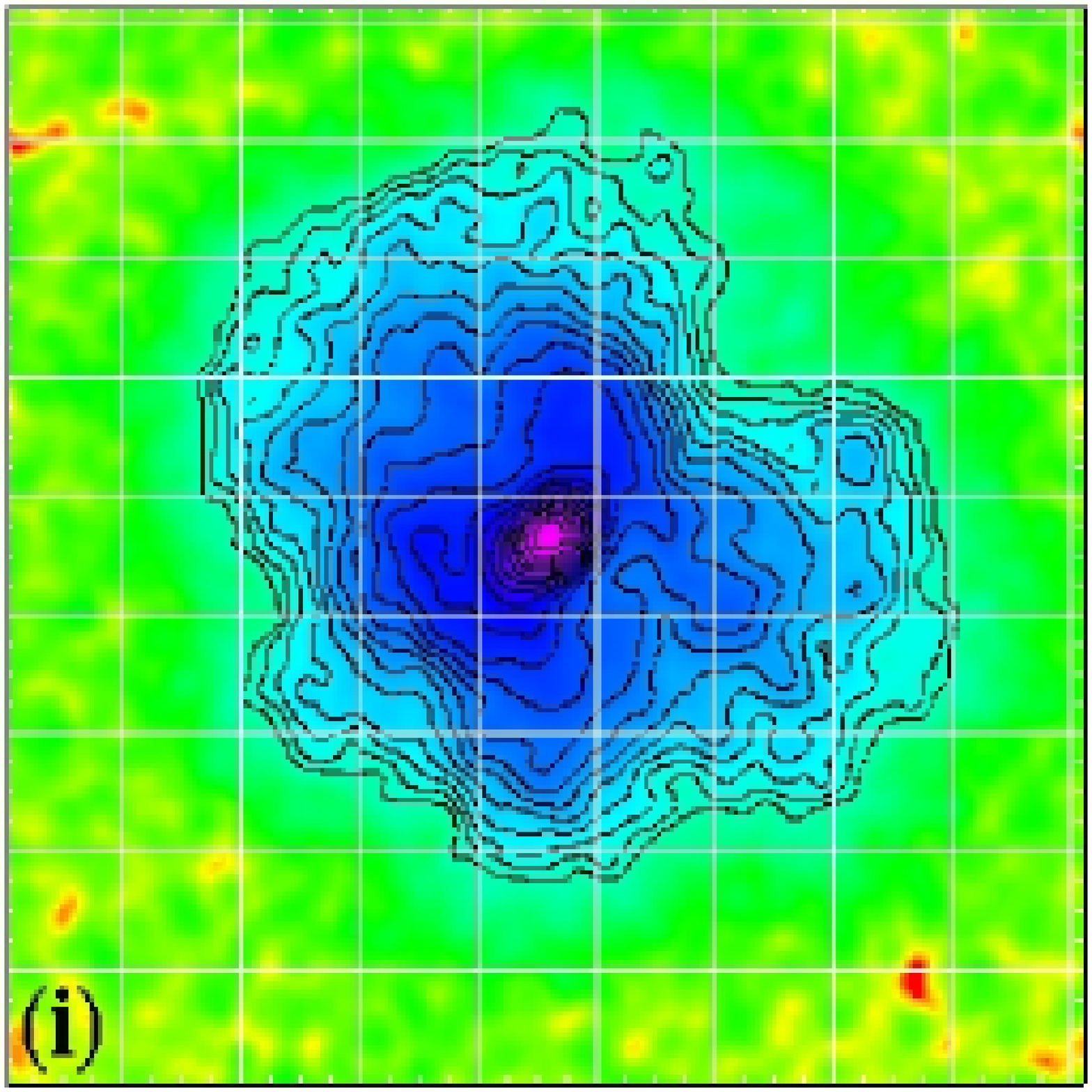}
\includegraphics[width=1.5in]{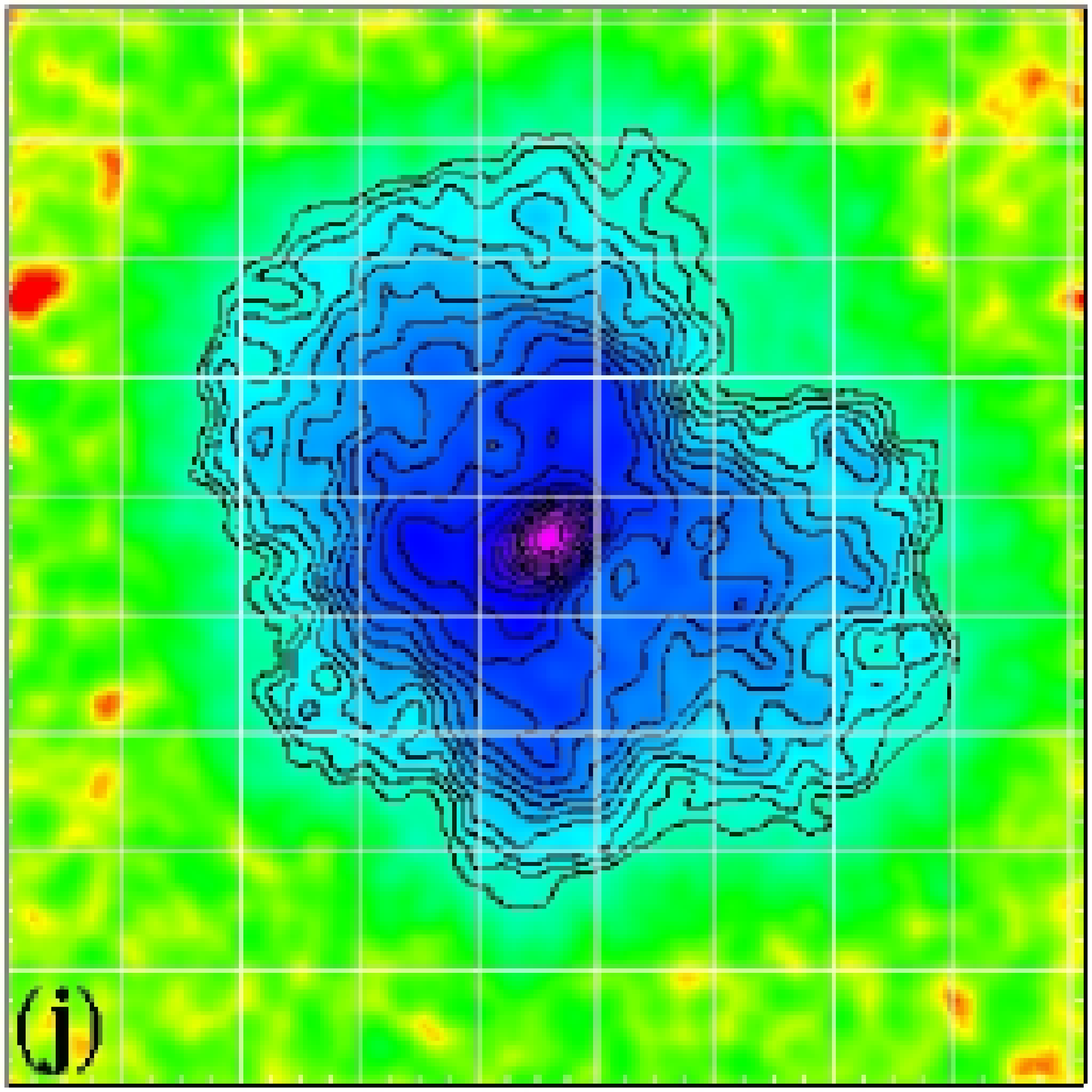}
\includegraphics[width=1.5in]{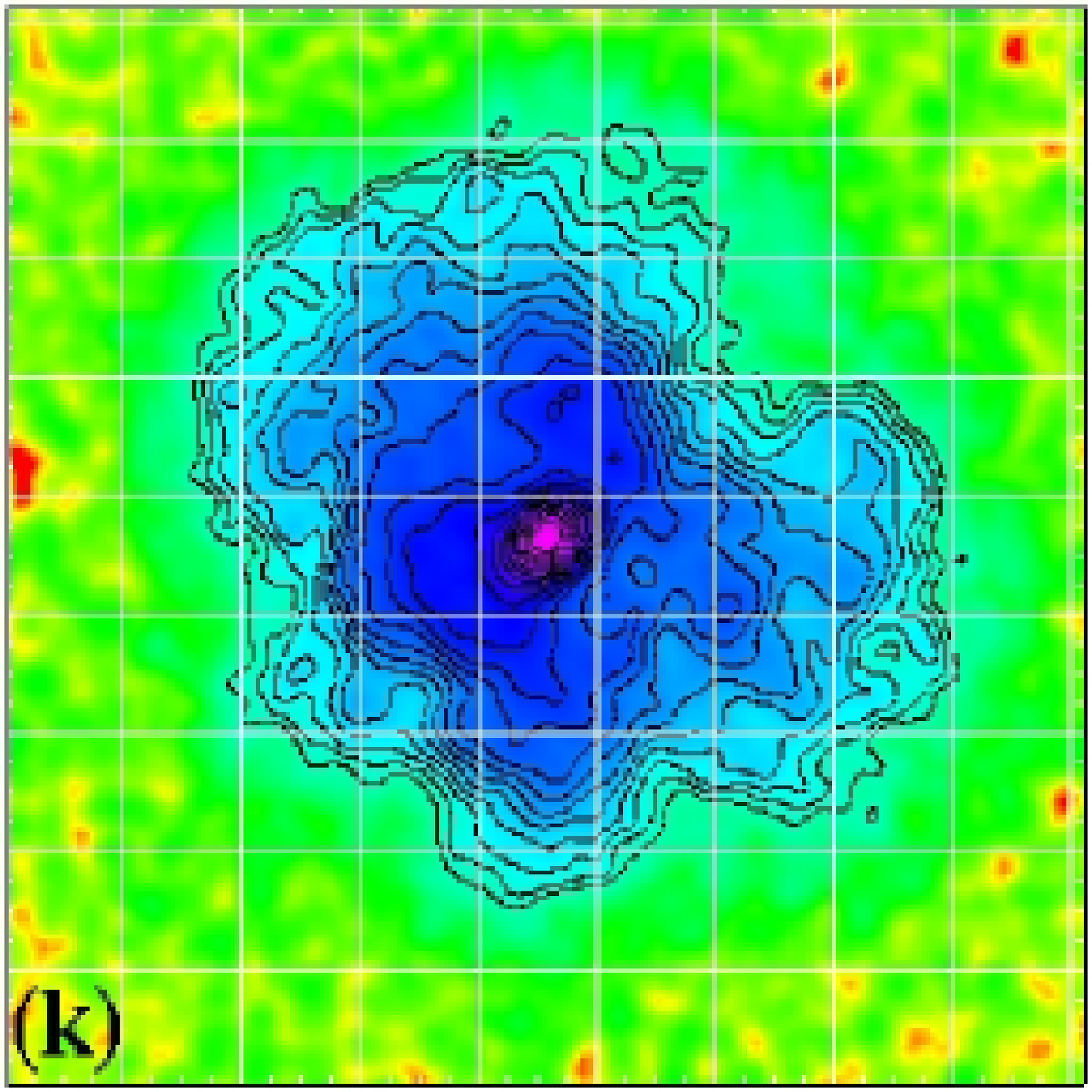}
\includegraphics[width=1.5in]{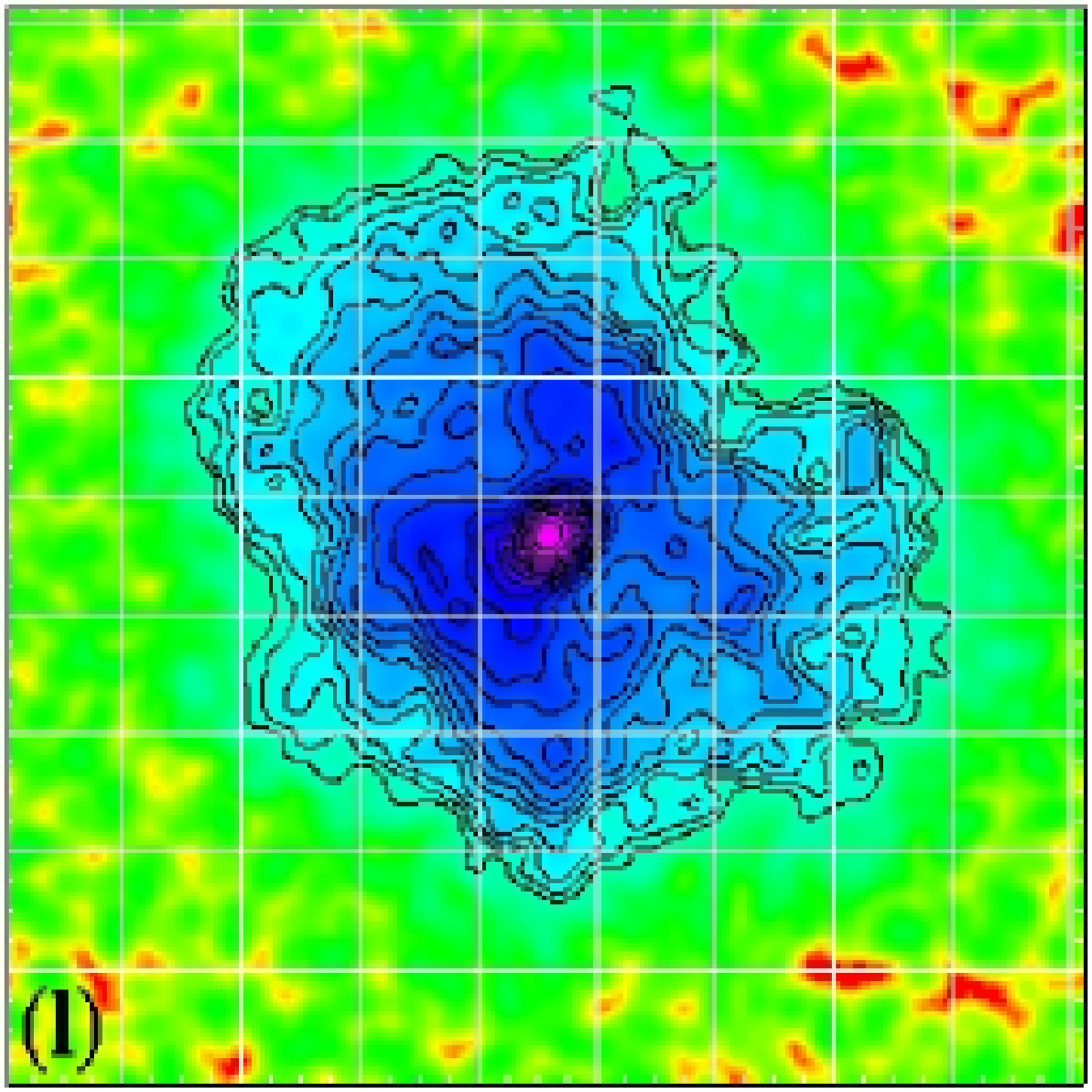}
\includegraphics[width=1.5in]{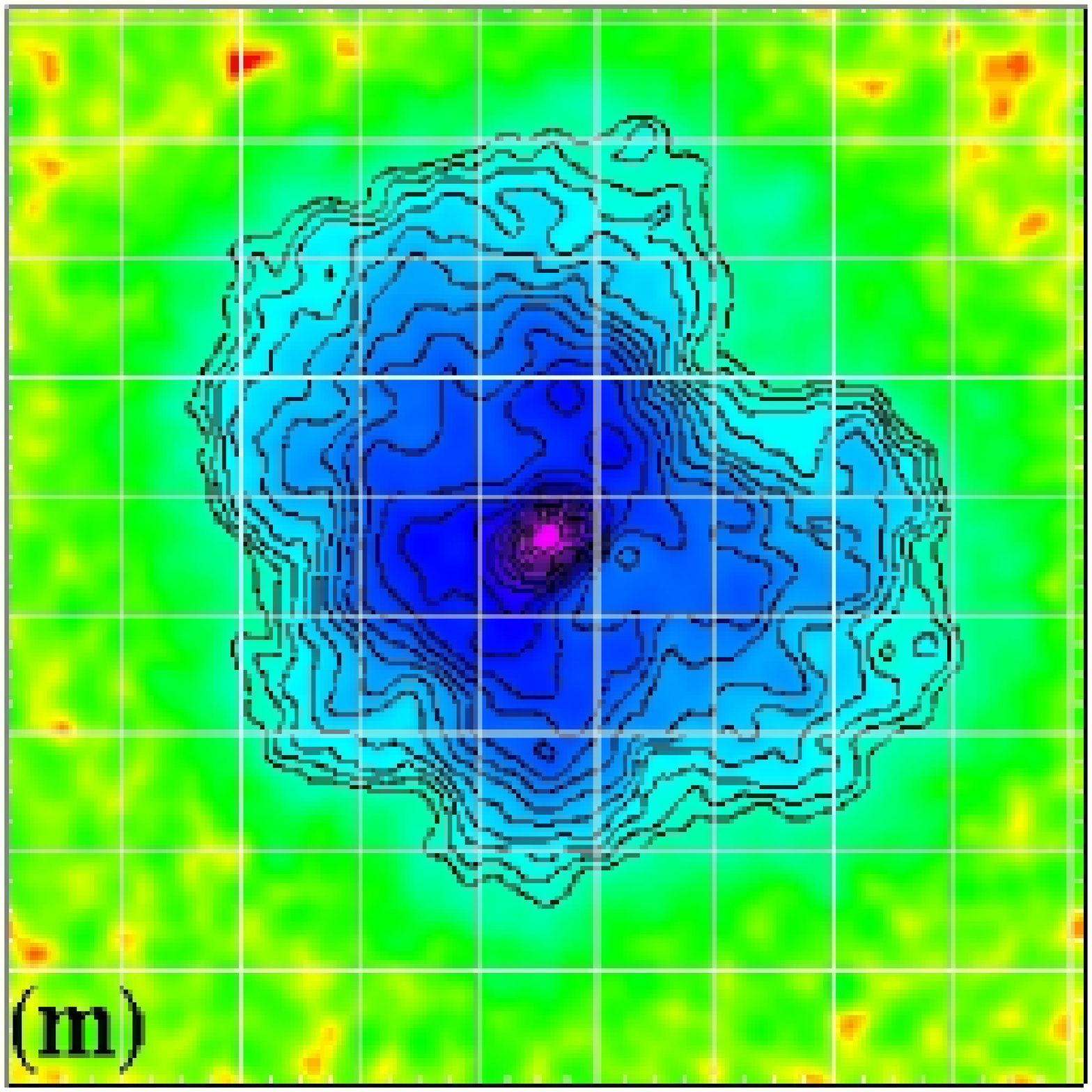}
\includegraphics[width=1.5in]{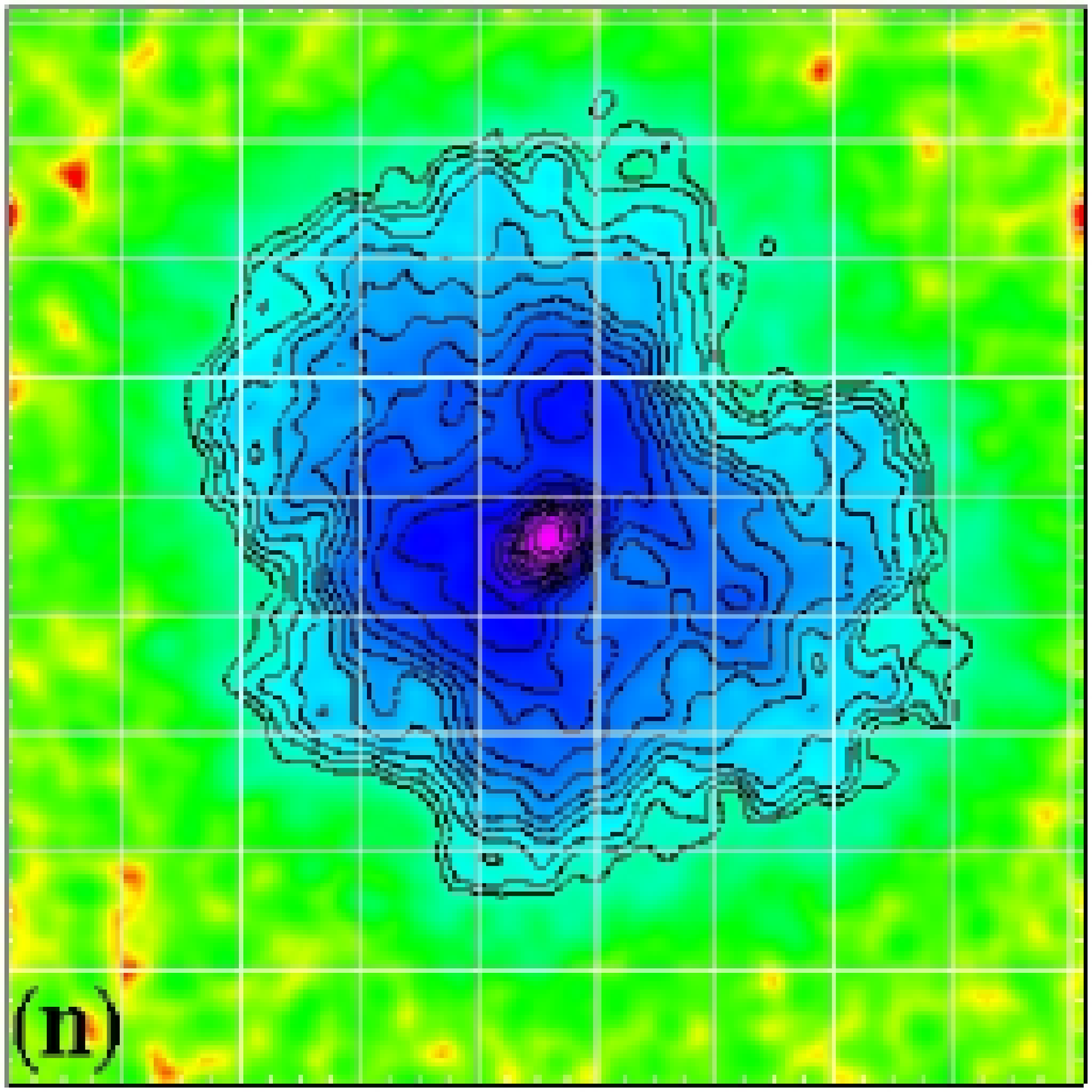}
\includegraphics[width=6.0in]{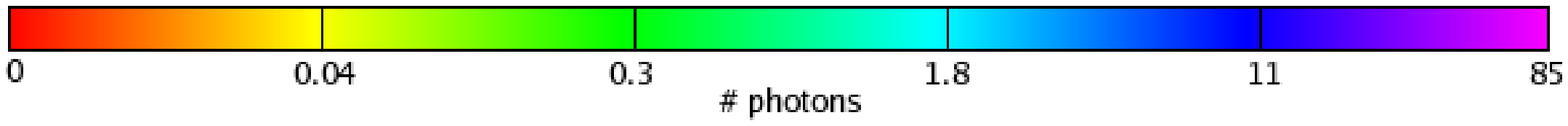}
\caption[Variability in \g21\ as seen with the ACIS.]{Combined ACIS 
  images of \g21\ for each  observing date.  All images are normalized to an effective 
  exposure of
  20~ks.  The colourbar used is a logarithmic scale and identical
  in all images.  Contours are at 1, 1.26, 1.60, 2.02, 2.55, 3.22,
  4.07, 5.14, 6.49, 8.20, 10.36, 13.09, 16.54, 20.90, 26.40, 33.36,
  42.15, 53.25, 67.28, and 85 photons per pixel.  Each image is 90\arcsec\ on a 
  side.  These images form the frames of a movie in the online version
  of the journal.  (a)~1999-08  (b)~1999-11 (c)~2000-05 (d)~2000-07 (e)~2000-09 (f)~2001-03
  (g)~2001-07 (h)~2002-09 (i)~2003-05 (j)~2003-11 (k)~2004-03
  (l)~2004-10 (m)~2005-02 (n)~2006-02.}
\label{figure:variable}
\end{center}
\end{figure}

\begin{figure}
\begin{center}
\includegraphics[width=1.5in]{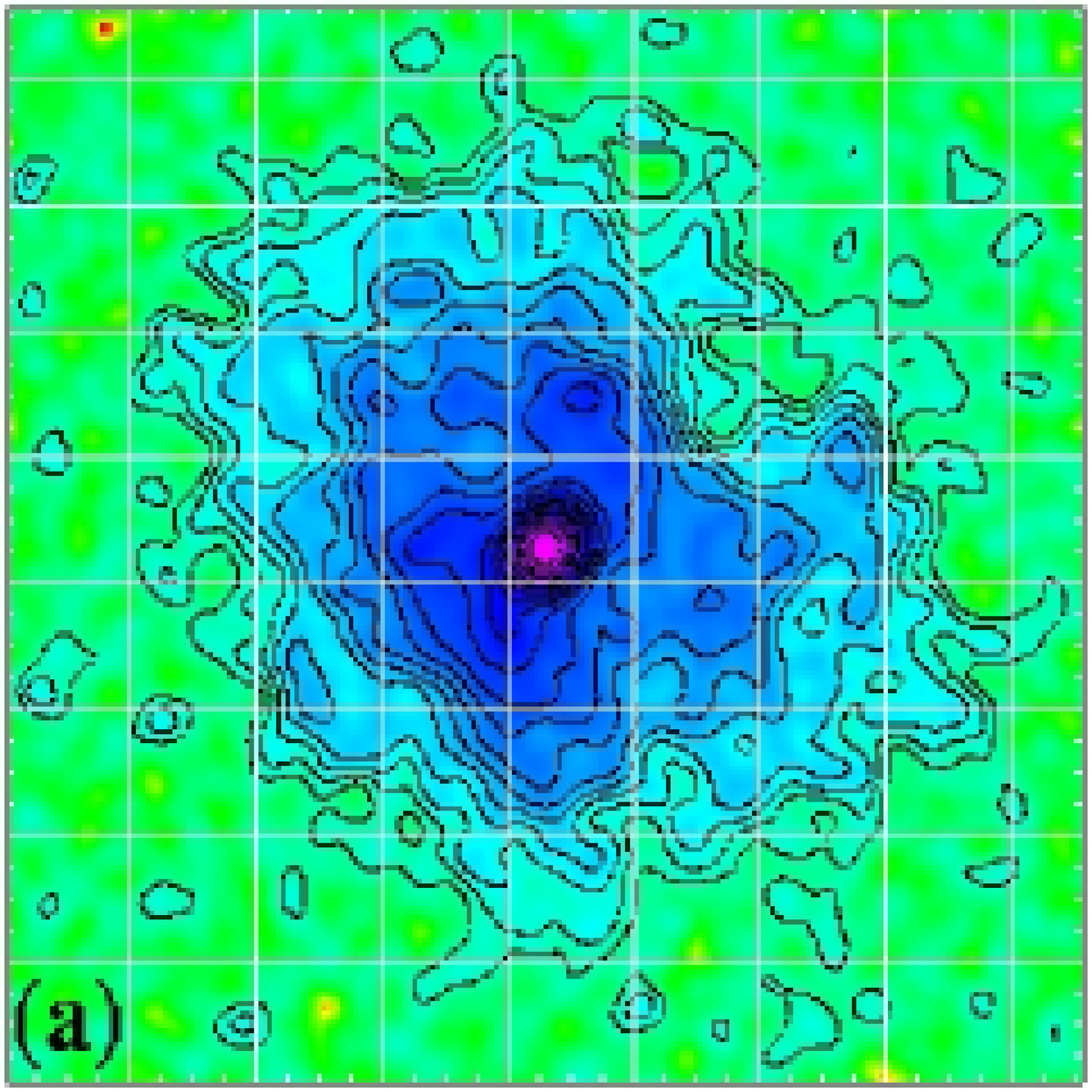}
\includegraphics[width=1.5in]{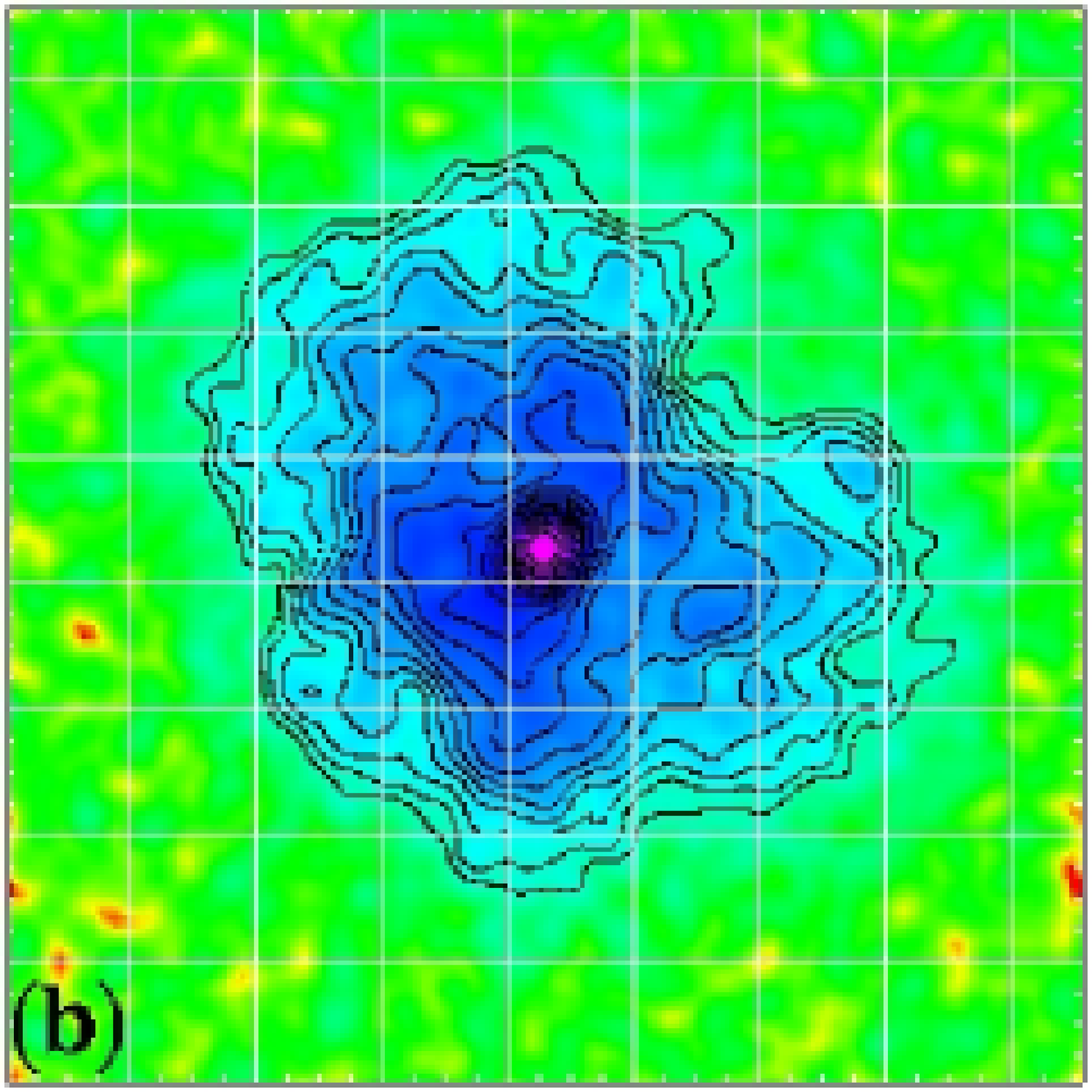}
\includegraphics[width=1.5in]{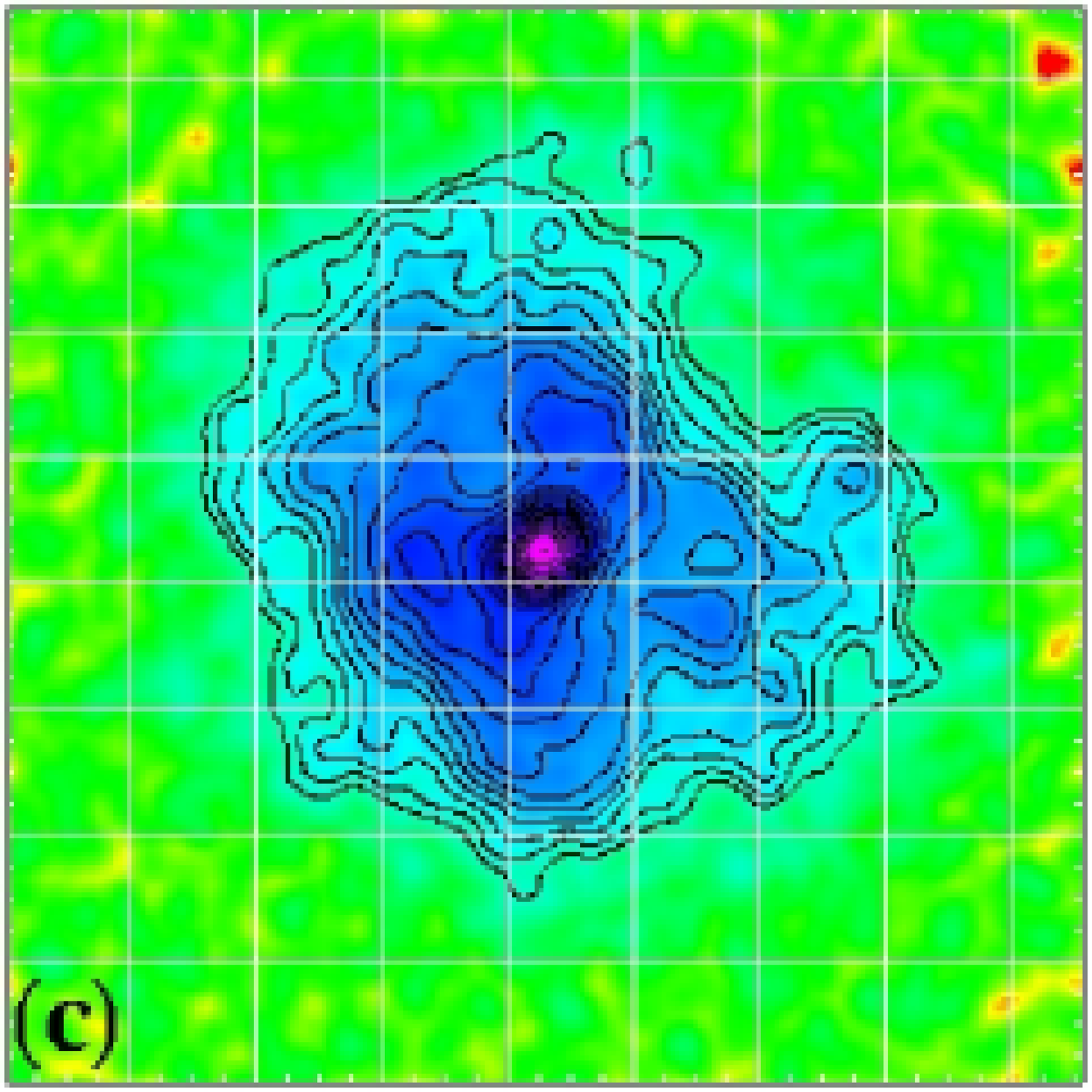}
\includegraphics[width=1.5in]{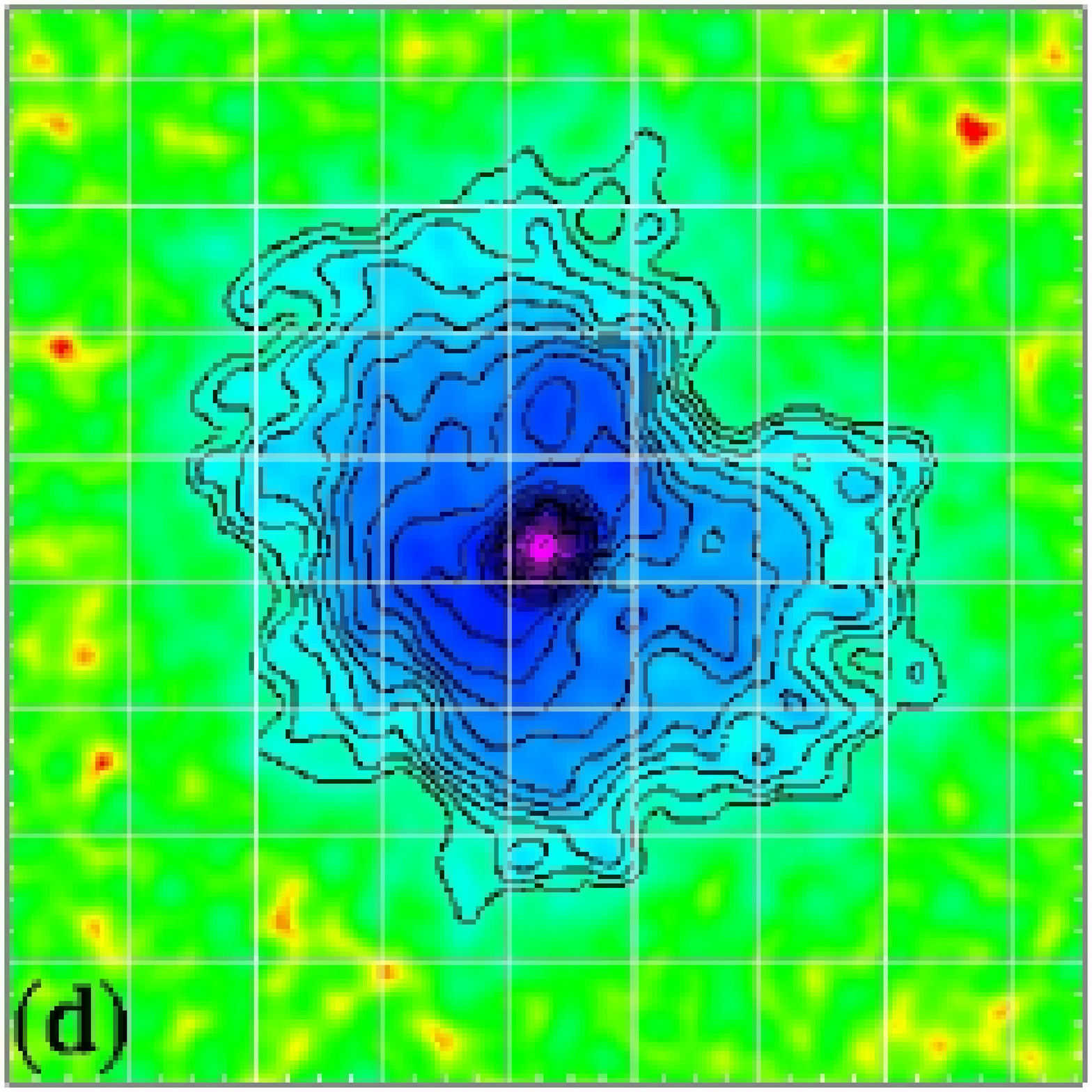}
\includegraphics[width=1.5in]{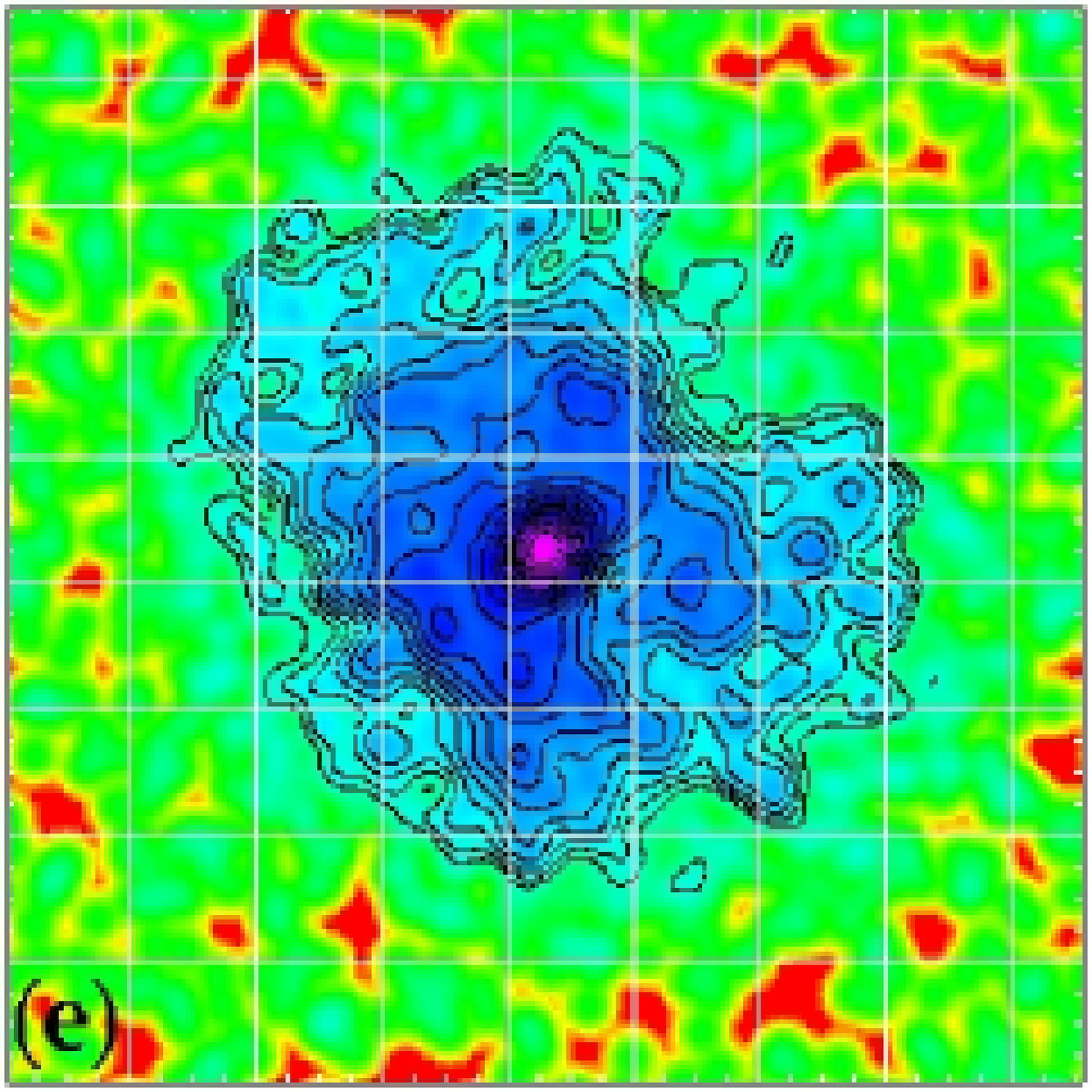}
\includegraphics[width=1.5in]{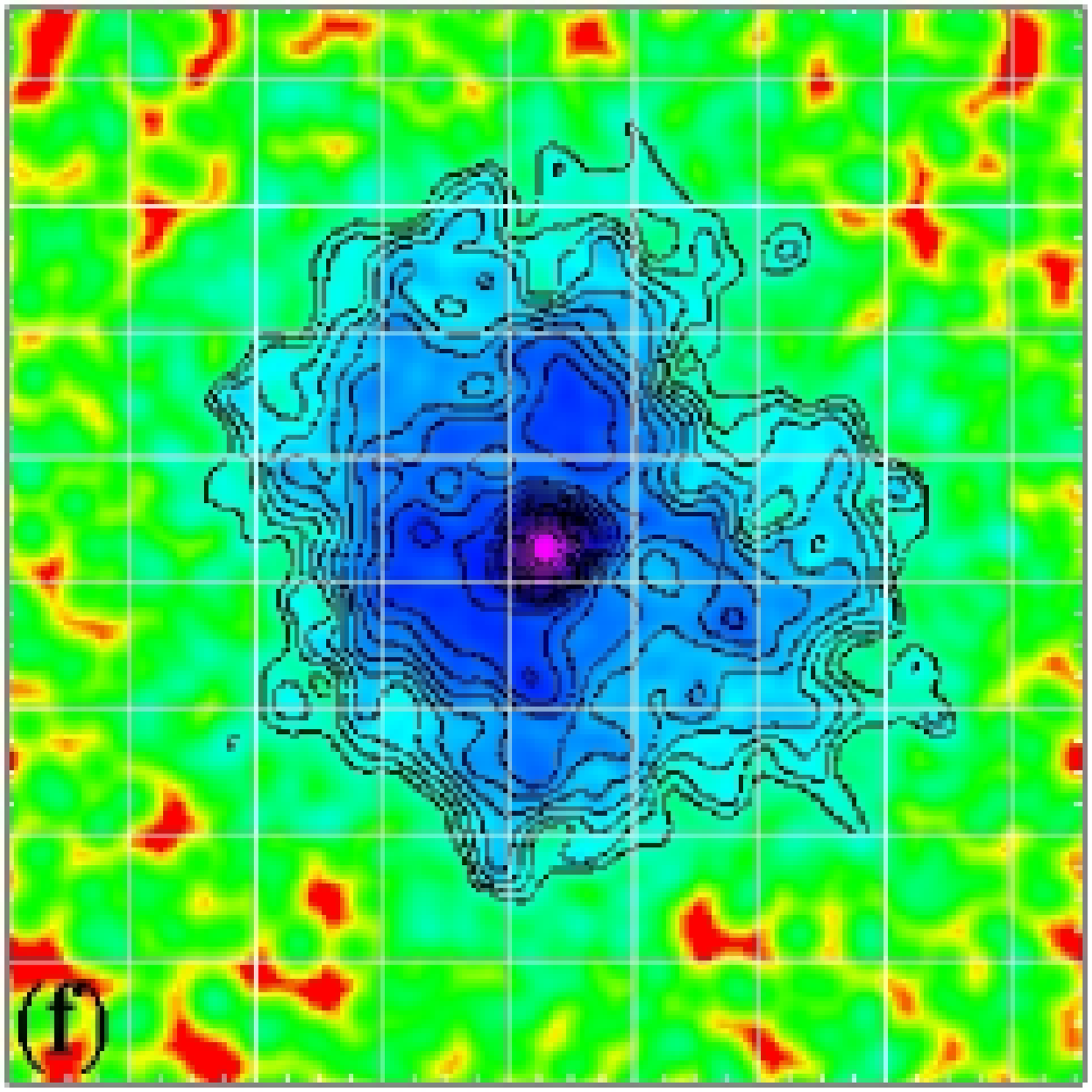}
\includegraphics[width=1.5in]{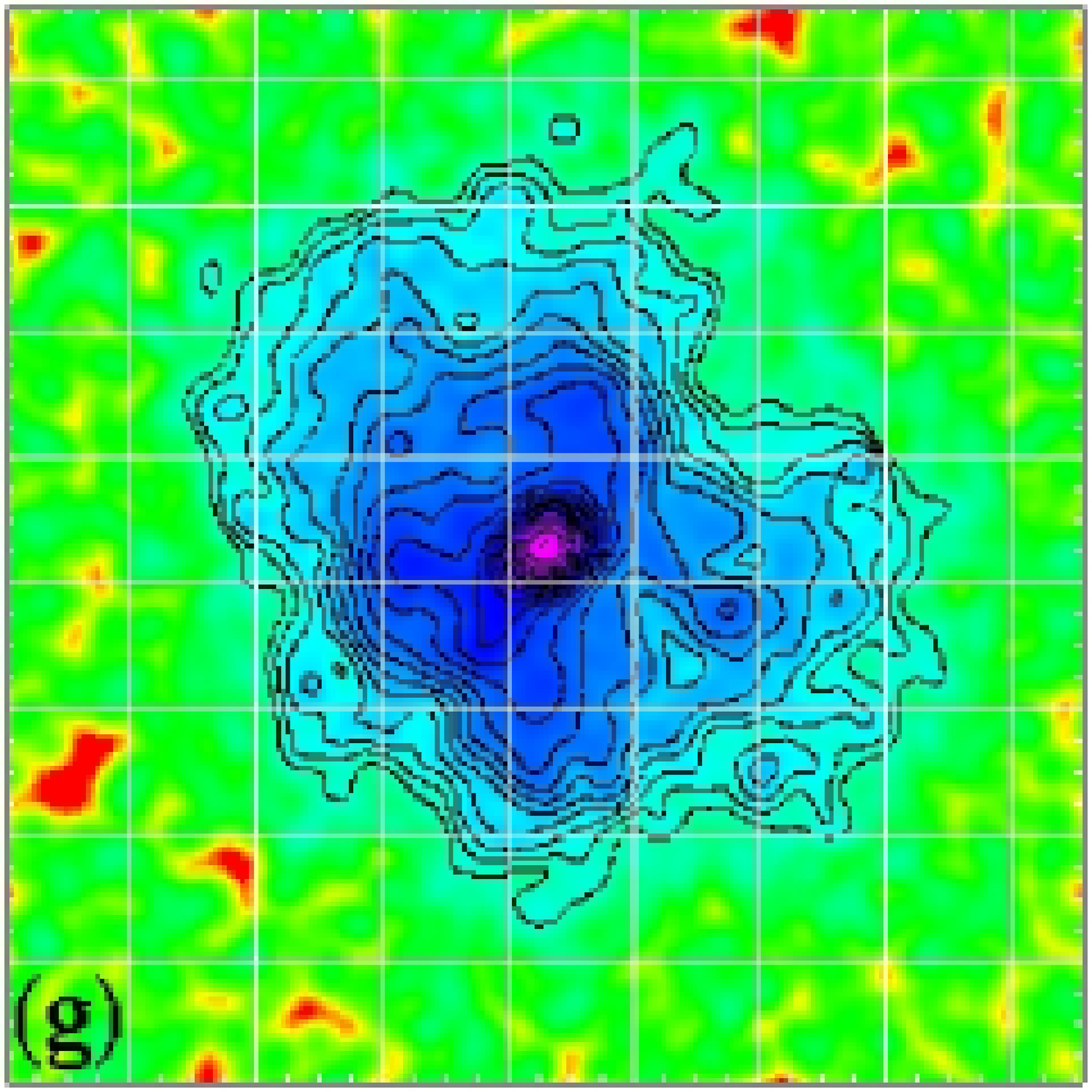}
\includegraphics[width=1.5in]{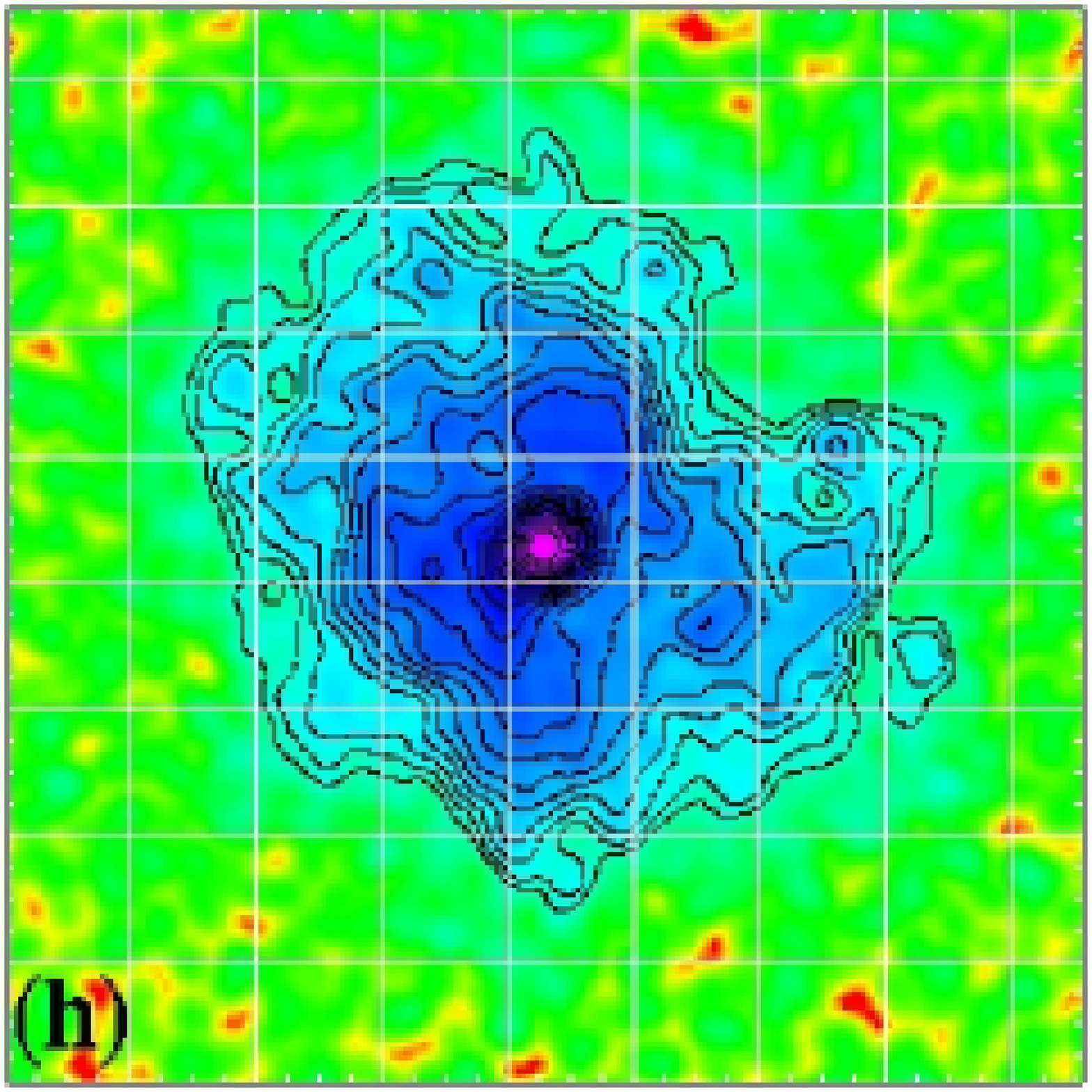}
\includegraphics[width=1.5in]{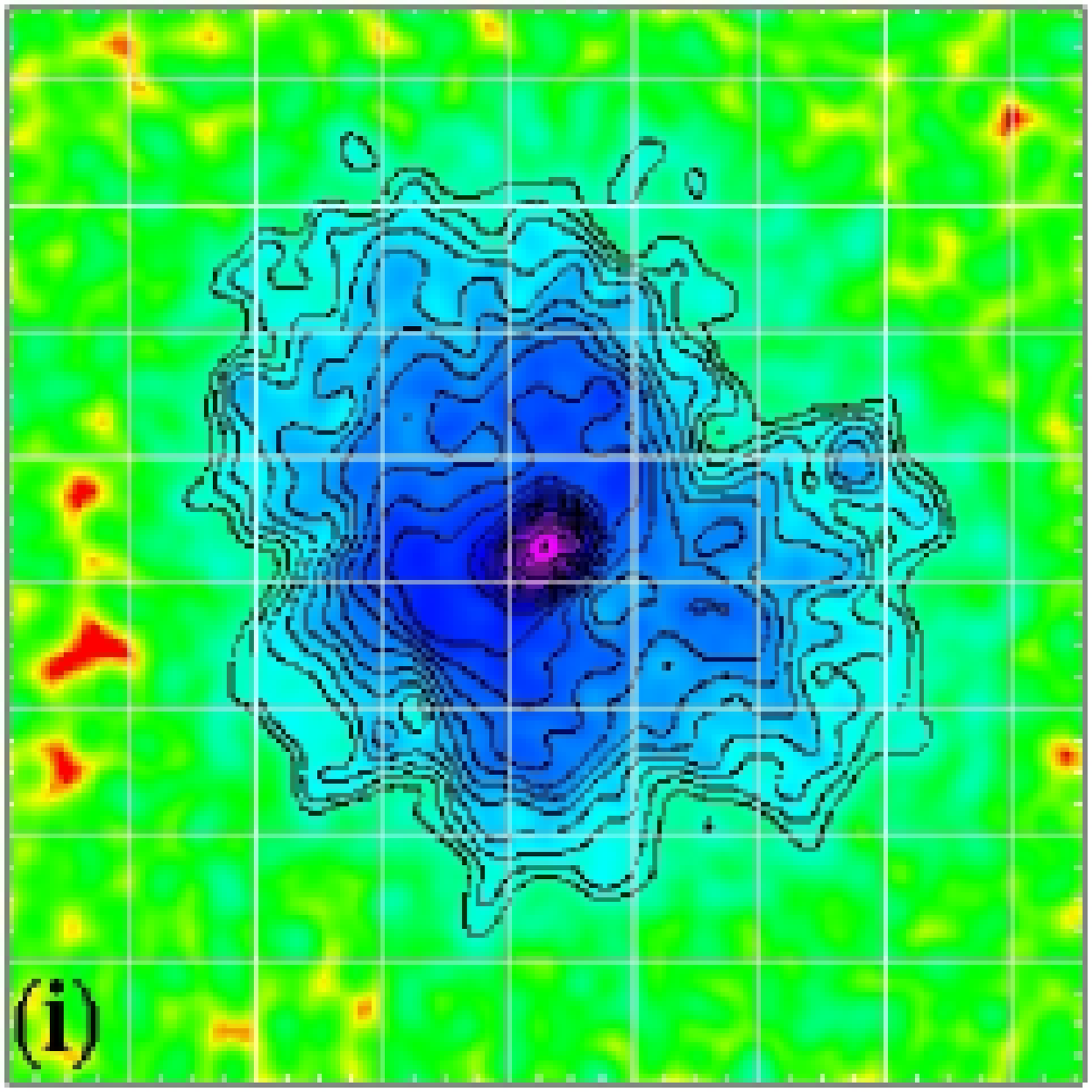}
\includegraphics[width=1.5in]{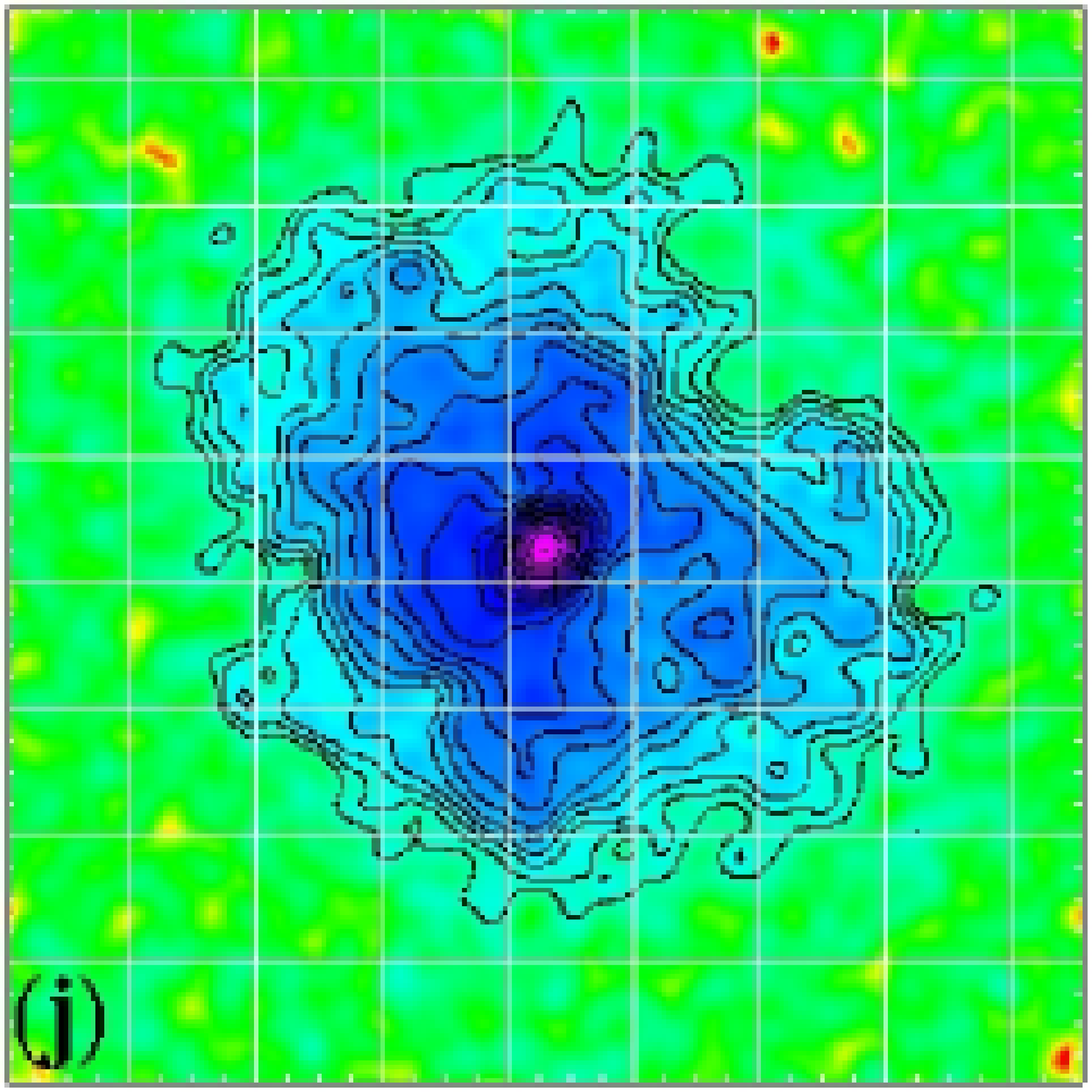}
\includegraphics[width=1.5in]{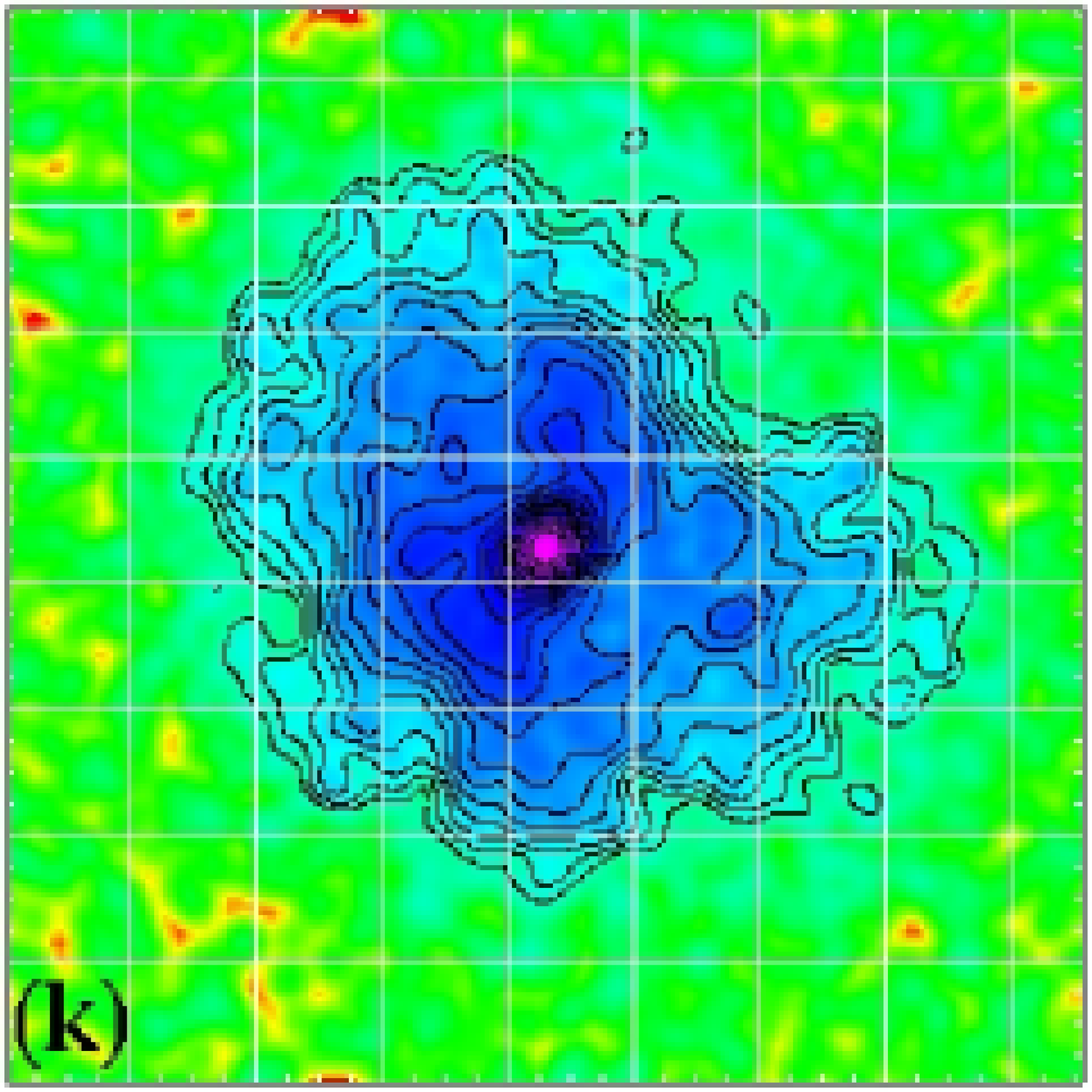}
\includegraphics[width=1.5in]{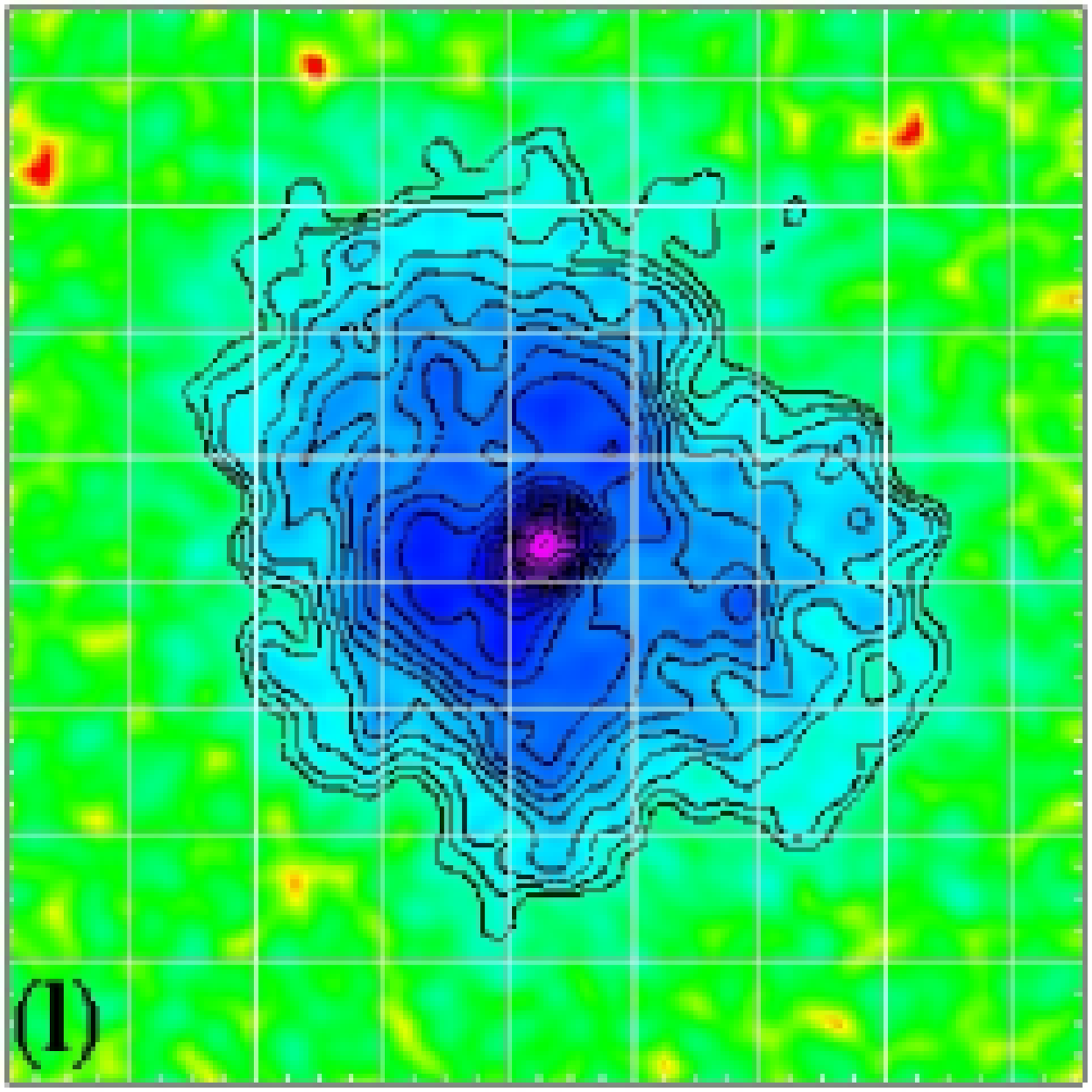}
\includegraphics[width=1.5in]{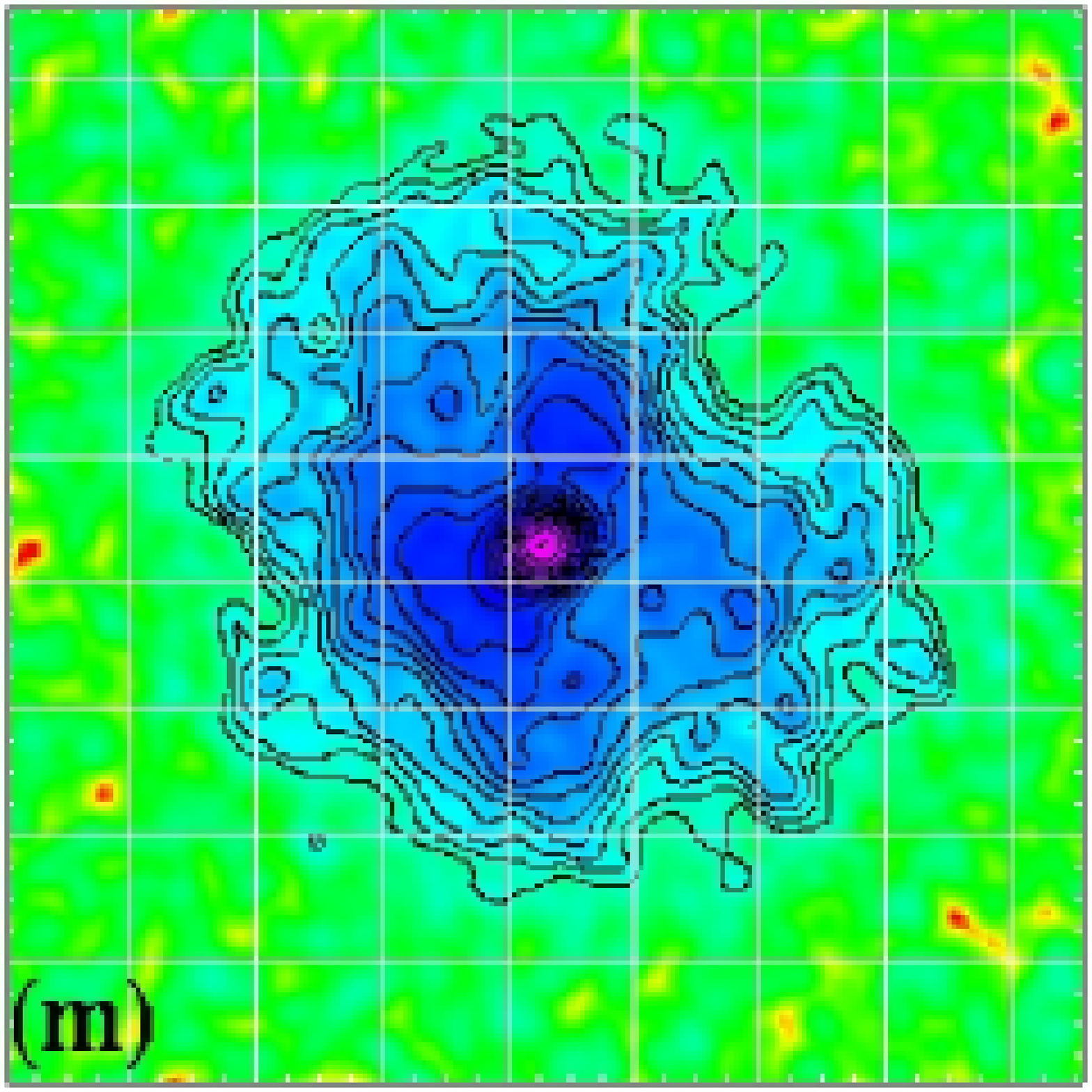}
\includegraphics[width=6.0in]{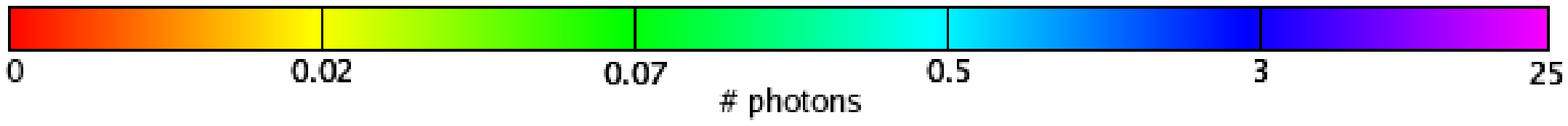}
\caption[Variability in \g21\ as seen with the HRC-I.]{Combined HRC-I 
  images of \g21\ for each  observing date.  All images are normalized to an effective 
  exposure of
  20~ks.  The colourbar used is a logarithmic scale and identical
  in all images.  Contours are at 0.3, 0.38, 0.48, 0.60, 0.76, 0.96,
  1.21, 1.53, 1.93, 2.43, 3.08, 3.88, 4.90, 6.19, 7.81, 9.85, 12.44,
  15.69, 19.81, and 25 photons per pixel.  Each image is 90\arcsec\ on a side.
  These images form the frames of a movie in the online version of the
  journal. 
  (a)~1999-09  (b)~1999-10 (c)~2000-02 (d)~2000-09 (e)~2001-03 (f)~2001-07
  (g)~2002-03 (h)~2002-07 (i)~2003-05 (j)~2003-11 (k)~2004-03
  (l)~2005-02 (m)~2006-02.}
\label{figure:variablehrc}
\end{center}
\end{figure}

\begin{figure}
\begin{center}
\plotone{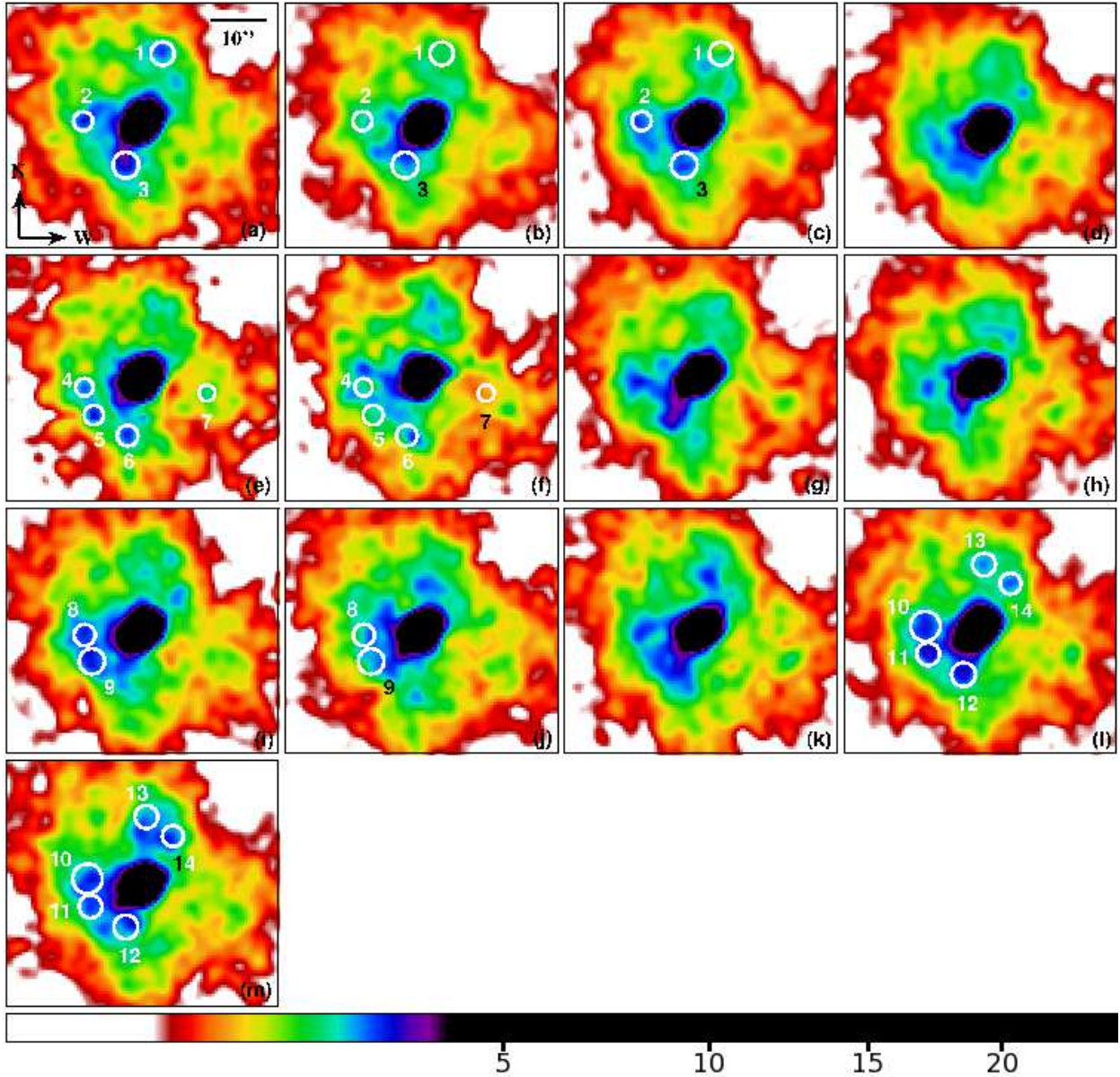}
\caption[Motion of knots in \g21.]{Variability in the PWN \g21.  HRC-I
  images from the dates (a)~1999-09  (b)~1999-10 (c)~2000-02
(d)~2000-09 (e)~2001-03 (f)~2001-07
  (g)~2002-03 (h)~2002-07 (i)~2003-05 (j)~2003-11 (k)~2004-03
  (l)~2005-02 (m)~2006-02.  The numbered circles mark the location of some
  knots of emission which either move or fade by the next image in the
  sequence.  See Section~\ref{section:variable} for details.}
\label{figure:hrc_se}
\end{center}
\end{figure}

\begin{figure}
\begin{center}
\plottwo{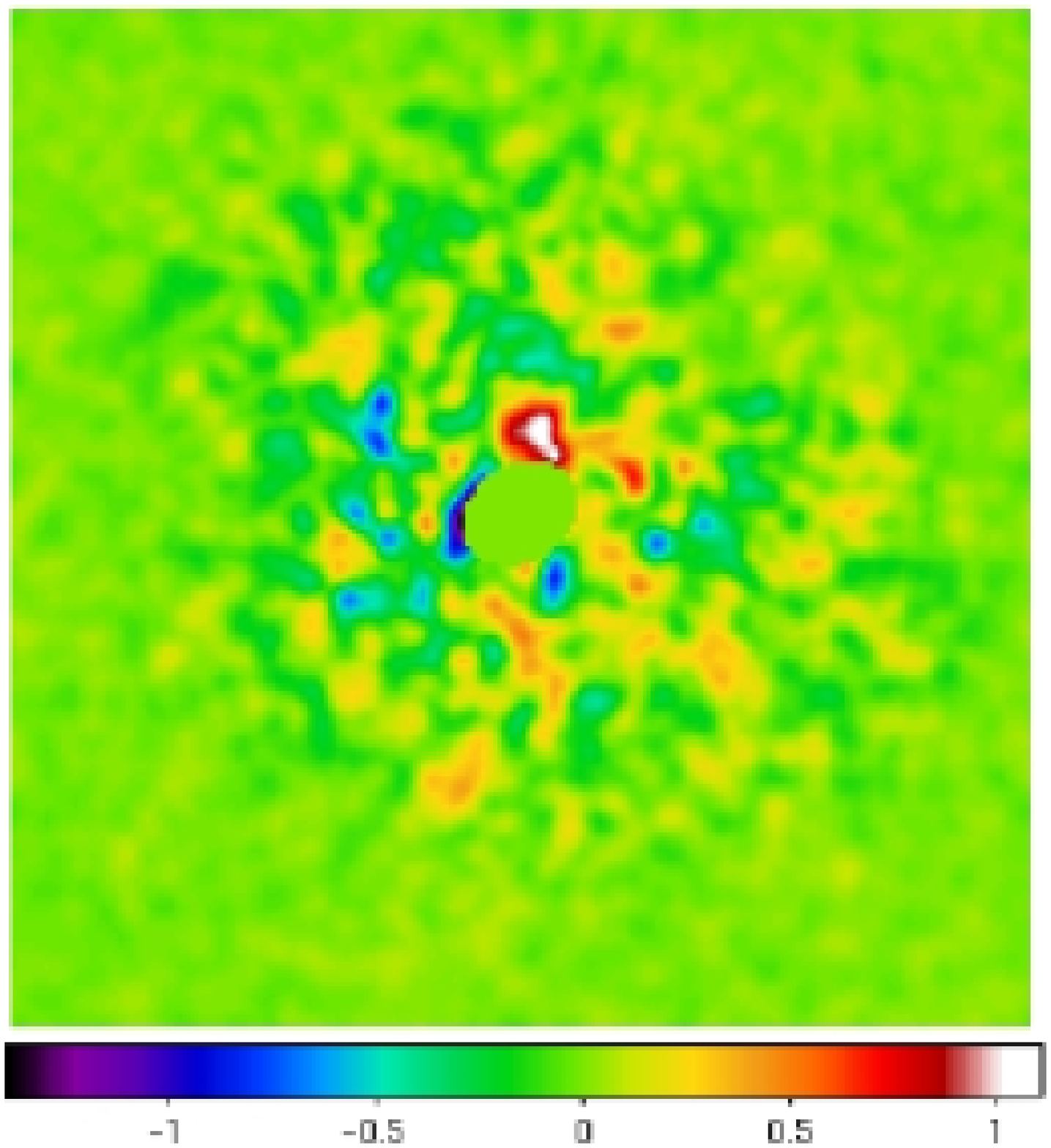}{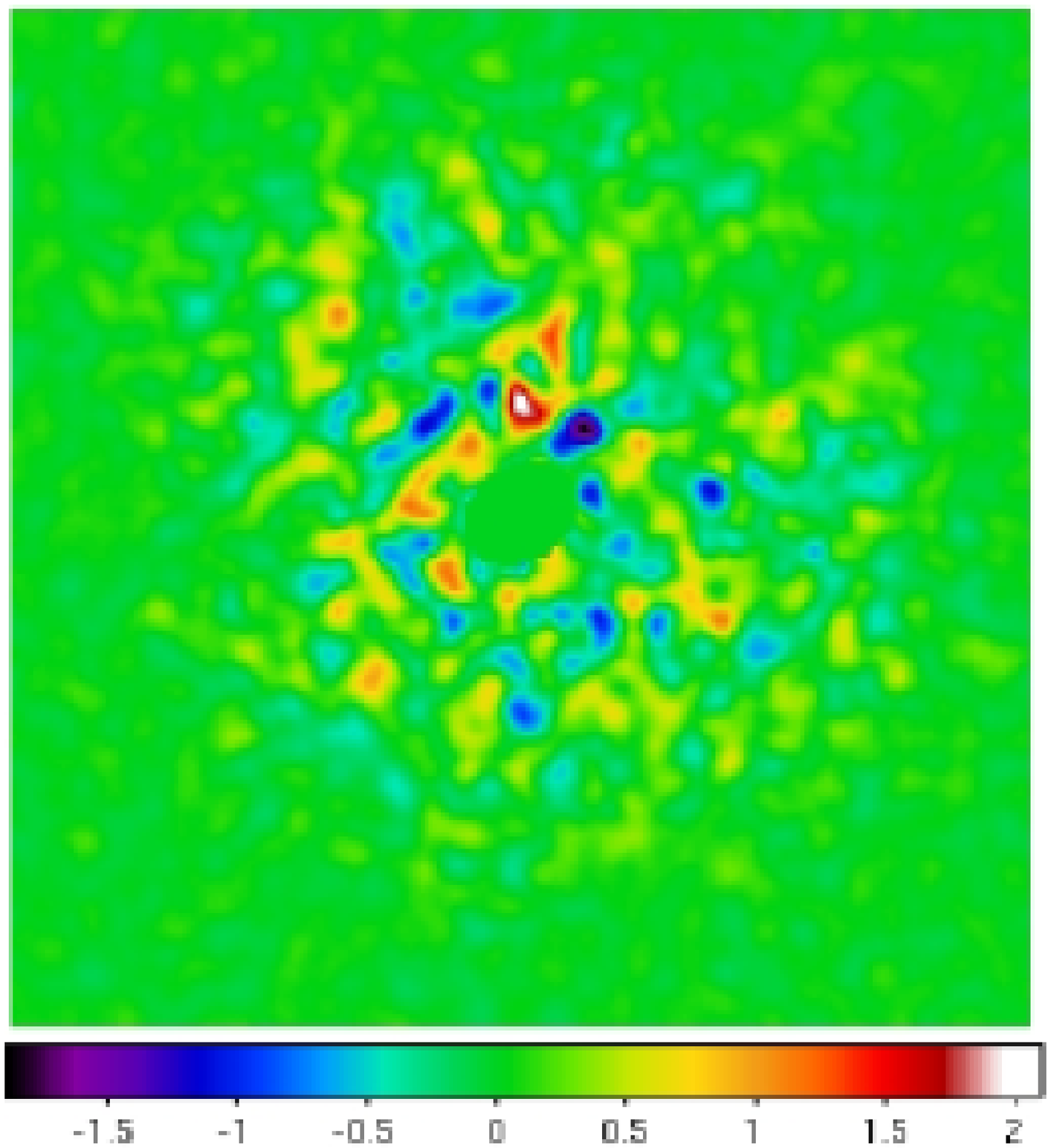}
\caption[Difference images of PWN]{Sample difference images created by
subtracting one of the images in Figure~\ref{figure:variable} from the
following image in the sequence.  The emission near the pulsar was
omitted to highlight the variability in the PWN. White and red
indicate regions that are brighter in the later image.  Black, purple, and blue
indicate regions that are brighter in the earlier image. Colourbars
are in units of counts per pixel.  (a)~2000-07 minus
2000-05 (b)~2006-02 minus 2005-02.}
\label{figure:differences}
\end{center}
\end{figure}

\begin{figure}
\begin{center}
\plotone{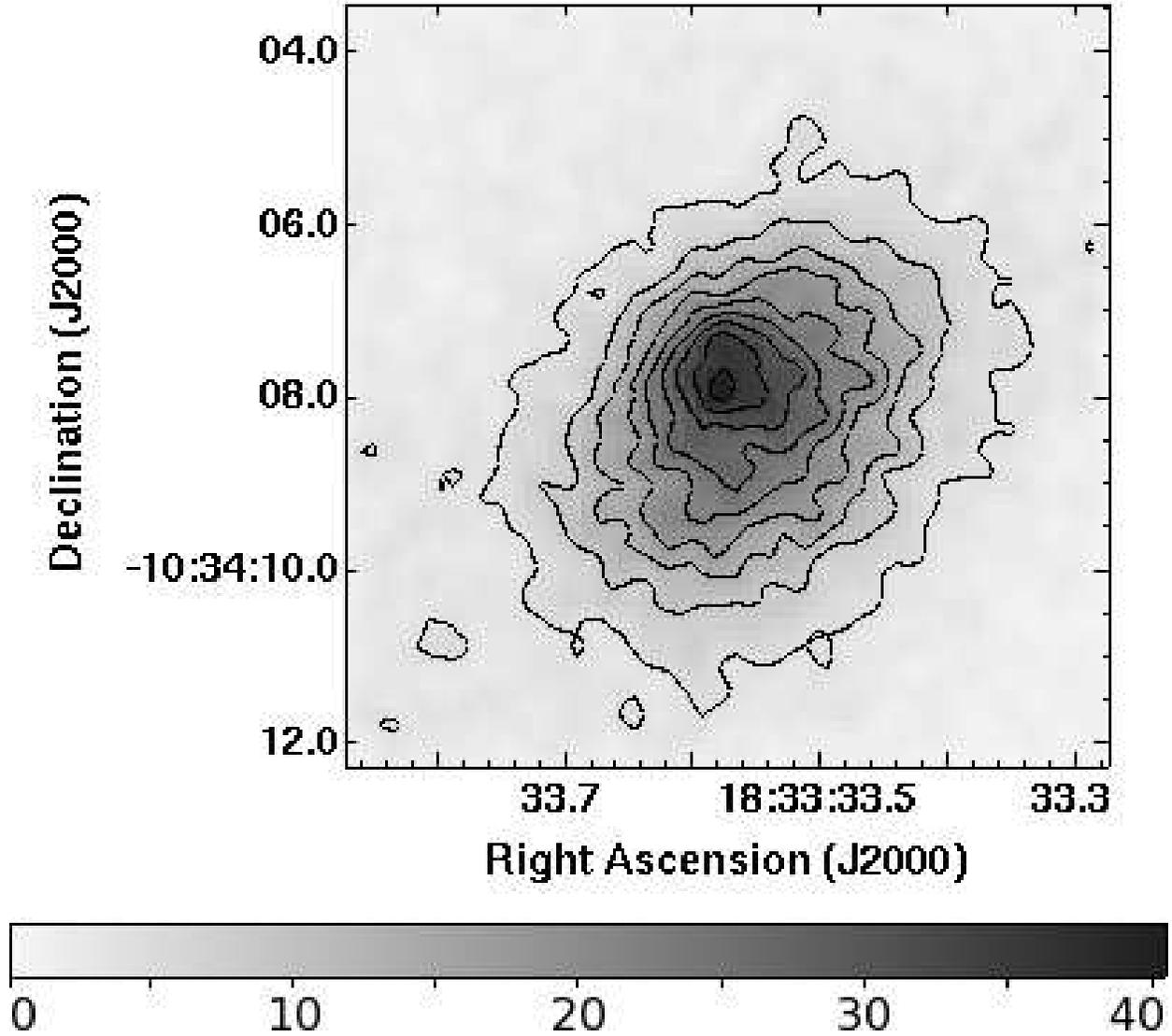}
\caption[HRC-I image of the central emission.]{HRC-I image of the
  central emission of \g21\ (smoothed by 2 pixels).  The pulsar appears offset to the
  northeast.  The image is 9\arcsec\ across.  Contours are linearly spaced at 4.3, 8.6, 12.8, 17.1, 21.4, 25.7, 29.9, 34.2, and 38.5 counts per pixel.}
\label{figure:psronly}
\end{center}
\end{figure}

\acknowledgments
Thanks to Roland Kothes for providing the 22.3 GHz radio data used in
Figure~\ref{figure:xrayradio}. Thanks also to Michael Bietenholz for
providing the 4.75 GHz VLA map of the PWN shown in
Figure~\ref{figure:xrayradio}, and for comments on the manuscript.  We thank the referee for comments on the manuscript.  
H.M. acknowledges the support of the Natural Sciences and 
Engineering Research Council of Canada (NSERC) in the form of a 
Canada Graduate Scholarship and the support of the Province of Manitoba
in the form of a Manitoba Graduate Scholarship.  S.S.H. is supported
by an NSERC Discovery Grant and the Canada Research Chairs program.

\clearpage
\begin{deluxetable}{crl}
\tabletypesize{\footnotesize}
\tablewidth{0pt} \tablecaption{Net exposure times and off-axis angles for \chan\ data} \tablehead{
\colhead{Detector} & \colhead{Time (ks)} & \colhead{Observations
  (observation date, off-axis angle (arcmin))} } \startdata
ACIS-I, -100$^\circ$C & 41.9 & 158 (1999 Aug 25, $5.8$), 160 (1999 Aug 25, $6.7$), 161 (1999 Aug 25, $5.9$),\\
                      &      & 162 (1999 Aug 27, $1.7$) \\
ACIS-I, -110$^\circ$C & 100.4 & 1233 (1999 Nov 5, $1.7$), 1441 (1999 Nov 15, $9.3$), 1442 (1999 Nov 15, $5.8$),\\
                      &       & 1443 (1999 Nov 15, $5.9$), 1772 (2000 Jul 5, $7.3$), 1773 (2000 Jul 5, $5.8$),\\
                      &       & 1774 (2000 Jul 5, $4.3$), 1775 (2000 Jul 5, $2.5$), 1776 (2000 Jul 6, $0.8$),\\
                      &       & 1777 (2000 Jul 6, $5.8$), 1778 (2000 Jul 6, $4.3$), 1779 (2000 Jul 6, $2.5$) \\
ACIS-I, -120$^\circ$C & 156.5 & 1551 (2001 Mar 8, $2.5$), 1552 (2001 Jul 13, $2.5$), 1719 (2000 May 23, $7.3$),\\
                      &       & 1720 (2000 May 23, $5.8$), 1721 (2000 May 23, $4.3$), 1722 (2000 May 23, $2.5$),\\
                      &       & 1723 (2000 May 23, $0.8$), 1724 (2000 May 24, $5.8$), 1725 (2000 May 24, $4.3$),\\
                      &       & 1726 (2000 May 24, $2.5$), 2872 (2002 Sep 13, $3.7$), 3473 (2002 May 16, $7.0$),\\
                      &       & 3692 (2003 May 16, $2.5$), 3699 (2003 Nov 9, $3.7$), 5158 (2005 Feb 26, $3.8$),\\
                      &       & 5165 (2004 Mar 26, $3.8$), 6070 (2005 Feb 26, $3.8$), 6740 (2006 Feb 21, $3.8$) \\
ACIS-S, -100$^\circ$C & 36.8 & 159 (1999 Aug 23, $0.3$), 165 (1999 Aug 23, $6.2$), 1230 (1999 Aug 23, $0.3$) \\
ACIS-S, -110$^\circ$C & 68.2 & 1433 (1999 Nov 15, $1.2$), 1434 (1999 Nov 16, $6.2$), 1769 (2000 Jul 5, $1.3$),\\
                      &      & 1770 (2000 Jul 5, $1.3$), 1771 (2000 Jul 5, $1.3$), 1780 (2000 Jul 5, $4.7$),\\
                      &      & 1781 (2000 Jul 5, $4.7$), 1782 (2000 Jul 5, $4.7$)\\
ACIS-S, -120$^\circ$C & 174.8 & 1553 (2001 Mar 18, $1.3$), 1554 (2001 Jul 21, $1.2$), 1716 (2000 May 23, $1.3$),\\
                      &       & 1717 (2000 May 23, $1.3$), 1718 (2000 May 23, $1.3$), 1727 (2000 May 24, $4.7$),\\
                      &       & 1728 (2000 May 24, $4.6$), 1729 (2000 May 24, $4.6$), 1838 (2000 Sep 2, $1.2$),\\
                      &       & 1839 (2000 Sep 2, $2.4$), 1840 (2000 Sep 2, $5.3$), 2873 (2002 Sep 14, $1.2$),\\
                      &       & 3474 (2002 May 16, $7.0$), 3693 (2003 May 16, $1.2$), 3700 (2003 Nov 9, $1.2$),\\
                      &       & 4353 (2003 May 15, $5.2$), 5159 (2004 Oct 27, $1.2$), 5166 (2004 Mar 14, $1.2$),\\
                      &       & 6071 (2005 Feb 26, $1.3$), 6741 (2006 Feb 22, $1.3$) \\
HRC-I & 278.4 & 142 (2000 Feb 16, $0.3$), 143 (1999 Sep 4, $0.3$), 144 (2000 Sep 1, $0.3$),\\
      &       & 1242 (1999 Sep 4, $0.4$), 1298 (1999 Sep 4, $0.3$), 1406 (1999 Oct 25, $0.3$),\\
      &       & 1555 (2001 Mar 9, $0.3$), 1556 (2001 Jul 13, $0.3$), 2867 (2002 Mar 13, $0.3$),\\
      &       & 2874 (2002 Jul 15, $0.3$), 3694 (2003 May 15, $0.3$), 3701 (2003 Nov 9, $0.2$),\\
      &       & 5167 (2004 Mar 25, $0.3$), 6072 (2005 Feb 26, $0.3$), 6742 (2006 Feb 21, $0.3$)
\enddata
\label{table:exptime}
\end{deluxetable}

\begin{deluxetable}{lccc}
\tabletypesize{\footnotesize} 
\tablewidth{0pt} \tablecaption{The regions studied in \g21\ (see Fig.~1)} \tablehead{ 
\colhead{Region}  
&\colhead{Background Subtracted Count Rate\tablenotemark{a}} 
&\colhead{Total Exposure Time}
&\colhead{Area} \\
\colhead{} &\colhead{(cts/s)} &\colhead{(ks)} 
&\colhead{(arcmin$^2$)} } \startdata
\j1833\  &0.175 $\pm$ 0.024 &238.9 & 0.0035 \\
Southern Halo & (1.14 $\pm$ 0.19) $\times$10$^{-1}$ &255.6 & 6.481 \\
Brightest Knot & (1.44 $\pm$ 0.24) $\times$10$^{-2}$  &360.6 &0.073 \\
Eastern Limb	& (5.03 $\pm$ 0.85) $\times$10$^{-2}$	&315.6	&2.260 \\
\enddata
\tablenotetext{a}{in the energy range 0.5--8.0 keV}
\label{table:regions}
\end{deluxetable}

\begin{deluxetable}{llc}
\tabletypesize{\scriptsize}
\tablewidth{0pt} \tablecaption{Spectral fitting results for \j1833\tablenotemark{a}.} \tablehead{
  \colhead{} &\colhead{} &\colhead{Compact Source}  } \startdata
min \# cts/bin  &                               &50                     \\
\hline
power-law   &$N_{\mathrm H}$ (10$^{22}$ atoms cm$^{-2}$)            &2.24 (2.14 - 2.33) \\
        &$\Gamma$						&1.47 (1.41 - 1.52) \\
        &norm (10$^{-4}$)\tablenotemark{b}			&7.73 (7.11 - 8.40) \\
        &$\chi^{2}_{\nu}$ ($\nu$)                               &1.230 (1047)       \\
        &flux~(10$^{-12}$)\tablenotemark{c}			&3.2 $\pm$ 0.3       \\
        &luminosity (10$^{34}$~\lumunits)	                &1.63 $\pm$ 0.14	\\
\hline
power-law + blackbody   &$N_{\mathrm H}$ (10$^{22}$ atoms cm$^{-2}$)      &2.24 (frozen)      \\
        &$\Gamma$						&1.14 (1.07 - 1.19) \\
        &norm (10$^{-4}$)\tablenotemark{b}			&4.45 (3.59 - 5.26) \\
        &kT (keV)                                               &0.52 (0.48 - 0.55)\\
        &norm (10$^{-6}$)\tablenotemark{d}                      &9.06 (5.42 - 12.48)\\
        &$\chi^{2}_{\nu}$ ($\nu$)				&1.216 (1046)       \\
        &flux~(10$^{-12}$)\tablenotemark{c}			&3.3 $\pm$ 0.7	  \\
        &luminosity (10$^{34}$~\lumunits)	                &1.6 $\pm$ 0.4	               \\
        &thermal flux~(10$^{-13}$)\tablenotemark{c}             & 2.6 $\pm$ 1.1           \\
        &non-thermal flux~(10$^{-12}$)\tablenotemark{c}         & 3.0 $\pm$ 0.6          
\enddata
\tablenotetext{a}{All confidence ranges are 90\%.  All models were fit to 
the data in the range 0.5--8.0~keV.  The values given for luminosity assume a distance of 5~kpc to \g21.}
\tablenotetext{b}{Units for the normalization factor on the power-law
  model are \pownormunits.} 
\tablenotetext{c}{Observed flux in units of  \fluxunits.}
\tablenotetext{d}{The normalization factor on the blackbody model is L$_{39}$/D$^2_{10}$ where L$_{39}$=L/(10$^{39}$ erg s$^{-1}$) and D$_{10}$=D/(10 kpc).}
\label{table:fitresults1}
\end{deluxetable}

\begin{deluxetable}{llcccc}
\rotate
\tabletypesize{\tiny}
\tablewidth{0pt} \tablecaption{Spectral fitting results for the X-ray halo 
of \g21\tablenotemark{a}.} \tablehead{
\colhead{} &\colhead{}                           &\colhead{Southern Halo}  & \colhead{Brightest Northern Knot} &\colhead{Eastern Limb}  } \startdata
min \# cts/bin  &                                             &50                   &10                   &20                  \\
\hline
power-law           &$\Gamma$                                          &2.50 (2.45 - 2.54)  &2.72 (2.63 - 2.80) &2.13 (1.94 - 2.33)   \\
                &norm (10$^{-4}$)\tablenotemark{b}                 &14.4 (13.7 - 15.1)   &1.99 (1.82 - 2.17) &1.8 (1.4 - 2.3)    \\
                &$\chi^{2}_{\nu}$ ($\nu$)                          &1.08 (865)            &1.18 (606)         &0.539 (910)     \\
                &flux~(10$^{-13}$)\tablenotemark{c}                &14.2 $\pm$ 0.7       &1.5 $\pm$ 0.2      &3.0 $\pm$ 0.7   \\
                &luminosity (10$^{33}$~\lumunits)                  &14.6 $\pm$ 0.7        &1.9 $\pm$ 0.2      &2.3 $\pm$ 0.6   \\
\hline
pshock           &$kT$ keV                                         &3.83 (3.62 - 4.10)   &4.89 (4.18 - 5.99)   &7.5 (5.0 - 14.4)    \\
                &$n_et$ (10$^{9}$ cm$^{-3}$ s)                     &6.8 (6.2 - 7.4)      &20.1 (16.5 - 25.8)   &8.7 (5.6 - 13.2)    \\
                &norm (10$^{-3}$ cm$^{-4}$)                        &2.51 (2.39 - 2.63)    &0.22 (0.19 - 0.24) &3.7 (3.0 - 4.4)   \\
                &$\chi^{2}_{\nu}$ ($\nu$)                          &1.30 (864)            &0.959 (605)         &0.538 (909)   \\
                &flux~(10$^{-12}$)\tablenotemark{c}                &1.4 $\pm$ 0.1         &0.17 $\pm$ 0.02    &0.32 $\pm$ 0.06    \\
                &luminosity (10$^{34}$~\lumunits)                  &8.8 $\pm$ 0.7          &0.83$\pm$ 0.09    &1.3 $\pm$ 0.3 \\
                &Mg                                                &                      &1.08 (0.95 - 1.21)   &0.83 (0.22 - 1.44)\\
                &Si                                                &                      &0.96 (0.79 - 1.13)   &0.86 (0.25 - 1.47)\\
                &S                                                 &                      &0.52 (0.12 - 0.93)   &0.94 (0 - 2.48)\\
\hline
power-law + pshock    &$\Gamma$                                          &2.25 (2.08 - 2.36)   &2.21 (2.06 - 2.36) &    \\
                &norm (10$^{-4}$)\tablenotemark{b}                 &10.4 (9.0 - 12.1)   &1.08 (0.89 - 1.31) &    \\
                &$kT$ keV                                          &0.33 (0.25 - 0.45)   &0.20 (0.14 - 0.24) &    \\
                &$n_et$ (10$^9$ cm$^{-3}$ s)                       &0.1 ($<$ 1.3)       &7.1 (3.7 - 14.2)   &     \\
                &norm (10$^{-2}$ cm$^{-2}$)                        &2.6 (1.1 - 7.3)      &5.2 (3.4 - 21.4) &   \\
                &$\chi^{2}_{\nu}$ ($\nu$)                          &1.04 (862)            &0.960 (603)         &  \\
                &flux~(10$^{-13}$)\tablenotemark{c}                &15 $\pm$ 3           &1.7 $\pm$ 0.4      &   \\
                &luminosity (10$^{35}$~\lumunits)                  &0.19 (0.13 - 0.34)       &1.6 $\pm$ 1.5      &  \\
                &non-thermal flux~(10$^{-13}$)\tablenotemark{c}    &14 $\pm$ 2            &1.6 $\pm$ 0.3      &    \\
                &thermal flux~(10$^{-15}$)\tablenotemark{c}        &46 (19 - 129)          &9.3 $\pm$ 3.2      &   \\
                &Mg                                                &                     &0.73 (0.40 - 1.06)  &\\
                &Si                                                &                      &0.84 (0.32 - 1.35)  &\\
                &S                                                 &                     &107.1 (3.9 - 210.3) &\\
\hline
srcut           &$\alpha$                                          &0.5 (frozen)       &0.5 (frozen)       &0.3 (frozen) / 0.5 (frozen) / 0.8 (frozen)    \\
                &$\nu_{\mathrm{rolloff}}$ (10$^{17}$ Hz)           &1.26 (1.11 - 1.42) &0.72 (0.60 - 0.93) &2.2 (1.2 - 4.7) / 4.3 (2.1 - 13.5) / 8.7 (5.4 - 2400)   \\
                &norm (10$^{-2}$ Jy)                               &7.1 (6.9 - 7.3)    &1.47 (1.40 - 1.54) &0.014 (0.011 - 0.016) / 0.49 (0.41 - 0.56) / 140 (120 - 170)   \\
                &$\chi^{2}_{\nu}$ ($\nu$)                          &1.14 (865)         &1.24               &0.542 (910) / 0.541 (910) / 0.542 (910)   \\
                &flux~(10$^{-13}$)\tablenotemark{c}                &13.8 $\pm$ 0.5     &1.5 $\pm$ 0.5      &2.9 $\pm$ 0.5 / 2.9 $\pm$ 0.4 / 2.7 $\pm$ 0.5  \\
                &luminosity (10$^{33}$~\lumunits)                  &12.9 $\pm$ 0.5     &1.6 $\pm$ 0.7      &2.0 $\pm$ 0.3 / 2.1 $\pm$ 0.3 / 2.2 $\pm$ 0.4  \\
\hline
srcut + pshock    &$\alpha$                                          &0.5 (frozen)             &0.3 (frozen) / 0.5 (frozen) / 0.8 (frozen)       &   \\
                &$\nu_{\mathrm{rolloff}}$ (10$^{17}$ Hz)           &4.2 (2.9 - 8.7)          &2.2 (1.4 - 3.1) / 4.3 (2.4 - 8.4) / 18 (8 - 65)    &   \\
                &norm (10$^{-3}$ Jy)                               &24.2 (23.4 - 25.0)       &0.074 (0.048 - 0.117) / 2.6 (1.7 - 4.1) / 580 (380 - 850)     &  \\
                &$kT$ keV                                          &0.38 (0.33 - 0.43)       &0.21 (0.18 - 0.25) / 0.21 (0.17 - 0.25)  / 0.20 (0.16 - 0.24) &    \\
                &$n_et$ (10$^{9}$ cm$^{-3}$ s)                    &0.1 ($<$0.98)             &5.6 (3.3 - 13.2) / 5.8 (4.0 - 12.9) / 6.3 (3.4 - 24.8)   &     \\
                &norm (10$^{-2}$ cm$^{-5}$)                        &2.2 (1.3 - 2.9)          &3.9 (2.1 - 8.7) / 4.1 (2.1 - 9.6) / 4.6 (2.2 - 10.8)    &  \\
                &$\chi^{2}_{\nu}$ ($\nu$)                          &1.04 (862)               &0.963 (603) / 0.963 (603) / 0.962 (603)         &  \\
                &flux~(10$^{-13}$)\tablenotemark{c}                &14.7 $\pm$ 0.8           &1.6 $\pm$ 0.9 / 1.6 $\pm$ 0.6 / 1.6 $\pm$ 0.9      &    \\
                &luminosity (10$^{34}$~\lumunits)                  &1.8 $\pm$ 0.3             &13.6 $\pm$ 5.9 / 14.0 $\pm$ 7 / 15.0 $\pm$ 7.7      &    \\
                &non-thermal flux~(10$^{-13}$)\tablenotemark{c}    &13.9 $\pm$ 0.5           &1.5 $\pm$ 0.9 / 1.5 $\pm$ 0.5 / 1.5 $\pm$ 0.9      &   \\
                &thermal flux~(10$^{-14}$)\tablenotemark{c}        &7.3 (4.3 - 9.8)           &1.1 $\pm$ 0.5 / 1.0 $\pm$ 0.5 / 1.0 $\pm$ 0.7     &   \\
                &Mg                                                &                          &0.74 (0.41 - 1.07) / 0.74 (0.41 - 1.07) / 0.73 (0.40 - 1.06) &\\
                &Si                                                &                          &0.84 (0.35 - 1.33) / 0.84 (0.34 - 1.34) / 0.84 (0.34 - 1.35) &\\
                &S                                                 &                          &96.7 (10.3 - 183.1) / 98.9 (9.0 - 188.7) / 102.6 (5.9 - 199.3)  &\\
\enddata
\tablenotetext{a}{The column density $N_{\mathrm H}$ was fixed at 2.24\nHunits, the best fit value obtained from the pulsar.  All confidence ranges are 90\%.  All models were fit to 
the data in the range 0.5--8.0~keV.  The values given for luminosity assume a distance of 5~kpc to \g21.}
\tablenotetext{b}{Units for the normalization factor on the power-law
  model are \pownormunits.}
\tablenotetext{c}{Observed flux in \fluxunits.}
\label{table:fitresults2}
\end{deluxetable}

\clearpage

\end{document}